\documentclass[iop]{emulateapj}

\begin{document}
\title{The Void Galaxy Survey: Optical Properties and H I Morphology and Kinematics}

\author{K. Kreckel\altaffilmark{1,2}, E. Platen\altaffilmark{3}, M. A. Arag\'on-Calvo\altaffilmark{4}, J. H. van Gorkom\altaffilmark{1}, R. van de Weygaert\altaffilmark{3}, J. M. van der Hulst\altaffilmark{3}, B. Beygu\altaffilmark{3}}

\altaffiltext{1}{Department of Astronomy, Columbia University, Mail Code 5246, 550 West 120th Street, New York, NY 10027, USA; email: kreckel@mpia.de}
\altaffiltext{2}{Current address: Max Planck Institute for Astronomy, K\"{o}nigstuhl 17, 69117 Heidelberg, Germany}
\altaffiltext{3}{Kapteyn Astronomical Institute, University of Groningen, PO Box 800, 9700 AV Groningen, the Netherlands}
\altaffiltext{4}{The Johns Hopkins University, 3701 San Martin Drive, Baltimore, MD 21218, USA}


\begin{abstract}
We have carefully selected a sample of 60 galaxies that reside in the deepest underdensities of geometrically identified voids within the SDSS.  H \textsc{i} imaging of 55 galaxies with the WSRT reveals morphological and kinematic signatures of ongoing interactions and gas accretion.  
We probe a total volume of 485 Mpc$^3$ within the voids, with an angular resolution of 8 kpc at an average distance of 85 Mpc. We reach column density sensitivities of $5 \times 10^{19}$ cm$^{-2}$, corresponding to an H \textsc{i} mass limit of $3 \times 10^8$ M$_\sun$.  We detect H \textsc{i} in 41 galaxies, with total masses ranging from $1.7 \times 10^8$ to $5.5 \times 10^9$ M$_\sun$.  The upper limits on the 14 non-detections are not inconsistent with their luminosities, given their expected H \textsc{i} mass to light ratios.  
We find that the void galaxies are generally gas rich, low luminosity, blue disk galaxies, with optical and H \textsc{i} properties that are not unusual for their luminosity and morphology.  The sample spans a range of absolute magnitudes (-16.1 $>$ M$_r$ $>$ -20.4) and colors (0.06 $< g-r <$ 0.87), and includes disk and irregular galaxies.  We also identify three as early type galaxies, all of which are not detected in H \textsc{i}.  All galaxies have stellar masses less than $3 \times 10^{10}$ M$_\sun$, and many have kinematic and morphological signs of ongoing gas accretion, suggesting that the void galaxy population is still in the process of assembling. 
The small scale clustering in the void, within 600 kpc and 200 km s$^{-1}$, is similar to that in higher density regions, and we identify 18 H \textsc{i} rich neighboring galaxies in the voids.   Most are within 100 kpc and 100 km s$^{-1}$ of the targeted galaxy, and we find no significant population of H \textsc{i} rich low luminosity galaxies filling the voids, contrary to what is predicted by simulations.
\end{abstract}

\section{Introduction}
Voids represent a unique environment for the study of galaxy evolution \citep{weyplaten2009}.  The lower density environment, equivalent to a lower $\Omega_m$ universe \citep{Goldberg2004}, should result in shorter merger histories and slower evolution of galaxies.  As different mechanisms of gas accretion are believed to dominate in low mass halos, voids present a unique environment in which to search for signs of ongoing cold mode accretion \citep{Keres2005}.
The nature of these void galaxies also provides a test of cold dark matter (CDM) cosmology theory, which predicts that these underdense regions should be filled with low mass dark matter halos that are not observed as low luminosity galaxies \citep{Peebles2001}. This may have implications for galaxy formation through the suppression of star formation in the lowest mass halos \citep{Hoeft2006}, and may provide general insights into how galaxies populate dark matter halos \citep{Tinker2009}  

Observations of void galaxies selected by different methods consistently show that they are typically low luminosity and late type disk galaxies \citep{Grogin1999,Rojas2004}.  They are blue with elevated specific star formation rates \citep{Grogin2000,Rojas2005}, though it is not clear if these characteristics are different from low luminosity disk galaxies in average density environments \citep{Patiri2006b, Park2007}.  As the observed shift in the luminosity function results naturally from CDM cosmology \citep{Hoyle2005, Aragon2007, Kreckel2011a}, 
it is possible that the large scale underdensity surrounding void galaxies has little effect on the evolution of these systems. 

\begin{figure*}[!htb]
\centering
\includegraphics[width=6in]{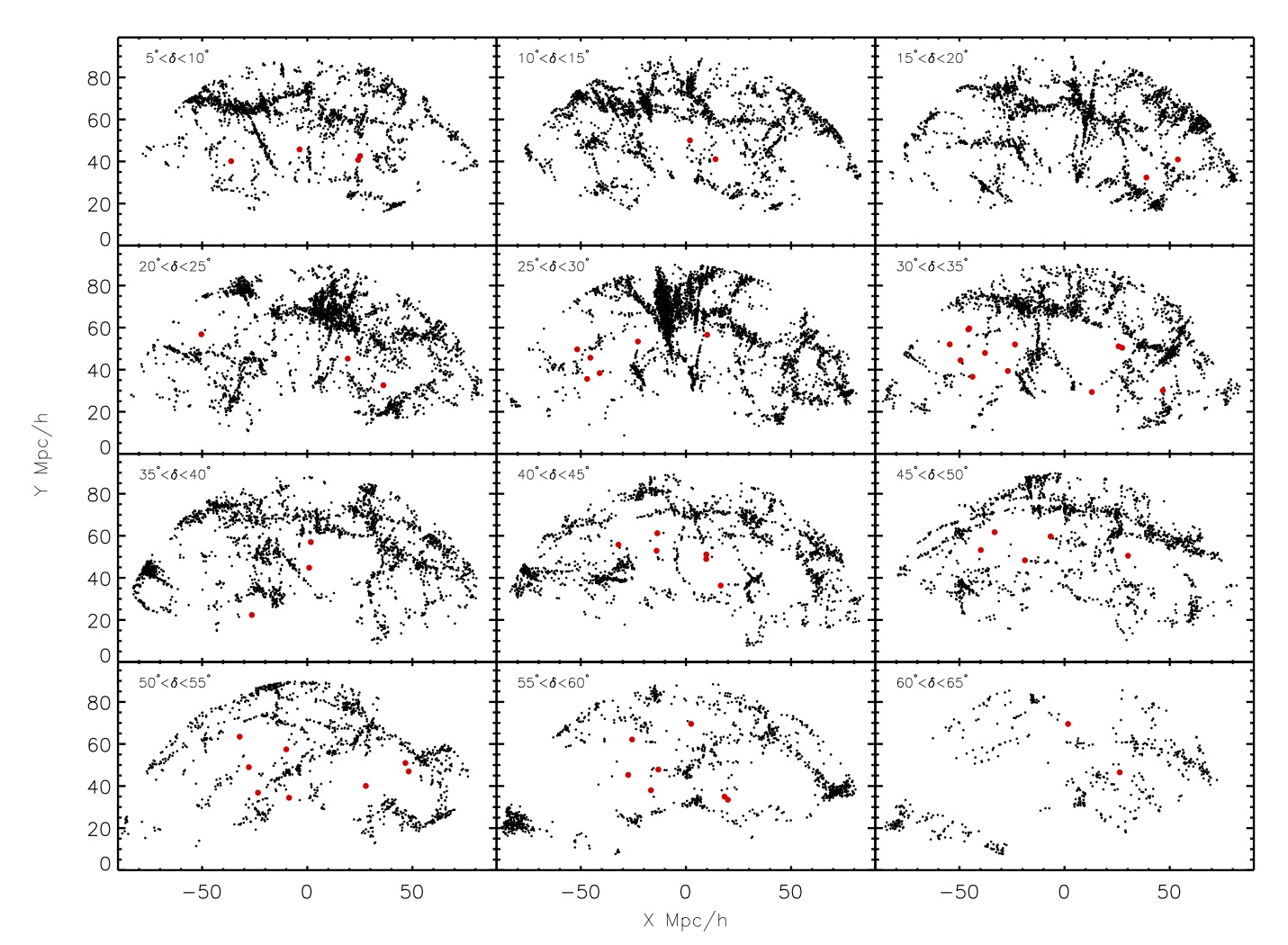}
\caption{The position of the VGS galaxies (red) within the SDSS galaxy distribution (black). Each panel shown a slice in declination of 5$^\circ$, with the specific range plotted given in the upper left corner.   
\label{fig:voids}}
\end{figure*}

We are undertaking a multi-wavelength study of 60 void galaxies as part of a new Void Galaxy Survey (VGS, see \citealt{Kreckel2011}). 
A variety of different techniques have been developed to identify voids and void galaxies within galaxy redshift surveys. \cite{Colberg2008} provides an interesting comparison of the results of different void-finding algorithms applied to a single set of simulation data.  These techniques generally agree in locating large scale underdensities, however the extent and shape of the identified voids differ dramatically.  We employ a powerful geometric method that makes no assumption about the size or shape of the voids within a galaxy distribution.  This unique approach allows us to robustly identify a void galaxy population suitable for systematic study.

The earliest void galaxy surveys were severely biased by the selection methods, targeting only emission-line \citep{Kirshner1981, Lee2000} or IRAS bright \citep{Szomoru1996} galaxies. 
Subsequent void galaxy studies based on more uniformly selected redshift surveys with denser coverage were also biased by the shallow depth available and select only the more luminous void galaxies \citep{Grogin1999}, which do not represent the bulk of the low luminosity void galaxy population. We use the Sloan Digital Sky Survey (SDSS) Data Release 7 (DR7), which contains uniform photometry and spectroscopic distances to all galaxies over a quarter of the sky that are brighter than 17.77 $r$-band magnitudes, and focus on relatively nearby galaxies (D $<$ 100 Mpc) allowing us to probe void galaxies to M$_r$ = -17.  
In this paper we discuss the SDSS optical properties of 60 void galaxies, carefully selected by their geometry to be centrally located within the deepest underdensities, and we discuss the neutral gas content in 55 of these systems observed with the Westerbork Synthesis Radio Telescope (WSRT).  

The WSRT employs a new backend that enables us to simultaneously image a wide velocity range encompassing the full radial extent of the void with high sensitivity, and good velocity and spatial resolution to trace gas morphology and kinematics of both the target and any neighboring galaxies within the void.  
Systematic H \textsc{i} imaging of void galaxies has been undertaken by \cite{Szomoru1996}, who identified 24 IRAS selected galaxies within the Bo\"{o}tes void.  Their targets were generally gas rich and late type, with H \textsc{i} morphologies consistent with similar galaxies in the field, though their sample suffers from contamination by galaxies in higher density substructures within the void \citep{Kreckel2011}.    Single dish observations of dwarf galaxies identify targets  in the void as having slightly higher H \textsc{i} mass to light ratio \citep{Huchtmeier1997, Pustilnik2002}, though the sample size is small.  H \textsc{i} blind observations of voids have not revealed a large number of uncataloged detections (\citealt{Weinberg1991}, \citealt{Saintonge2008}). 

In Section \ref{sec:selection} we describe our methods for identifying voids and void galaxies.  Section \ref{sec:observations} outlines the H \textsc{i} observations undertaken at the WSRT, and Section \ref{sec:results} details our results.  We discuss the significance and implications of this work in Section \ref{sec:discussion}, and include a full catalog of the H \textsc{i} and SDSS optical data for the sample in Appendix \ref{sec:appendix}.  We assume H$_0 = 70$ km s$^{-1}$ Mpc$^{-1}$.

\section{Sample selection}
\label{sec:selection}
A complete description of our void finding method and void galaxy selection criteria are presented in the pilot study for this project \citep{Kreckel2011}.  We summarize here the main points of our technique and refer the reader to the more detailed description previously published and references therein.

We apply the Delaunay Tessellation Field Estimator (DTFE\footnote{Public software available at http://www.astro.rug.nl/$\sim$voronoi/DTFE/dtfe.html \citep{Cautun2011}}; \citealt{Schaap2000, Schaap2007, weyschaap2009}) to the SDSS redshift survey.  We use these galaxies, independent of their luminosity, as tracers to recover the underlying galaxy density field within the local universe.  
Using redshift as a proxy for distance, we construct the Delaunay tessellation of the
       three-dimensional galaxy distribution and estimate the local density at each galaxy
       location on the basis of the inverse of the corresponding contiguous Voronoi cell. The
       density field throughout the sample volume consists of the linear piecewise interpolation
       of the density values over the corresponding dual Delaunay tessellation.  After weighing
       the resulting density field for the radial selection function of the SDSS survey, i.e. the
       decrease in luminosity sensitivity as a function of redshift, we apply a $1h^{-1}$Mpc
       Gaussian smoothing filter to  obtained galaxy density field. 

The void identification proceeds by treating the reconstructed density field as a landscape of hills and valleys that is slowly flooded.  As neighboring valleys fill and intervening hills are submerged we identify void substructure and boundaries at these higher density regions.  In this way we outline the void shapes, without imposing requirements of spherical voids or selection of a local minimum, and are able to unambiguously identify galaxies that reside within the voids.  This watershed transform results in basins that are identified as voids by the Watershed Voidfinder algorithm \citep{Platen2007}, and the corresponding boundaries are subsequently identified as walls and filaments using the Cosmic Spine formalism \citep{Aragon2010}.

We have selected a sample of 60 galaxies located in the local universe using voids identified within the SDSS DR7.  A redshift limit of $z < 0.025$ is imposed to maximize the H \textsc{i} sensitivity and spatial resolution considering the $\sim 20^{\prime\prime}$ resolution of the WSRT. We also exclude any galaxies within $\sim$750 km s$^{-1}$ of the line of sight of a cluster to avoid  contamination by redshift distortions, such as the finger of god effect.  An initial sample of galaxies is selected to have a density contrast,  $\delta \equiv \rho / \rho_u -1$, of less than -0.5, where $\rho_u$ is the mean density estimated for the SDSS.  We then choose only those galaxies most centrally located within their void and away from the SDSS boundaries so the surrounding void is well contained within the redshift survey.  All galaxies were selected purely on the grounds of their location within the geometrically determined outline of the voids in the SDSS galaxy distribution, with no bias for morphology, luminosity or color outside of the inherent SDSS redshift survey limitations.  This is clear when observing the range of galaxy type in the SDSS color images as scaled to the same physical size (see Figure \ref{fig:poststamps}).  The SDSS positions of all 60 galaxies are shown in Figure \ref{fig:voids}.

\section{Observations}
\label{sec:observations}
 We observed a total of 55 of the 60 VGS galaxies at the WSRT between 2006-2010, with the instrumental parameters given in Table \ref{tab:obsparams}. We list the properties measured from the SDSS in Table \ref{tab:params}.  All targets were observed in maxi-short configuration, with shortest baselines of  36, 54, 72 and 90 meters to maximize surface brightness sensitivity and a longest baseline of 2754 meters to achieve angular resolutions of $19^{\prime\prime} \times 19^{\prime\prime}/cos(\delta)$ using an optimal u,v taper for the detection of low surface brightness emission.  This corresponds to spatial scales of $\sim$8 kpc at 85 Mpc, the average distance for our sample. As the WSRT is an east-west array, we scheduled full 12 hour track observations to achieve complete UV coverage of each target, however a number of our observations were interrupted by priority target of opportunity observations.  Some, but not all, of our incomplete observations  were completed at later times.  The final total integration time for all galaxies not detected in H \textsc{i} is at least 10.5 hours.  The final total integration time for detected galaxies is at least 8 hours, with only VGS\_18 having just 5 hours.

\begin{table}[t!]
\begin{center}
\tabletypesize{\scriptsize}
\caption{Parameters of the WSRT observations
\label{tab:obsparams}}
\begin{tabular}{l l}
\tableline
\tableline
Configuration 		& 	Maxi-short \\
Date 			& 	2006-2010 \\
No. telescopes 		& 	13 \\
Total bandwidth 	& 	40 MHz \\
No. channels 		& 	4 $\times$ 512 \\
Shortest spacing 	& 	36 m\\
Longest spacing 	& 	2754 m\\
FWHM primary beam & 	36$^{\prime}$ \\
Synthesized beam 	&  	19$^{\prime\prime} \times$ 19$^{\prime\prime}$/sin($\delta$) \\
rms (robust = 1)	& 	0.4 mJy Beam$^{-1}$ \\
velocity resolution 	&	8.6 km s$^{-1}$ \\
\tableline
\end{tabular}
\end{center}
\end{table}

\begin{deluxetable*}{ l l l l r r r r r }
\centering
\tabletypesize{\tiny}
\tablecaption{Parameters of selected void galaxies taken from the SDSS catalog. 
\label{tab:params}}
\tablewidth{0pt}
\tablehead{
\colhead{Name} & \colhead{SDSS ID} & 
\colhead{ra} & \colhead{dec} & 
\colhead{$z$} &   \colhead{$r$} & \colhead{$g-r$} & \colhead{M$_r$} &  \colhead{$\delta$} \\
\colhead{} & \colhead{} & 
\colhead{(J2000)} & \colhead{(J2000)} & \colhead{} & 
\colhead{} &  \colhead{} & \colhead{} & \colhead{} 
}
\startdata
VGS\_01 & J083707.48+323340.8 &  08 37 07.5  & +32 33 41 &   0.018531 & 17.17 & 
 0.49 & -17.4 & -0.77 \\
VGS\_02 & J085453.60+181924.6 &   08 54 53.6  & +18 19 25
 & 0.022558 & 17.39 &  0.40 &  -17.6 & -0.83 \\
VGS\_03 & J090500.99+183128.4 &   09 05 01.0  & +18 31 28
 & 0.016868 & 17.51 &  0.46 &  -16.9 & -0.80 \\
VGS\_04 & J091355.47+244552.3 &   09 13 55.5  & +24 45 52
 & 0.016228 & 16.55 &  0.53 &  -17.8 & -0.75 \\
VGS\_05 & J092252.91+513243.7 &   09 22 52.9  & +51 32 44
 & 0.022437 & 15.03 &  0.76 &  -19.9 & -0.93 \\
VGS\_06 & J093602.69+515638.6 &   09 36 02.7  & +51 56 39
 & 0.023032 & 17.40 &  0.31 &  -17.6 & -0.85 \\
VGS\_07 & J100642.44+511623.9 &  10 06 42.5  & +51 16 24 &   0.016261 & 17.30 & 
 0.06 & -16.9 & -0.84 \\
VGS\_08 & J102235.27+453821.2 &   10 22 35.3  & +45 38 21
 & 0.019613 & 17.72 &  0.33 &  -16.9 & -0.82 \\
VGS\_09 & J102250.68+561932.1 &  10 22 50.7  & +56 19 32 &   0.012995 & 17.52 & 
 0.26 & -16.2 & -0.78 \\
VGS\_10 & J102316.63+091330.1 &   10 23 16.6  & +09 13 30
 & 0.015781 & 16.86 &  0.34 &  -17.4 & -0.84 \\
VGS\_11 & J102351.86+095914.8 &   10 23 51.9  & +09 59 15
 & 0.016486 & 16.05 &  0.51 &  -18.3 & -0.61 \\
VGS\_12 & J102819.23+623502.6 &  10 28 19.2  & +62 35 03 &   0.017804 & 17.41 & 
 0.26 & -17.0 & -0.74 \\
VGS\_13 & J103152.68+315034.6 &   10 31 52.7  & +31 50 35
 & 0.019126 & 16.35 &  0.48 &  -18.3 & -0.71 \\
VGS\_14 & J103506.46+550847.5 &  10 35 06.5  & +55 08 48 &   0.013154 & 16.72 & 
 0.28 & -17.1 & -0.81 \\
VGS\_15 & J103913.14+310650.4 &   10 39 13.1  & +31 06 50
 & 0.019071 & 15.40 &  0.56 &  -19.2 & -0.76 \\
VGS\_16 & J104807.05+430525.4 &   10 48 07.1  & +43 05 25
 & 0.013308 & 17.57 &  0.28 &  -16.2 & -0.88 \\
VGS\_17 & J105042.23+315119.5 &   10 50 42.2  & +31 51 20
 & 0.010695 & 16.95 &  0.18 &  -16.4 & -0.77 \\
VGS\_18 & J105352.26+214549.6 &   10 53 52.3  & +21 45 50
 & 0.016383 & 17.47 &  0.39 &  -16.8 & -0.78 \\
VGS\_19 & J111029.61+134558.1 &   11 10 29.6  & +13 45 58
 & 0.014467 & 16.52 &  0.39 &  -17.5 & -0.52 \\
VGS\_20 & J114124.92+415221.9 &   11 41 24.9  & +41 52 22
 & 0.016630 & 18.04 &  0.17 &  -16.3 & -0.92 \\
VGS\_21 & J114303.00+404939.1 &   11 43 03.0  & +40 49 39
 & 0.017344 & 15.02 &  0.83 &  -19.4 & -0.87 \\
VGS\_22 & J114535.20+270742.5 &   11 45 35.2  & +27 07 43
 & 0.019162 & 17.57 &  0.37 &  -17.1 & -0.44 \\
VGS\_23 & J121716.54+124742.8 &   12 17 16.5  & +12 47 43
 & 0.016688 & 15.24 &  0.41 &  -19.1 & -0.82 \\
VGS\_24 & J121754.98+583935.6 &   12 17 55.0  & +58 39 36
 & 0.023207 & 14.90 &  0.55 &  -20.1 & -0.80 \\
VGS\_25 & J121908.24+372644.1 &   12 19 08.2  & +37 26 44
 & 0.019014 & 17.64 &  0.26 &  -17.0 & -0.86 \\
VGS\_26 & J122032.40+604958.4 &   12 20 32.4  & +60 49 58
 & 0.023186 & 16.38 &  0.46 &  -18.7 & -0.84 \\
VGS\_27 & J122123.12+393659.4 &   12 21 23.1  & +39 36 59
 & 0.014941 & 17.77 &  0.35 &  -16.3 & -0.71 \\
VGS\_28 & J124420.87+082412.5 &   12 44 20.9  & +08 24 13
 & 0.015286 & 17.61 &  0.69 &  -16.5 & -0.82 \\
VGS\_29 & J125119.35+480144.3 &   12 51 19.4  & +48 01 44
 & 0.020022 & 16.28 &  0.36 &  -18.4 & -0.65 \\
VGS\_30 & J130526.08+544551.9 &  13 05 26.1  & +54 45 52 &   0.019435 & 18.05 & 
 0.22 & -16.6 & -0.89 \\
VGS\_31 & J131606.19+413004.2 &   13 16 06.2  & +41 30 04
 & 0.020903 & 14.75 &  0.32 &  -20.1 & -0.64 \\
VGS\_32 & J132232.48+544905.5 &  13 22 32.5  & +54 49 06 &   0.011835 & 14.17 & 
 0.53 & -19.4 & -0.80 \\
VGS\_33 & J132505.06+430405.9 &   13 25 05.1  & +43 04 06
 & 0.018236 & 17.61 &  0.45 &  -16.9 & -0.82 \\
VGS\_34 & J132718.56+593010.2 &  13 27 18.6  & +59 30 10 &   0.016539 & 15.22 & 
 0.87 & -19.1 & -0.81 \\
VGS\_35 & J135113.62+453509.2 &  13 51 13.6  & +45 35 09 &   0.017299 & 16.36 & 
 0.41 & -18.0 & -0.70 \\
VGS\_36 & J135535.46+593041.3 &  13 55 35.5  & +59 30 41 &   0.022398 & 16.46 & 
 0.36 & -18.5 & -0.75 \\
VGS\_37 & J135836.29+292121.4 &   13 58 36.3  & +29 21 21
 & 0.019354 & 17.06 &  0.35 &  -17.6 & -0.87 \\
VGS\_38 & J140034.49+551515.1 &  14 00 34.5  & +55 15 15 &   0.013820 & 16.95 & 
 0.35 & -16.9 & -0.64 \\
 VGS\_39 & J140328.54+324151.7 &   14 03 28.5  & +32 41 52
 & 0.019037 & 15.25 &  0.81 &  -19.4 & -0.74 \\
VGS\_40 & J141326.46+503841.6 &   14 13 26.5  & +50 38 42
 & 0.023717 & 16.90 &  0.50 &  -18.2 & -0.89 \\
VGS\_41 & J141916.95+472839.1 &   14 19 17.0  & +47 28 39
 & 0.023371 & 17.21 &  0.43 &  -17.8 & -0.92 \\
VGS\_42 & J142416.41+523208.3 &  14 24 16.4  & +52 32 08 &   0.018762 & 15.88 & 
 0.61 & -18.7 & -0.68 \\
VGS\_43 & J142540.61+443835.2 &   14 25 40.6  & +44 38 35
 & 0.021455 & 17.85 &  0.36 &  -17.0 & -0.86 \\
VGS\_44 & J143052.33+551440.0 &  14 30 52.3  & +55 14 40 &   0.017656 & 14.92 & 
 0.46 & -19.5 & -0.89 \\
VGS\_45 & J143553.77+524400.6 &  14 35 53.8  & +52 44 01 &   0.014553 & 17.33 & 
 0.26 & -16.7 & -0.62 \\
VGS\_46 & J144338.46+322002.7 &   14 43 38.5  & +32 20 03
 & 0.015915 & 16.80 &  0.35 &  -17.4 & -0.63 \\
VGS\_47 & J145314.59+462910.8 &   14 53 14.6  & +46 29 11
 & 0.022155 & 14.56 &  0.75 &  -20.4 & -0.89 \\
VGS\_48 & J145450.57+305729.0 &   14 54 50.6  & +30 57 29
 & 0.024966 & 17.13 &  0.64 &  -18.1 & -0.88 \\
VGS\_49 & J145659.94+313308.5 &   14 56 59.9  & +31 33 09
 & 0.024920 & 15.51 &  0.41 &  -19.7 & -0.85 \\
VGS\_50 & J145909.32+324756.3 &   14 59 09.3  & +32 47 56
 & 0.020358 & 15.35 &  0.72 &  -19.4 & -0.85 \\
VGS\_51 & J151211.61+243344.1 &   15 12 11.6  & +24 33 44
 & 0.025307 & 17.02 &  0.23 &  -18.3 & -0.93 \\
VGS\_52 & J151410.95+064449.0 &   15 14 11.0  & +06 44 49
 & 0.018018 & 17.61 &  0.27 &  -16.9 & -0.90 \\
VGS\_53 & J152523.40+291018.8 &   15 25 23.4  & +29 10 19
 & 0.021473 & 15.65 &  0.54 &  -19.2 & -0.79 \\
VGS\_54 & J153035.83+264408.5 &   15 30 35.8  & +26 44 09
 & 0.023895 & 16.28 &  0.70 &  -18.9 & -0.68 \\
VGS\_55 & J153132.44+343055.8 &   15 31 32.4  & +34 30 56
 & 0.025146 & 16.27 &  0.46 &  -19.0 & -0.94 \\
VGS\_56 & J153341.47+280843.5 &   15 33 41.5  & +28 08 44
 & 0.018715 & 15.79 &  0.74 &  -18.8 & -0.82 \\
VGS\_57 & J153821.22+331105.1 &   15 38 21.2  & +33 11 05
 & 0.022171 & 14.59 &  0.53 &  -20.4 & -0.69 \\
VGS\_58 & J154452.18+362845.6 &  15 44 52.2  & +36 28 46 &   0.011522 & 15.70 & 
 0.35 & -17.8 & -0.88 \\
VGS\_59 & J154615.07+332017.8 &   15 46 15.1  & +33 20 18
 & 0.019028 & 17.81 &  0.39 &  -16.8 & -0.67 \\
VGS\_60 & J155721.55+254718.9 &   15 57 21.6  & +25 47 19
 & 0.019646 & 15.85 &  0.81 &  -19.0 & -0.78 \\

\enddata
\tablecomments{Units of right ascension are hours, minutes, and 
seconds, and units of declination are degrees, arcminutes, and arcseconds.   $g$ and $g-r$ are drawn from the apparent model magnitudes as measured by the SDSS DR7. Absolute magnitudes have been corrected for galactic extinction.  $\delta$ gives the filtered density contrast at R$_f$ = 1 h$^{-1}$ Mpc.}
\end{deluxetable*}

Target observations were done with two polarizations and 512 channels of 19.5 kHz each, for a total bandwidth of 10 MHz.  
All data reductions were done using AIPS.  The absolute flux scale was determined using \cite{Baars1977} and corrected to the observing frequency. Two of the flux calibrators 3C48, 3C286, 3C147 and CTD93 were observed with 15 minute snapshots, one before and one after the target observation, which were also used for phase and bandpass calibration.  For the non-standard calibrator CTD93 a 20 cm flux of 4.83 Jy was assumed, as given in the VLA calibrator manual.

Continuum emission is subtracted in the UV plane using the emission free channels of each cube.  Image cubes with a robust parameter of 1 were CLEANed down to 0.5 mJy beam$^{-1}$ ($\sim$1 $\sigma$) based on a box around the emission region of each cube, and were Hanning smoothed to a typical velocity resolution of 8.6 km s$^{-1}$.  Zeroth and first moment maps were made using masking by a data cube Hanning and Gaussian smoothed to 25 km s$^{-1}$ and 30$^{\prime\prime}$, respectively, and 1.5$\sigma$ clipping.  We reach typical column density sensitivities of 5 $\times$ 10$^{19}$ cm$^{-2}$.

Given the 36$^\prime$ full width half maximum of the WSRT primary beam and the 10 MHz bandwidth, with each observation we probe a total volume covering $\sim$1.2 Mpc and $\sim$1,200 km s$^{-1}$ at 85 Mpc, the average distance of our sample.  
Blind H \textsc{i} detections were identified using data cubes smoothed to 20 km s$^{-1}$ and 30$^{\prime\prime}$, increasing the 1$\sigma$ sensitivity to $\sim$0.25 mJy beam$^{-1}$, and 5$\sigma$ clipping in forming a zeroth moment map.  We consider only detections within 25$^\prime$ of the beam center, which corresponds to a factor of five reduced sensitivity. Candidate objects were then confirmed by eye to be extended both spatially and in velocity, and those confirmed to be environmentally located within the void are included here as companion void galaxies (Table \ref{tab:company}).

\subsection{HI properties}
H \textsc{i} parameters of these galaxies are calculated following the methods described in the pilot study \citep{Kreckel2011}. H \textsc{i} properties for all target and companion galaxies are listed in Table \ref{tab:vgs}. H \textsc{i} mass is calculated as M$_{\textrm{H \textsc{i}}} = 2.36 \times 10^5~d^2 \int S~dv$ M$_\sun$, where d is distance in Mpc, S is flux density in Jy and $dv$ is in km s$^{-1}$.   For targets at an average distance of 85 Mpc and assumed velocity width of 100 km s$^{-1}$ the 3$\sigma$ upper limit on detections is $\sim3 \times 10^8$ M$_\sun$.  We measure the 20\% and 50\% H \textsc{i} line width, W$_{20}$ and W$_{50}$, of each galaxy by constructing a global H \textsc{i} profile, summing the flux in each channel for a boxed area surrounding the galaxy. Errors of 15 km s$^{-1}$ reflect uncertainties of one channel on either side.

\begin{deluxetable*}{ l l l l r r r r c c c c }[htb!]
\tabletypesize{\scriptsize}
\tablecaption{Companion galaxy parameters taken from the SDSS catalog
\label{tab:company}}
\tablewidth{0pt}
\tablehead{
\colhead{name} & \colhead{SDSS ID} &
\colhead{ra} & \colhead{dec} & 
\colhead{$r$} & \colhead{$g - r$} &
\colhead{M$_r$} & 
\colhead{$\Delta \theta$} & \colhead{$\Delta$d} &
\colhead{$\Delta$ v}  & \colhead{$\delta$} \\
\colhead{} & \colhead{} & 
\colhead{(J2000.0)} & \colhead{(J2000.0)} & 
\colhead{} & \colhead{} & \colhead{} &
\colhead{($^\prime$)} & \colhead{(kpc)} &
\colhead{(km s$^{-1}$)} & \colhead{}
}
\startdata
VGS\_07a &   J100519.69+511038.3 &  10 05 19.7  & +51 10 38 &  20.12 &  0.03 &  -14.1 &  14.2 &   288 &          -21 & -0.84 \\
VGS\_09a &   J102241.41+561208.5  &  10 22 41.4  & +56 12 09 &  22.26 &  0.02 &  -11.4 &   7.5 &   121 &          -85 & -0.78\\
VGS\_10a &   J102308.72+085847.1 &  10 23 08.7  & +08 58 47  &  16.31
 &  0.31 &  -17.9 &  14.8 &   292 &           20 & -0.84 \\
VGS\_26a &   J122105.48+610514.2 &  12 21 05.5  & +61 05 14  &  13.79
 &  0.52 &  -21.0 &  15.8 &   455 &         -856 & -0.55 \\
VGS\_30a &   J130531.13+544553.8  &  13 05 31.1  & +54 45 54 &  18.39 &  0.29 &  -16.2 &   0.7 &    17 &          -74 & -0.89\\
VGS\_31a &   J131614.69+412940.0 &  13 16 14.7  & +41 29 40  &  14.38
 &  0.50 &  -20.4 &   1.6 &    42 &          -10 & -0.64 \\
VGS\_31b &   J131559.18+412955.9 &  13 15 59.2  & +41 29 56  &  16.78
 &  0.21 &  -18.0 &   1.3 &    34 &           50 & -0.64 \\
VGS\_34a &   J132640.92+593202.5  &  13 26 40.9  & +59 32 03 &  20.31 &  0.10 &  -14.0 &   5.1 &   104 &           42 & -0.81\\
VGS\_36a & J135533.54+593110.7 & 13 55 33.5 & +59 31 11 & 19.10 & 0.14 & -15.8 & 0.6  &  17 & 0 & -0.75\\
VGS\_37a &   J135836.06+292321.3 &  13 58 36.1  & +29 23 21  &  16.26
 &  0.36 &  -18.3 &   2.0 &    48 &         -114 & -0.82 \\
VGS\_38a &   J140032.44+551445.9  &  14 00 32.4  & +55 14 46 &  17.59 &  0.15 &  -16.1 &   0.6 &     9 &          -21 & -0.64 \\
VGS\_38b &   J140025.68+551318.5  &  14 00 25.7  & +55 13 19 &  18.82 &  0.20 &  -14.9 &   2.3 &    37 &           10 & -0.64 \\
VGS\_39a &   J140321.80+324530.3 &  14 03 21.8  & +32 45 30  &  19.48
 & -0.04 &  -15.2 &   3.9 &    92 &           73 & -0.74 \\
VGS\_51a &   J151230.45+243352.1 &  15 12 30.4  & +24 33 52  &  20.66
 &  0.05 &  -14.6 &   4.3 &   133 &           19 & -0.85 \\
VGS\_53a &   J152641.63+292359.8 &  15 26 41.6  & +29 24 00  &  16.78
 &  0.24 &  -18.1 &  21.9 &   586 &          -26 & -0.79 \\
VGS\_54a &   J153035.99+264446.1 &  15 30 36.0  & +26 44 46  &  19.05
 &  0.42 &  -16.1 &   0.6 &    18 &          -98 & -0.65 \\
VGS\_56a &   J153443.07+281309.6 &  15 34 43.1  & +28 13 10  &  18.91
 &  0.29 &  -15.7 &  14.3 &   333 &           47 & -0.83 \\
VGS\_57a &   J153836.08+331637.6 &  15 38 36.1  & +33 16 38  &  17.83
 &  0.13 &  -17.1 &   6.4 &   175 &          -72 & -0.61 \\

\enddata
\tablecomments{Units of right ascension are hours, minutes, and 
seconds, and units of declination are degrees, arcminutes, and arcseconds.   $g$ and $g-r$ are drawn from the apparent model magnitudes as measured by the SDSS DR7. Absolute magnitudes have been corrected for galactic extinction.   $\Delta \theta$, $\Delta$d and $\Delta$v list the displacement from the beam center, projected sky separation, and velocity separation, respectively, between the target and companion galaxy.  $\delta$ gives the filtered density contrast at R$_f$ = 1 h$^{-1}$ Mpc. }
\end{deluxetable*}

As in our pilot study, the majority of the targets are incompletely resolved by the WSRT, many extending no more than 2 or 3 times the H \textsc{i} beam. To estimate the radial extent of the H \textsc{i} we apply the iterative deconvolution method described by \cite{Lucy1974}, which sums the H \textsc{i} total intensity along the disk minor axis to create a one dimensional strip profile of the disk.  This profile is then iteratively fit assuming an axisymmetric disk to recover the face-on H \textsc{i} radial surface density profile (see Figure \ref{fig:surfdens}).  We measure the H \textsc{i} radius, r$_{\textrm{H \textsc{i}}}$, as that position where the surface density falls below 1 M$_\sun$ pc$^{-2}$, and quote an upper limit where we are resolution limited, with typical errors of 5$^{\prime\prime} \sim$~2 kpc.  

\begin{figure}[t!]
\centering
\includegraphics[width=3.2in,clip=true]{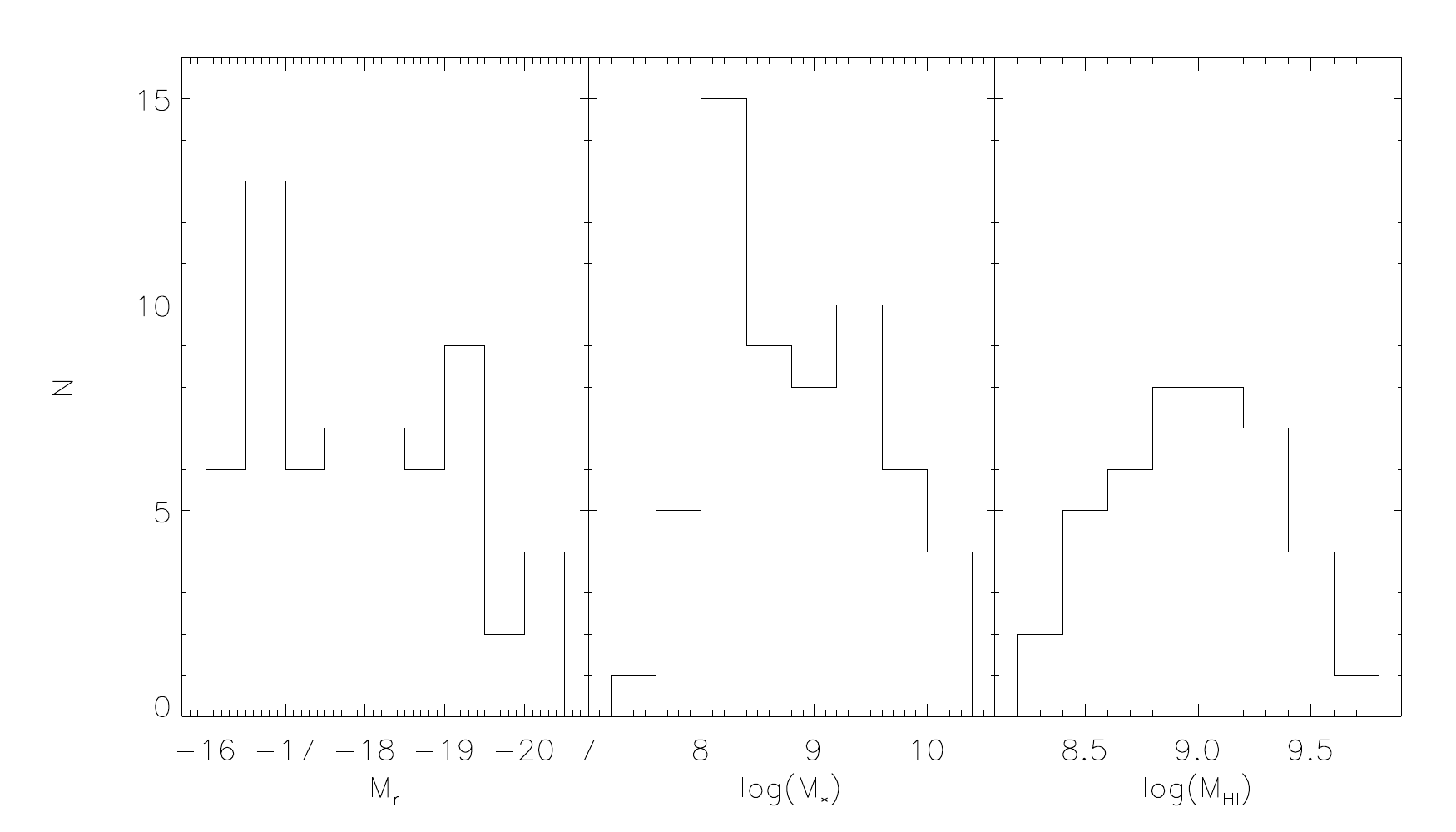}
\caption{Distributions for the absolute magnitude, stellar mass and H \textsc{i} masses in the VGS.
\label{fig:hists}}
\end{figure}

Inclinations are calculated from SDSS $r$-band isophotal radii under the assumption that galaxy disks are intrinsically oblate and axisymmetric with a three dimensional axis ratio of q$_o = 0.19$ \citep{Geha2006}, as
\begin{equation}
\textrm{sin}(i) = \sqrt{\frac{1-(b/a)^2}{1-q_o^2}}.
\end{equation}
Adopting a value of q$_o = 0.3$ changes the resulting inclinations by less than 10\%.  Position angles are determined by considering both the $r$-band optical and H \textsc{i} kinematic major axes. In cases where they disagree significantly (VGS\_12) we choose the orientation of the H \textsc{i} disk.

\subsection{SDSS optical parameters}
\label{sec:sdssprop}
As part of the SDSS survey, optical spectra were obtained for each galaxy using a 3$^{\prime\prime}$ fiber, which in addition to being used to determine redshift, may be used to estimate other stellar properties.  We use the publicly available MPA-JHU catalog for SDSS DR7 \footnote{http://www.mpa-garching.mpg.de/SDSS/DR7/} to obtain estimates of the star formation rate (SFR) and stellar mass in all void and control galaxies (Table \ref{tab:sf}).
Star formation rates are calculated from the emission line features and aperture corrected to estimate the total SFR for the galaxy, using the method developed by \cite{Brinchmann2004}.  Stellar masses are obtained using fits to the \textit{ugriz} photometry following the philosophy of \cite{Kauffmann2003} and \cite{Salim2007}, and correcting for the contribution by emission lines to the broadband magnitudes using the fiber spectra.  Errors in the SFR are relatively large, roughly a factor of two.  Errors on the stellar mass estimate are typically $\sim$0.1 dex, with significantly larger scatter for the lowest (M$_* \sim 10^8$ M$_\sun$) stellar masses.

\begin{deluxetable*}{ l r r r r r r r r r r r r}
\tabletypesize{\tiny}
\tablecaption{H \textsc{i} properties of targeted void galaxies and companions. 
\label{tab:vgs}} 
\tablewidth{0pt}
\tablehead{
\colhead{Name} & \colhead{t} & \colhead{$M_{\textrm{H \textsc{I}}}$} & \colhead{$V_{sys}$} & \colhead{D} & 
\colhead{r$_{90}$} & \colhead{r$_{\textrm{H \textsc{I}}}$} & 
\colhead{$W_{50}$} & \colhead{$W_{20}$} &\colhead{i} & \colhead{P.A.} & \colhead{M$_{dyn}$}  & \colhead{M$_{\textrm{H \textsc{I}}}$/L$_r$}  \\
\colhead{} & \colhead{(hr)} & \colhead{($10^8$ M$_\sun$)} & \colhead{(km s$^{-1}$)} & \colhead{(Mpc)} &
\colhead{(kpc)} & \colhead{(kpc)} & \colhead{(km s$^{-1}$)} & \colhead{(km s$^{-1}$)} & \colhead{($^\circ$)} & \colhead{($^\circ$)} &  
\colhead{($10^{10}$ M$_\sun$)} & \colhead{M$_\sun$/L$_\sun$} 
}
\startdata

VGS\_01 &   12.0 &  $<$  2.1  &            - &   - &  2.4 &  - &            - &           - & 60 & - & - &   - \\
VGS\_02 &  12.0 &  6.49 $\pm$ 1.85 &         6637 &           95 &   3.4 & $<$   7.6 & 
  111 &   136 &           47 &  80 & $<$  1.00 &   0.80 \\
VGS\_03 &  12.0 &  $<$ 1.85 &  -  &  -  &   1.8 &  -  &  -  &  -  &           55 &  -  & 
 -  &  -  \\
VGS\_04 &  12.0 &  $<$ 1.71 &  -  &  -  &   1.5 &  -  &  -  &  -  &           46 &  -  & 
 -  &  -  \\
VGS\_05 &  12.0 &  $<$ 3.27 &  -  &  -  &   6.1 &  -  &  -  &  -  &           37 &  -  & 
 -  &  -  \\
VGS\_06 &  12.0 & 15.63 $\pm$ 2.52 &         6883 &           98 &   3.9 &  12.9 &   180
 &   187 &           71 & 282 &  2.71 &   1.90 \\
VGS\_07 &   12.0 & 8.6 $\pm$  0.8 &         4901 &  70 &  2.6 &  9.2 &          119 &          144 &           50 & 270 &    1.28 &   1.9 \\
VGS\_07a & 12.0 & 3.23 $\pm$ 0.65 &     4880 &       69 & 3.5 &  -  &           41 &           49 &  -  &  - & -  &  9.7 \\ 
VGS\_08 &  12.0 &  3.95 $\pm$ 1.51 &         5853 &           84 &   4.4 & $<$   5.7 & 
   84 &   126 &           51 &  12 & $<$  0.39 &   0.89 \\
VGS\_09 & 12.0 &  10.3 $\pm$  1.0 &         3881 &  55 &  3.1 &  9.1 &          110 &          126 &           68 & 105 &    0.74 &   4.4 \\
VGS\_09a & 12.0 & 0.57 $\pm$ 0.14 &     3796 &       54 & 1.5 &  -  &           41 &           49 &  -  &  - & -  & 20.2 \\ 
VGS\_10 &  12.0 & 15.01 $\pm$ 2.00 &         4737 &           68 &   5.4 &  10.2 &   152
 &   176 &           87 & 310 &  1.36 &   2.21 \\
VGS\_10a &  12.0 &  16.73 $\pm$   1.91 &         6637 &           95 &   6.5 & 
 -  &  177 & 193   &  -  &  -  &  -  &   1.5 \\
VGS\_11 &  12.0 & 21.73 $\pm$ 2.11 &         4907 &           70 &   4.2 &  17.0 &    75
 &   100 &           18 &   5 &  6.16 &   1.38 \\
VGS\_12 &  12.0 & 29.9 $\pm$  4.2 &         5316 &  76 &  2.4 & 18.0 &          154 &          179 &           29 &   5 & 10.55 &   6.2 \\
VGS\_13 &  12.0 & 12.35 $\pm$ 2.45 &         5691 &           81 &   3.5 &  12.8 &   213
 &   229 &           68 & 275 &  3.89 &   0.82 \\
VGS\_14 &   12.0 & 6.4 $\pm$  1.0 &         3933 &  56 &  3.3 &  8.7 &          110 &          134 &           67 & 40 &    0.72 &   1.3 \\
VGS\_15 &   0.0 &  -  &  -  &  -  &   7.6 &  -  &  -  &  -  &           53 &  -  &  -  & 
 -  \\
VGS\_16 &  12.0 &  $<$ 1.16 &  -  &  -  &   1.7 &  -  &  -  &  -  &           60 &  -  & 
 -  &  -  \\
VGS\_17 &   0.0 &  -  &  -  &  -  &   3.4 &  -  &  -  &  -  &           79 &  -  &  -  & 
 -  \\
VGS\_18 &   5.0 &  3.91 $\pm$ 1.17 &         4850 &           69 &   3.5 & $<$   7.1 & 
  118 &   134 &           82 & 312 & $<$  0.58 &   0.99 \\
VGS\_19 &  12.0 &  3.16 $\pm$ 0.76 &         4318 &           62 &   1.7 & $<$   1.5 & 
   32 &   125 &           60 & 230 & $<$  0.01 &   0.43 \\
VGS\_20 &   0.0 &  -  &  -  &  -  &   1.1 &  -  &  -  &  -  &           43 &  -  &  -  & 
 -  \\
VGS\_21 &   8.0 & 21.45 $\pm$ 3.58 &         5171 &           74 &   8.8 &  15.8 &   314
 &   330 &           79 & 288 &  9.38 &   0.51 \\
VGS\_22 &  12.0 &  $<$ 2.62 &  -  &  -  &   1.6 &  -  &  -  &  -  &           63 &  -  & 
 -  &  -  \\
VGS\_23 &   8.5 & 38.52 $\pm$ 4.57 &         4990 &           71 &   4.5 &  16.9 &   203
 &   236 &           49 &  75 &  7.11 &   1.16 \\
VGS\_24 &  12.0 &  $<$ 3.35 &  -  &  -  &   4.6 &  -  &  -  &  -  &           35 &  -  & 
 -  &  -  \\
VGS\_25 &  12.0 &  1.69 $\pm$ 0.64 &         5680 &           81 &   2.3 & $<$   2.0 & 
  135 &   135 &           43 & 325 & $<$  0.46 &   0.37 \\
VGS\_26 &  10.5 & 14.90 $\pm$ 3.23 &         6942 &           99 &   5.8 &  12.5 &   206
 &   231 &           69 &  75 &  3.55 &   0.69 \\
VGS\_26a &  10.5 & 100.25 $\pm$  18.28 &         4907 &           70 &   6.8 & 
 -  &  171  & 257  &  -  &  -  &  -  &   0.6 \\
VGS\_27 &  12.0 &  3.32 $\pm$ 0.75 &         4421 &           63 &   1.7 & $<$   3.7 & 
   41 &    91 &           55 &  55 & $<$  0.05 &   1.31 \\
VGS\_28 &  12.0 &  $<$ 1.67 &  -  &  -  &   4.2 &  -  &  -  &  -  &           54 &  -  & 
 -  &  -  \\
VGS\_29 &   0.0 &  -  &  -  &  -  &   3.8 &  -  &  -  &  -  &           65 &  -  &  -  & 
 -  \\
VGS\_30 &   12.0 & 5.5 $\pm$  1.0 &         5666 &  81 &  3.7 &  $<$  7.5 &           50 &           93 &           77 &  180 & $<$     0.12 &   1.7 \\
VGS\_30a & 12.0 & 4.52 $\pm$ 0.79 &     5592 &       79 & 2.1 &  -  &           24 &           41 &  -  &  - & -  &  2.1 \\ 
VGS\_31 &  12.0 & 19.89 $\pm$ 2.90 &         6247 &           89 &   4.0 & $<$  10.4 & 
  153 &   178 &           52 & 155 & $<$  2.31 &   0.25 \\
VGS\_31a &  12.0 &  14.63 $\pm$   1.97 &         5691 &           81 &   6.5 & 
 -  &  196 & 298  &  -  &  -  &  -  &   0.1 \\
VGS\_31b &  12.0 &   1.66 $\pm$   0.95 &         4850 &           69 &   2.1 & 
 -  &  110  &  170  &  -  &  -  &  -  &   0.1 \\
VGS\_32 &  12.0 & 38.0 $\pm$  4.5 &         3522 &  50 &  4.1 & 19.0 &          171 &          187 &           46 &   310 &  6.29 &   0.9 \\
VGS\_33 &  12.0 &  8.39 $\pm$ 1.81 &         5443 &           78 &   2.0 & $<$   8.3 & 
  110 &   117 &           60 & 285 & $<$  0.77 &   1.97 \\
VGS\_34 &  12.0 & 23.9 $\pm$  2.9 &         4917 &  70 &  3.6 & 10.2 &          231 &          299 &           50 & 280 &     5.34 &   0.7 \\
VGS\_34a & 12.0 & 0.50 $\pm$ 0.16 &     4959 &       70 & 1.1 &  -  &           33 &           58 &  -  &  - & -  &  1.7 \\ 
VGS\_35 &  12.0 & 10.7 $\pm$  1.3 &         5191 &  74 &  3.0 & 10.8 &          145 &          187 &           65 &  340 &    1.61 &   0.9 \\
VGS\_36 &  12.0 & 19.8 $\pm$  2.7 &         6684 &  95 &  5.0 & 13.0 &          190 &          224 &           79 &    190 &  2.82 &   1.1 \\
VGS\_36a & 12.0 & - & 6684 & 95 & 1.6 & - & - &  - & - & - & - & - \\
VGS\_37 &  12.0 & 13.46 $\pm$ 1.89 &         5787 &           83 &   5.6 &  12.8 &   153
 &   186 &           67 & 195 &  2.05 &   1.70 \\
VGS\_37a &  12.0 &   8.57 $\pm$   1.67 &         4318 &           62 &   3.1 & 
 -  &  50  &  92 &  -  &  -  &  -  &   0.5 \\
VGS\_38 &   12.0 & 9.1 $\pm$  0.7 &         3853 &  55 &  2.8 &  6.9 &           50 &           92 &           39 & 130 &     0.26 &   2.0 \\
VGS\_38a & 12.0 & 0.86 $\pm$ 0.14 &     3832 &       54 & 1.3 &  -  &           67 &           75 &  -  &  - & -  &  0.4 \\ 
VGS\_38b & 12.0 & 1.39 $\pm$ 0.22 &     3863 &       55 & 1.4 &  -  &           41 &           66 &  -  & - &   -  &  2.0 \\ 
VGS\_39 &  11.0 &  $<$ 2.59 &  -  &  -  &   3.7 &  -  &  -  &  -  &           66 &  -  & 
 -  &  -  \\
VGS\_39a &  11.0 &   2.56 $\pm$   0.39 &         5171 &           74 &   1.1 & 
 -  & 16  &  32  &  -  &  -  &  -  &   3.0 \\
VGS\_40 &  12.0 &  6.00 $\pm$ 1.40 &         7091 &          101 &   2.8 & $<$   4.2 & 
  206 &   231 &           42 &  77 & $<$  2.32 &   0.43 \\
VGS\_41 &  10.5 &  $<$ 2.86 &  -  &  -  &   1.8 &  -  &  -  &  -  &           29 &  -  & 
 -  &  -  \\
VGS\_42 &   12.0 & 4.0 $\pm$  1.5 &         5601 &  80 &  3.5 &  $<$  8.1 &          128 &          171 &           58 &  310 & $<$     1.08 &   0.2 \\
VGS\_43 &  14.0 &  $<$ 2.87 &  -  &  -  &   2.0 &  -  &  -  &  -  &           30 &  -  & 
 -  &  -  \\
VGS\_44 &   12.0 & 4.9 $\pm$  1.1 &         5295 &  76 &  3.6 &  $<$  7.0 &           76 &          110 &           31 &  145 & $<$     0.90 &   0.1 \\
VGS\_45 &   12.0 & 3.5 $\pm$  1.3 &         4316 &  62 &  4.0 &  $<$  6.9 &           68 &          102 &           66 &  320 & $<$     0.22 &   1.0 \\
VGS\_46 &  12.0 &  5.55 $\pm$ 1.66 &         4751 &           68 &   2.9 & $<$   4.9 & 
  186 &   202 &           71 & 260 & $<$  1.10 &   0.80 \\
VGS\_47 &  12.0 & 13.02 $\pm$ 2.71 &         6630 &           95 &   8.0 &  12.4 &   317
 &   325 &           72 & 340 &  7.97 &   0.12 \\
VGS\_48 &   0.0 &  -  &  -  &  -  &   3.6 &  -  &  -  &  -  &           68 &  -  &  -  & 
 -  \\
VGS\_49 &  12.0 &  $<$ 3.60 &  -  &  -  &   3.9 &  -  &  -  &  -  &           40 &  -  & 
 -  &  -  \\
VGS\_50 &  12.0 & 55.27 $\pm$ 7.30 &         6129 &           88 &   3.2 &  16.1 &   287
 &   330 &           83 & 349 &  7.85 &   1.29 \\
VGS\_51 &  12.0 & 19.01 $\pm$ 2.46 &         7476 &          107 &   4.3 & $<$  10.1 & 
  172 &   206 &           63 & 285 & $<$  2.18 &   1.27 \\
VGS\_51a &  12.0 &   1.00 $\pm$   0.21 &         4990 &           71 &   1.4 & 
 -  &  68  &  68  &  -  &  -  &  -  &   2.0 \\
VGS\_52 &  12.0 &  9.00 $\pm$ 2.25 &         5426 &           78 &   3.8 &  10.0 &   118
 &   117 &           70 & 275 &  0.92 &   2.07 \\
VGS\_53 &  13.0 &  5.19 $\pm$ 1.66 &         6450 &           92 &   5.4 & $<$   2.2 & 
  162 &   178 &           64 & 155 & $<$  0.42 &   0.14 \\
VGS\_53a &  13.0 &   4.38 $\pm$   2.16 &         6942 &           99 &   6.2 & 
 -  &  119  & 119  &  -  &  -  &  -  &   0.3 \\
VGS\_54 &   9.0 & 33.18 $\pm$ 4.00 &         7028 &          100 &   6.7 &  16.5 &   240
 &   265 &           80 &  38 &  5.72 &   1.21 \\
VGS\_54a &   9.0 &   - &         4421 &           63 &   3.0 & 
 -  &  -  &  -  &  -  &  -  &  -  &   - \\
VGS\_55 &   8.5 & 17.33 $\pm$ 3.56 &         7485 &          107 &   5.7 &  13.2 &   154
 &   197 &           55 & 145 &  2.71 &   0.60 \\
VGS\_56 &  12.0 &  $<$ 2.67 &  -  &  -  &   3.3 &  -  &  -  &  -  &           59 &  -  & 
 -  &  -  \\
VGS\_56a &  12.0 &   2.15 $\pm$   0.66 &         6247 &           89 &   2.1 & 
 -  &  101  &  109  &  -  &  -  &  -  &   1.5 \\
VGS\_57 &  13.0 &  6.37 $\pm$ 1.48 &         6630 &           95 &   5.7 & $<$   9.0 & 
  119 &   187 &           30 & 330 & $<$  3.01 &   0.06 \\
VGS\_57a &  13.0 &   1.72 $\pm$   0.49 &         5443 &           78 &   1.9 & 
 -  &  59  &  101  &  -  &  -  &  -  &   0.3 \\
VGS\_58 &   12.0 & 7.0 $\pm$  0.6 &         3351 &  48 &  2.6 &  7.4 &          143 &          151 &           38 &    280 &  2.39 &   0.7 \\
VGS\_59 &  12.0 &  $<$ 2.59 &  -  &  -  &   3.1 &  -  &  -  &  -  &           67 &  -  & 
 -  &  -  \\
VGS\_60 &  12.0 &  2.51 $\pm$ 1.61 &         5890 &           84 &   4.9 & $<$   5.7 & 
  273 &   281 &           81 & 106 & $<$  2.53 &   0.09 
\enddata
\tablecomments{Non-detections list the 3$\sigma$ upper limit on the H \textsc{i} mass. t lists the total integration time for each object. $V_{sys}$ is the systemic H \textsc{i} velocity, using the optical definition.  D is the distance to the target galaxy.  r$_{90}$ and r$_{\textrm{H \textsc{I}}}$ list the optical and H \textsc{i} radius, respectively.  When possible, poorly resolved systems list the upper limit for r$_{\textrm{H \textsc{I}}}$ and $M_{dyn}$.  $W_{50}$ and $W_{20}$ are the 50\% and 20\% H \textsc{i} line widths, respectively, corrected for instrumental broadening.  The inclination, $i$,  is calculated such that 90$^\circ$ is edge-on.}
\end{deluxetable*}

\begin{deluxetable*}{l r r r r }
\tabletypesize{\tiny}
\tablecaption{Stellar and star formation parameters for void galaxies
\label{tab:sf}}
\tablewidth{0pt}
\tablehead{
\colhead{Name} & \colhead{$M_{*}$} & 
 \colhead{SFR$_{H\alpha}$} &  
 \colhead{SFR$_{H\alpha}/M_{*}$} & \colhead{SFR$_{H\alpha}$/$M_{\textrm{H \textsc{I}}}$} \\
\colhead{} & 
\colhead{($10^8 M_\odot$)} & 
\colhead{($M_\odot$ yr$^{-1}$)} & 
\colhead{($10^{-11}$ yr$^{-1}$)} & \colhead{($10^{-11}$ yr$^{-1}$)}
}
\startdata

VGS\_01 &   4.5 &  0.08 &  15.3 &    - \\
VGS\_02 &   4.5 &  0.08 &  16.2 &   13.1 \\
VGS\_03 &   2.4 &  0.04 &  15.3 &    - \\
VGS\_04 &   7.4 &  0.14 &  17.0 &    - \\
VGS\_05 & 129.1 &  0.03 &   0.2 &    - \\
VGS\_06 &   2.6 &  0.17 &  55.4 &   10.7 \\
VGS\_07 &   0.5 &  0.18 & 278.1 &   20.5 \\
VGS\_08 &   1.9 &  0.06 &  24.8 &   14.1 \\
VGS\_09 &   0.7 &  0.04 &  48.9 &    3.9 \\
VGS\_10 &   2.2 &  0.09 &  36.7 &    6.1 \\
VGS\_11 &  11.5 &  0.14 &  10.9 &    6.4 \\
VGS\_12 &   1.6 &  0.08 &  42.6 &    2.6 \\
VGS\_13 &  10.6 &  0.11 &   9.4 &    9.2 \\
VGS\_14 &   1.6 &  0.07 &  40.1 &   11.3 \\
VGS\_15 &  26.0 &  0.04 &   1.4 &    - \\
VGS\_16 &   0.9 &  0.03 &  32.6 &    - \\
VGS\_17 &   0.2 &  0.08 & 277.6 &    - \\
VGS\_18 &   2.1 &  0.04 &  15.1 &    9.3 \\
VGS\_19 &   3.7 &  0.12 &  28.1 &   36.9 \\
VGS\_20 &   0.5 &  0.07 & 139.8 &    - \\
VGS\_21 &  93.6 &  0.11 &   1.0 &    4.9 \\
VGS\_22 &   1.9 &  0.04 &  19.8 &    - \\
VGS\_23 &  19.3 &  0.44 &  19.9 &   11.3 \\
VGS\_24 &   - &  - & -  &    - \\
VGS\_25 &   1.4 &  0.10 &  61.1 &   56.5 \\
VGS\_26 &  19.9 &  0.34 &  15.6 &   22.7 \\
VGS\_27 &   1.0 &  0.03 &  26.3 &    8.7 \\
VGS\_28 &   - &  - &  -  &    - \\
VGS\_29 &   8.1 &  0.21 &  22.1 &    - \\
VGS\_30 &   1.0 &  0.06 &  48.0 &   10.0 \\
VGS\_31 &  35.1 &  1.42 &  35.6 &   71.5 \\
VGS\_32 &  27.9 &  0.26 &   8.2 &    6.9 \\
VGS\_33 &   1.7 &  0.02 &  10.8 &    2.7 \\
VGS\_34 &  75.4 &  0.87 &  10.9 &   36.4 \\
VGS\_35 &   6.6 &  0.11 &  13.8 &   10.0 \\
VGS\_36 &   8.9 &  0.26 &  26.2 &   13.3 \\
VGS\_37 &   4.1 &  0.11 &  24.1 &    8.3 \\
VGS\_38 &   0.7 &  0.07 &  88.2 &    7.8 \\
VGS\_39 & 102.3 &  0.20 &   1.8 &    - \\
VGS\_40 &  11.6 &  0.20 &  15.1 &   32.9 \\
VGS\_41 &   6.1 &  0.10 &  15.0 &    - \\
VGS\_42 &  25.3 &  0.16 &   5.7 &   40.8 \\
VGS\_43 &   1.9 &  0.05 &  21.7 &    - \\
VGS\_44 &  32.5 &  0.79 &  21.9 &  161.5 \\
VGS\_45 &   1.1 &  0.05 &  39.3 &   13.1 \\
VGS\_46 &   3.3 &  0.09 &  24.3 &   16.5 \\
VGS\_47 & 213.4 &  0.59 &   2.5 &   45.5 \\
VGS\_48 &  19.4 &  0.12 &   5.3 &    - \\
VGS\_49 &  33.1 &  0.76 &  19.8 &    - \\
VGS\_50 &  83.5 &  0.36 &   4.0 &    6.5 \\
VGS\_51 &   3.6 &  0.27 &  65.7 &   14.0 \\
VGS\_52 &   1.3 &  0.06 &  37.5 &    6.3 \\
VGS\_53 &  32.0 &  0.27 &   7.6 &   52.7 \\
VGS\_54 &  42.5 &  0.18 &   3.8 &    5.6 \\
VGS\_55 &  15.7 &  0.17 &   9.2 &    9.6 \\
VGS\_56 &  49.0 &  0.27 &   4.9 &    - \\
VGS\_57 & 113.5 &  1.73 &  13.3 &  271.7 \\
VGS\_58 &   3.9 &  0.11 &  25.1 &   16.1 \\
VGS\_59 &   1.8 &  0.05 &  22.4 &    - \\
VGS\_60 &  56.5 &  0.13 &   2.0 &   51.0 
\enddata
\end{deluxetable*}

\section{Results}
\label{sec:results}

\begin{figure}[b!]
\centering
\includegraphics[width=3.2in,angle=0]{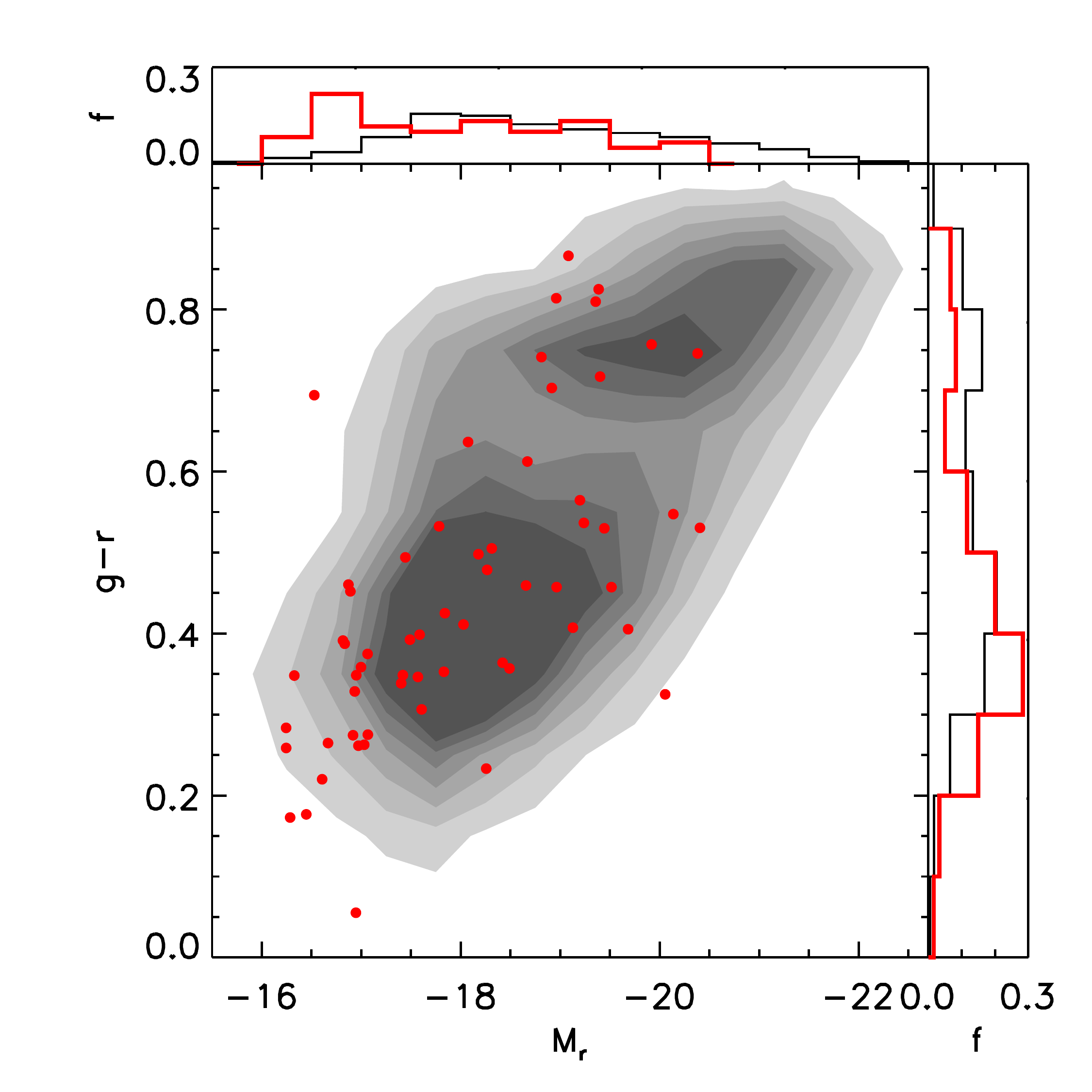}
\caption{Color magnitude diagram of the VGS (red) compared with a magnitude limited sample of galaxies from SDSS selected in a similar redshift range $0.01 < z < 0.03$.  Our void galaxies are typically blue and faint, but do span a range of colors and luminosities.
\label{fig:cmd}}
\end{figure}

Despite differing void finding methods, there is general agreement that void galaxies are usually faint, late type, blue disk galaxies, but the details of their properties at fixed luminosity and morphological type are less clear, and may depend on an accurate identification of the void galaxy population \citep{Grogin1999, Grogin2000, Rojas2004, Rojas2005, Patiri2006b, Park2007, BendaBeck2008}.    
From our sample of 60 void galaxies, 55 were observed in H \textsc{i} and 14 were not detected.  Measurements of the first 15 galaxies are discussed as a pilot study to this project by \cite{Kreckel2011}, and are included again here for completeness.  The VGS includes a range of luminosities, from -16.2 $>$ M$_r$ $>$ -20.4, corresponding to a stellar mass range of  $2 \times 10^7$ M$_\sun$ to $2 \times 10^{10}$ M$_\sun$. We measure H \textsc{i} masses ranging from $1.7 \times 10^8$ M$_\sun$ to $5.5 \times 10^9$ M$_\sun$. Distributions for the absolute magnitude, stellar mass and H \textsc{i} masses are shown in Figure \ref{fig:hists}.
Along with the selected galaxy sample, we have detected by their H \textsc{i} emission 18 other void galaxies within $\sim$ 800 km s$^{-1}$ and 25$^{\prime}$ of the VGS targets (Table \ref{tab:company}).  As these are H \textsc{i} selected they probe to fainter luminosities, as faint as M$_r \sim -14$, with H \textsc{i} masses as low as $\sim 5 \times 10^7$ M$_\sun$.

We find that our sample consists almost exclusively of gas-rich blue disk galaxies.  Though many of our targets suffer from an H~\textsc{i} spatial resolution that is limited to a few beams per galaxy, combined with the kinematic information we judge that about half have strongly irregular or disturbed H \textsc{i} morphologies and kinematics.
In the following sections we compare the VGS H \textsc{i} observations at the WSRT with two catalogs of resolved H \textsc{i} observations in the literature. The first is drawn from the Westerbork observations of neutral Hydrogen in Irregular and SPiral galaxies (WHISP), which presents an atlas of the H \textsc{i} morphology and kinematics for a large number of galaxies.  This sample is biased towards gas rich targets but unconstrained environmentally.  We use the atlas of 73 galaxies presented in \cite{Swaters2002}.  In addition, we also compare with an atlas of 43  spiral galaxies in the Ursa Major cluster, also observed with the WSRT \citep{Verheijen2001}. This sample is located within a poor cluster of galaxies, and extends to higher luminosities than \cite{Swaters2002}.  These two atlases together encompass a large range of galaxy H \textsc{i} morphologies and kinematics, all observed with the same radio interferometer as the VGS.

\subsection{Optical color and morphology}
Though we have selected the VGS purely on the basis of their location within the geometrically reconstructed density field, it spans a  range of luminosities and colors. In Figure \ref{fig:cmd} we compare the color-magnitude distribution with a magnitude limited ($z < 0.03$, m$_r < 17.77$) sample of SDSS galaxies.  Our void galaxies exhibit a wide range of colors, however there are significantly fewer red galaxies and more blue galaxies in the void. 
Our sample of void galaxies is also shifted towards fainter galaxies, with no galaxies brighter than M$_r = -20.4$.  This matches the generally observed shift in the luminosity function found in underdense regions and predictions from CDM cosmology \citep{Hoyle2005, Aragon2007, Kreckel2011a}. None of the VGS galaxies have a stellar masses above 3 $\times 10^{10}$ M$_\sun$, the observationally identified transition mass below which galaxies are typically younger, and still in the process of assembling \citep{Kauffmann2003c}.

\begin{figure}[b!]
\centering
\includegraphics[height=1.3in]{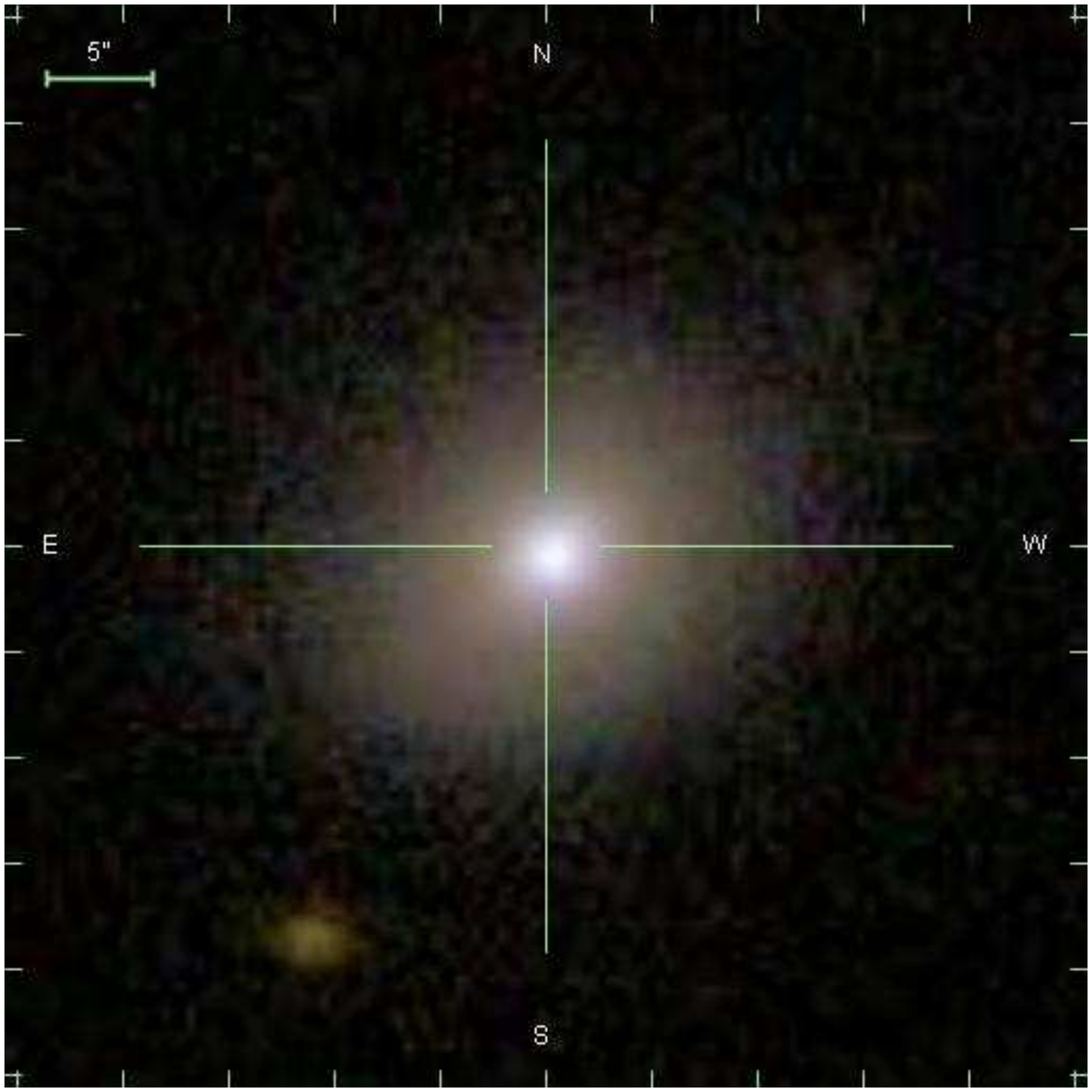}
\includegraphics[height=1.3in]{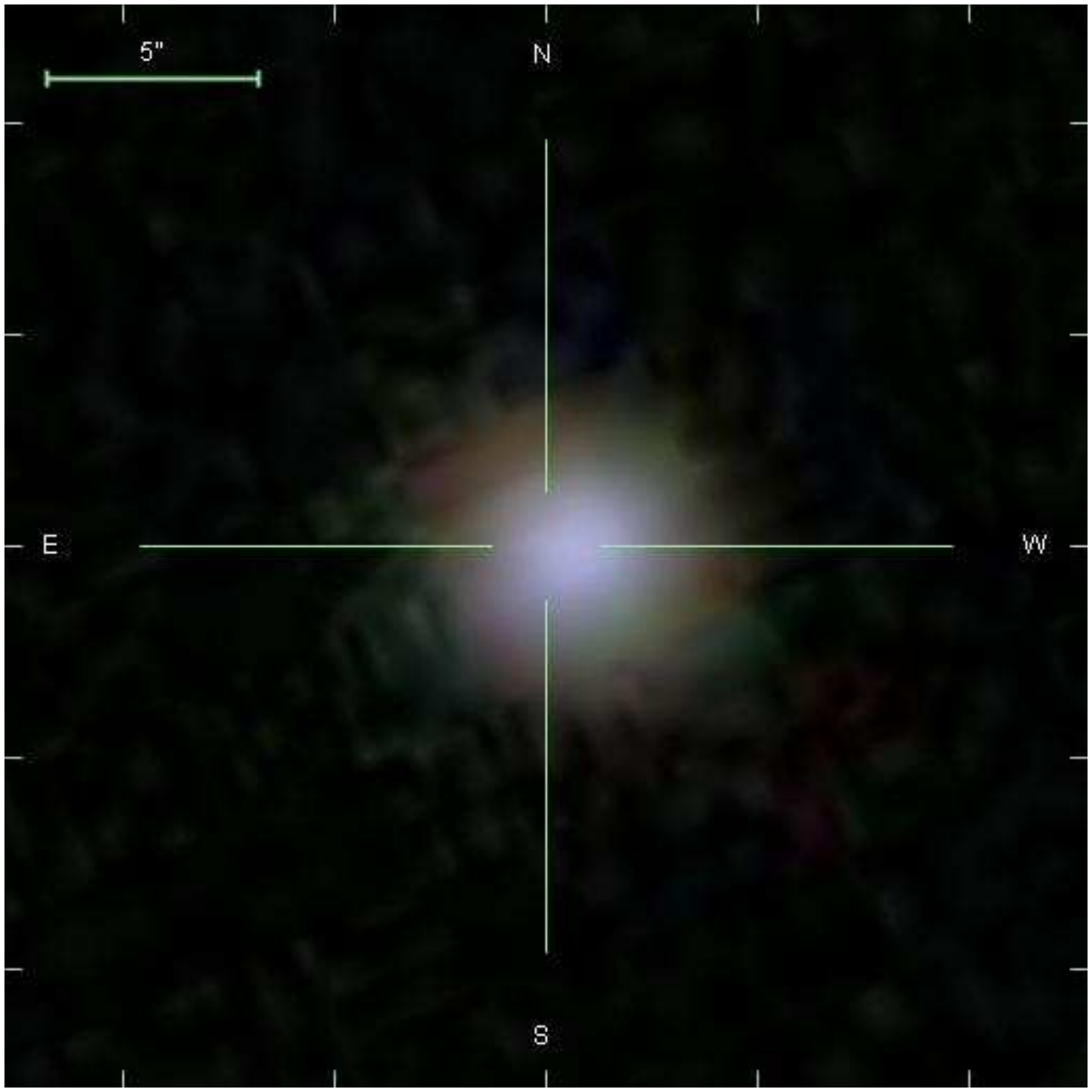}
\includegraphics[height=1.3in]{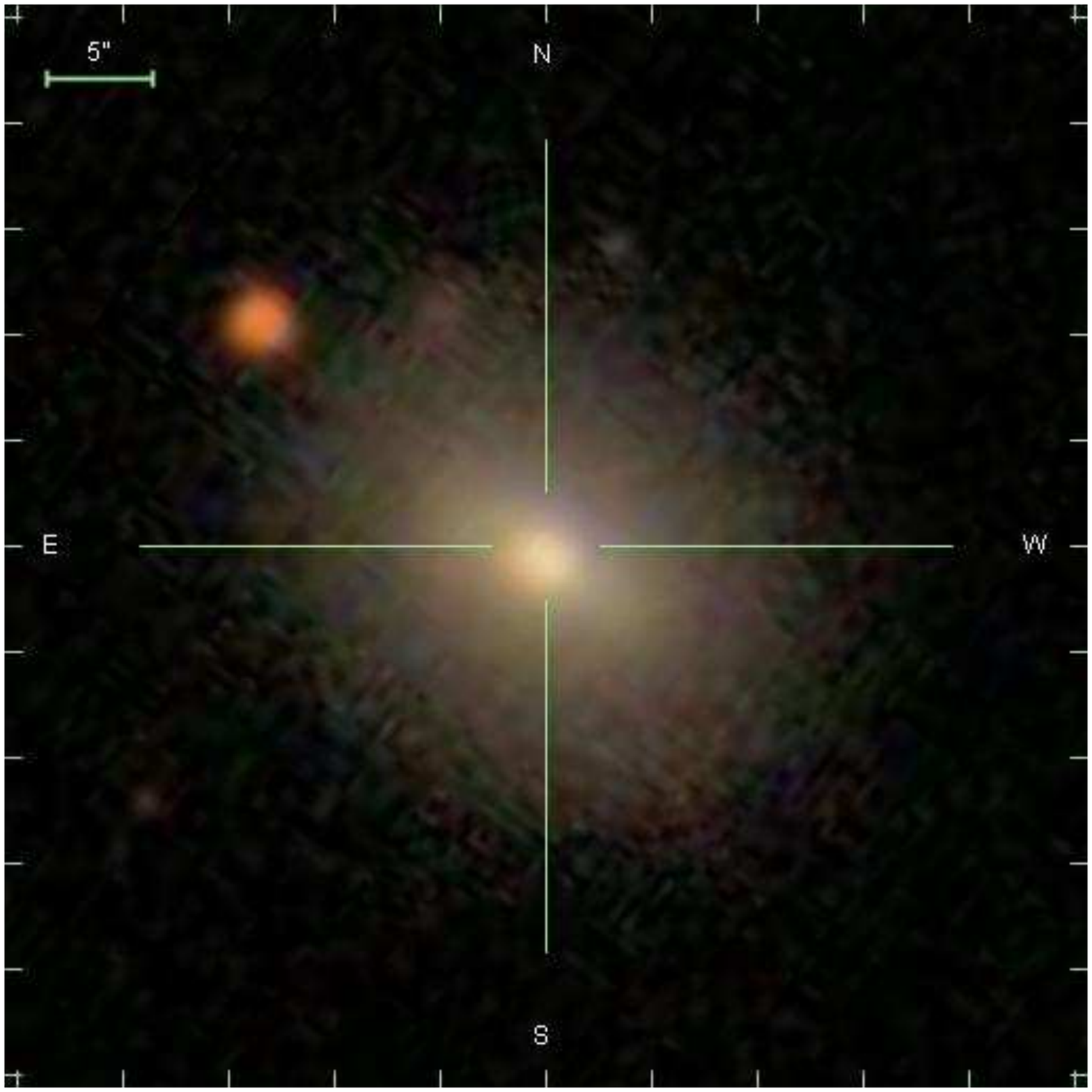} \\
\includegraphics[height=1.3in]{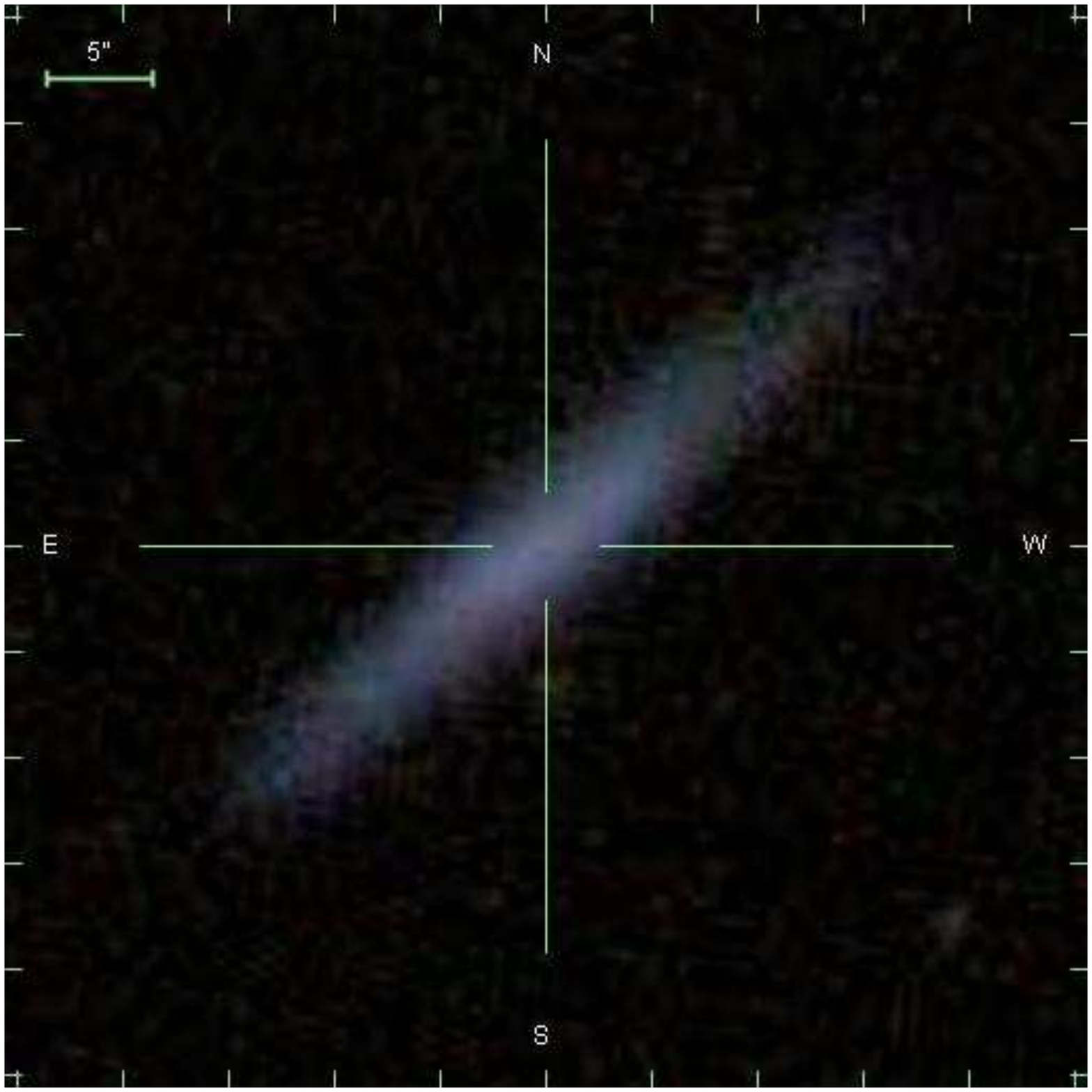}
\includegraphics[height=.9in]{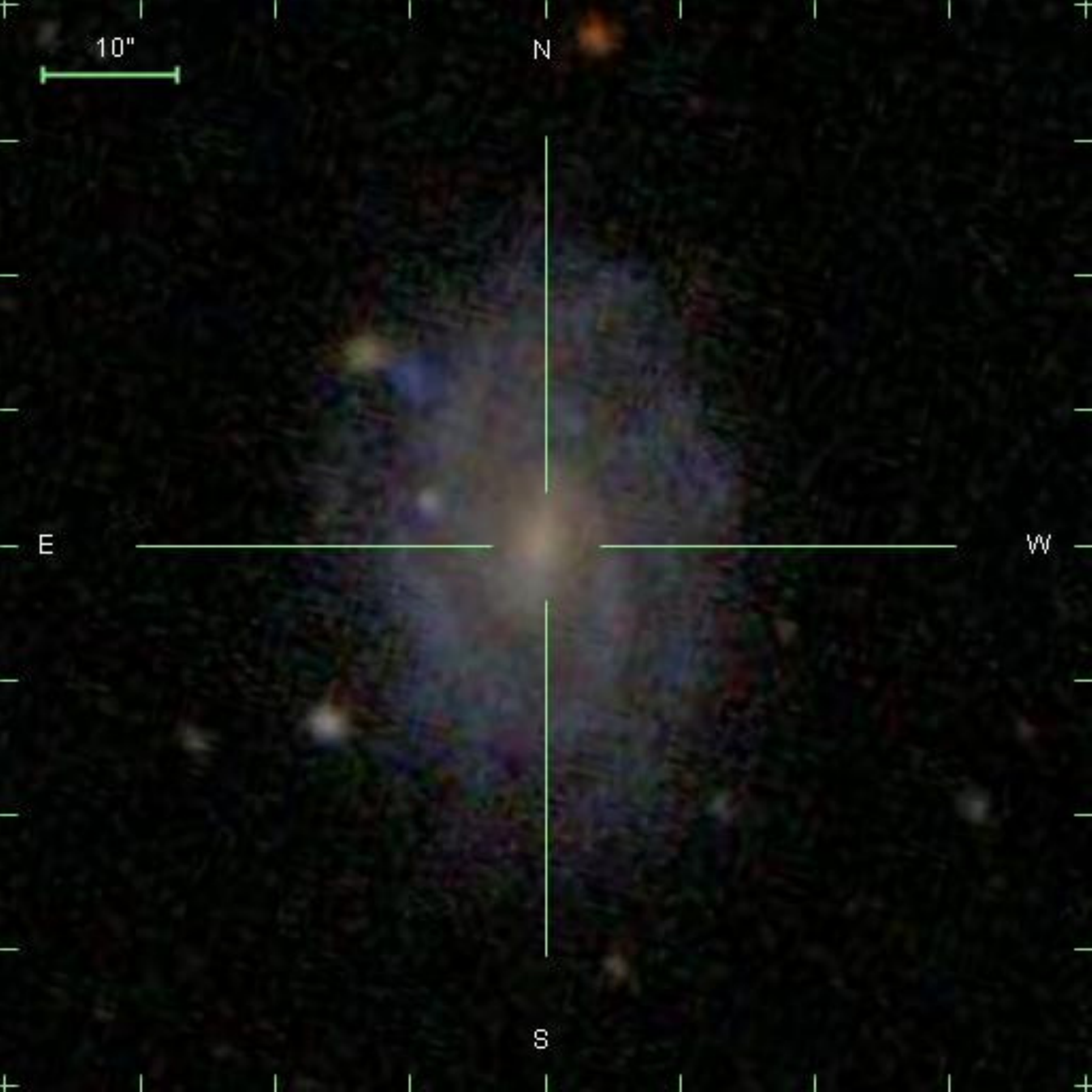}
\includegraphics[height=1.3in]{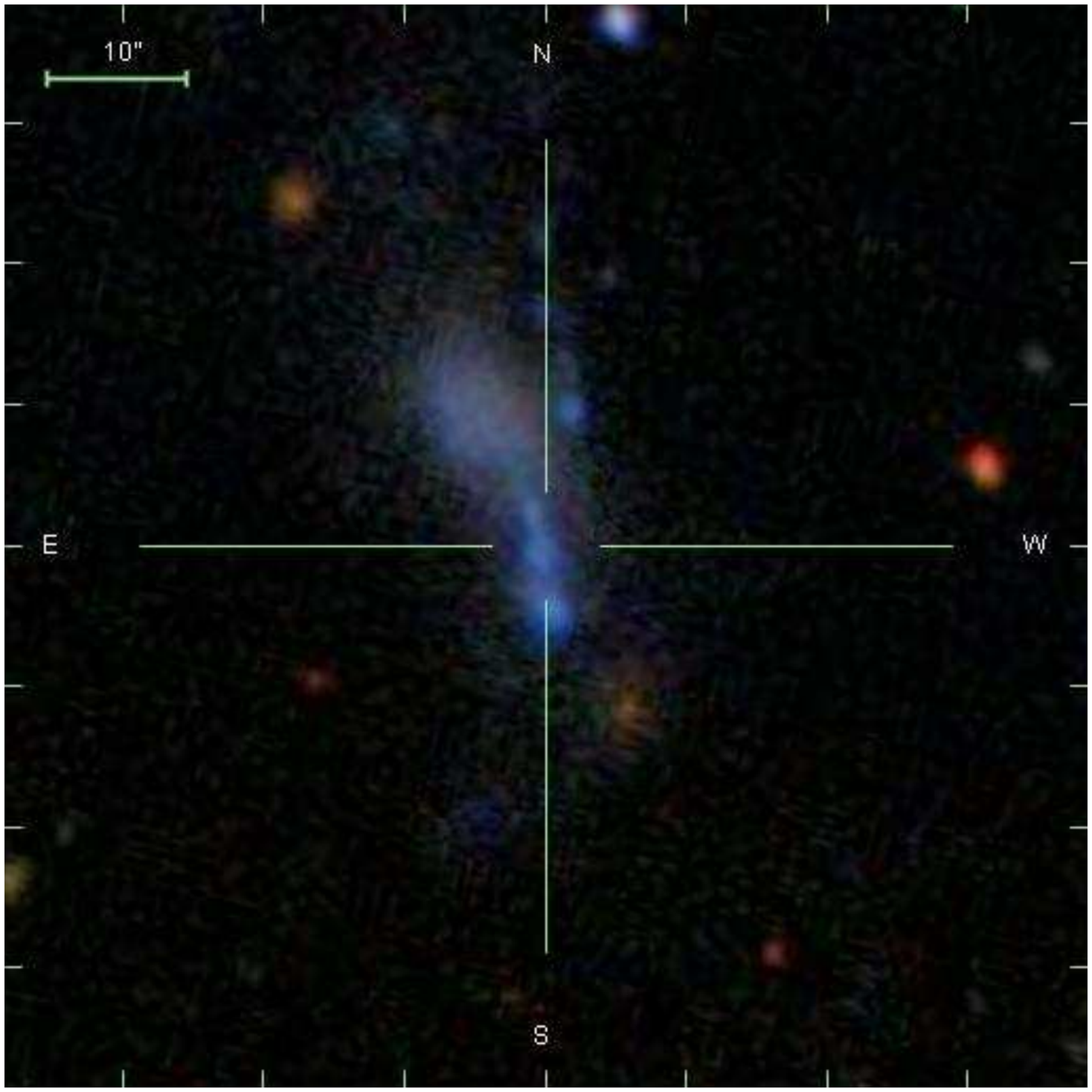}
\caption{The VGS includes a range of stellar morphologies, with elliptical (VGS\_24, top left; VGS\_41, top center), lenticular (VGS\_05, top right), bulge free (VGS\_10, bottom left), spiral (VGS\_15, bottom center), and irregular (VGS\_17, bottom right) galaxies.
\label{fig:optmorph}}
\end{figure}

The VGS spans a range of stellar morphologies, however the distance to this sample makes morphological subtypes difficult to distinguish by eye, and we find that it is also very difficult to constrain with simple parameterization.  In considering whether any of the VGS are elliptical galaxies, we note that most of the red galaxies with $g-r > 0.6$ appear to be edge-on disks or bulge-dominated systems.  Classification by the galaxy concentration index, defined as the ratio of the Petrosian radius containing 90\% of the light to that containing 50\% (r$_{90}$/r$_{50}$),  is also not perfect, as half of those with a high concentration index appear to be bulge dominated disk galaxies. 
Judging  the SDSS images by eye (see Figure \ref{fig:poststamps}), only three might be classified as early type galaxies. VGS\_24 and VGS\_41 have very smooth stellar distributions but are blue,  while VGS\_05 is red but appears to have a bar and may be an SB0 galaxy.  All three are not detected in H \textsc{i}.  We show these three, as well as examples of bulge free (VGS\_10), spiral (VGS\_15) and irregular (VGS\_17) galaxies in Figure \ref{fig:optmorph}.

\begin{figure}[t!]
\centering
\includegraphics[height=1.8in]{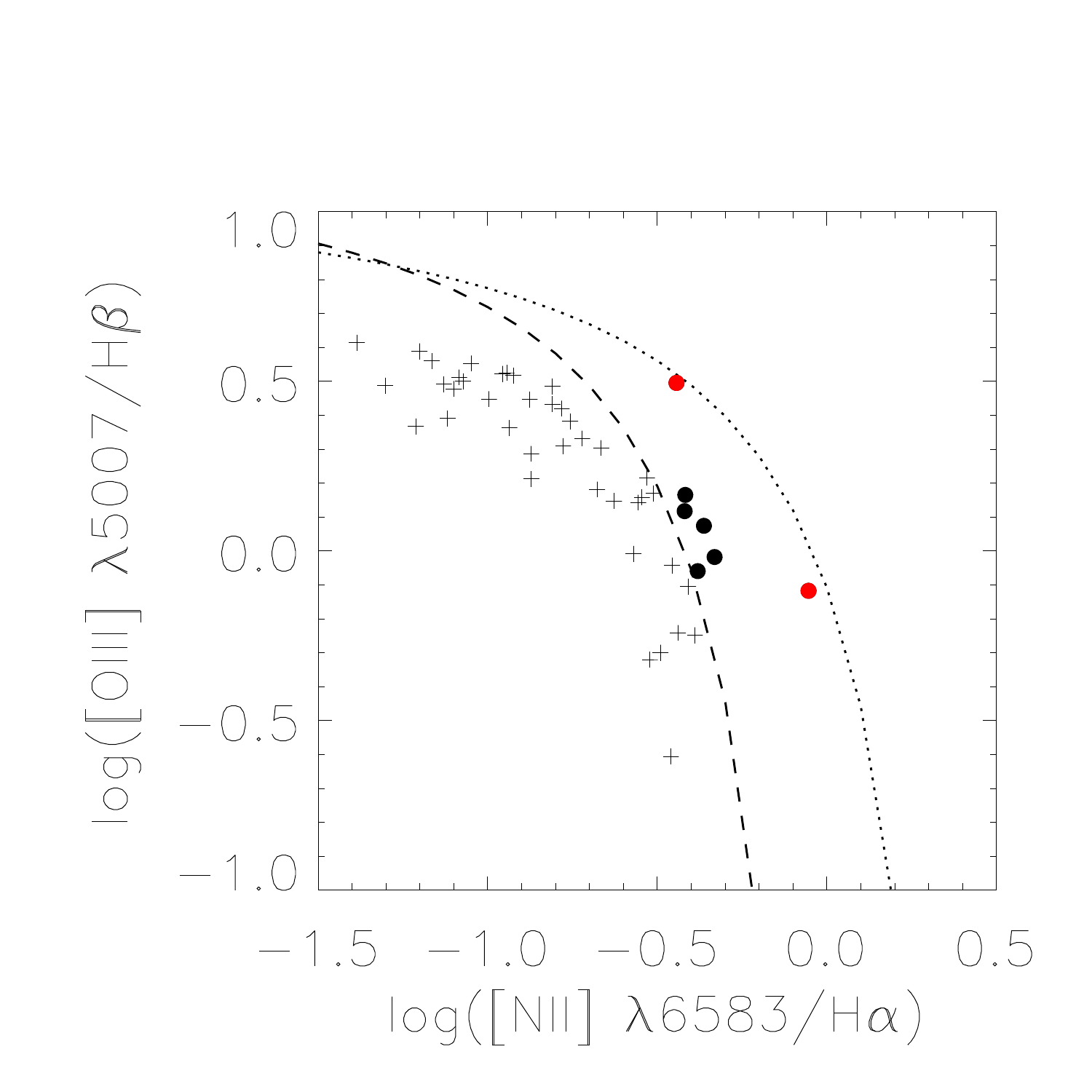}~\hspace{-.5cm}~\includegraphics[height=1.6in]{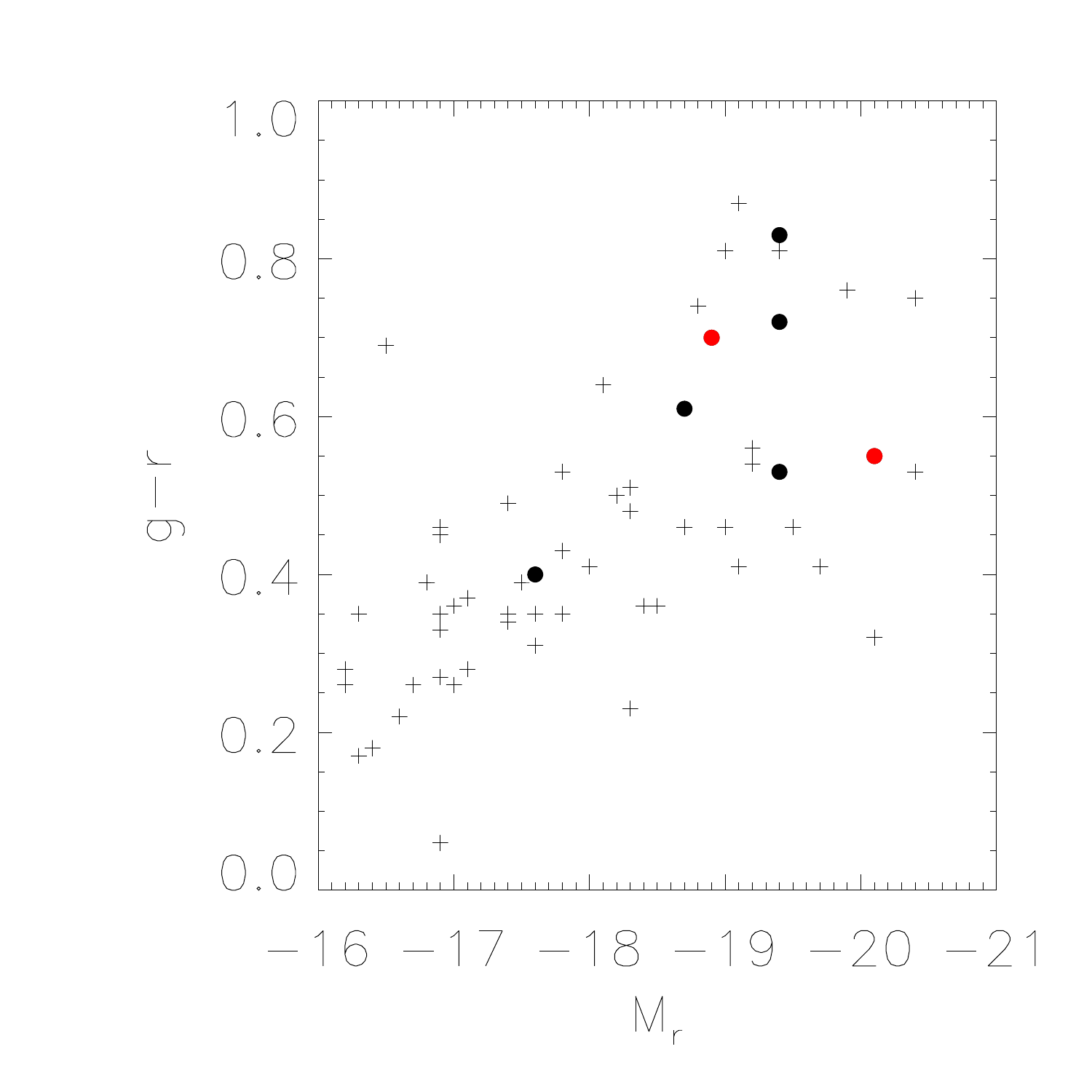}
\caption{Left: The BPT \citep{Baldwin1981} diagram for the VGS, showing the ratio of emission-line flux in [O \textsc{iii}]/H $\beta$ to [N \textsc{iii}]/H $\alpha$.  No galaxies fall above the AGN demarkation determined by \cite{Kewley2001} (dotted line).  Seven galaxies (circles; VGS\_02, VGS\_21, VGS\_24, VGS\_42, VGS\_50, VGS\_54) are classified as AGNs by relation in \cite{Kauffmann2003b} (dashed line), with two falling significantly away from the star-forming galaxies (red circles). Right: The location of these seven galaxies within the color magnitude distribution for the VGS.  Most of the AGN candidates are among the redder and more massive galaxies in the sample.  Colors and symbols are as in the BPT diagram. 
\label{fig:agn}}
\end{figure}

None of the VGS sample shows strong evidence of AGN activity.  We detected 1.4 GHz radio continuum emission in 15 VGS galaxies at above $\sim$1 mJy, which corresponds to luminosities of less than $5 \times 10^{21}$ W Hz$^{-1}$, consistent with star formation \citep{Smolcic2008}.  This can also be constrained optically following a line index classification scheme using the NII $\lambda$6583, OIII $\lambda$5007, H$\alpha$ and H$\beta$ emission features  \citep{Baldwin1981} measured in the SDSS spectra (Figure \ref{fig:agn}, left).  Using only targets where the signal to noise ratio is greater than three, our galaxies are not classified as AGNs according to the demarcation determined by \cite{Kewley2001}. The less strict demarcation determined by \cite{Kauffmann2003b} allows seven of our targets as possible AGNs, two of which (VGS\_24 and VGS\_54) do fall significantly away from  the region containing star-forming galaxies.  These seven AGN candidates are among the brighter and redder galaxies in our sample (Figure \ref{fig:agn}, right).  VGS\_24 is one of three elliptical galaxies in this sample, and none of these or the AGN candidates have strong radio continuum emission at 1.4 GHz.

\subsection{Size}
\begin{figure}[t!]
\centering
\includegraphics[height=3.2in,angle=90]{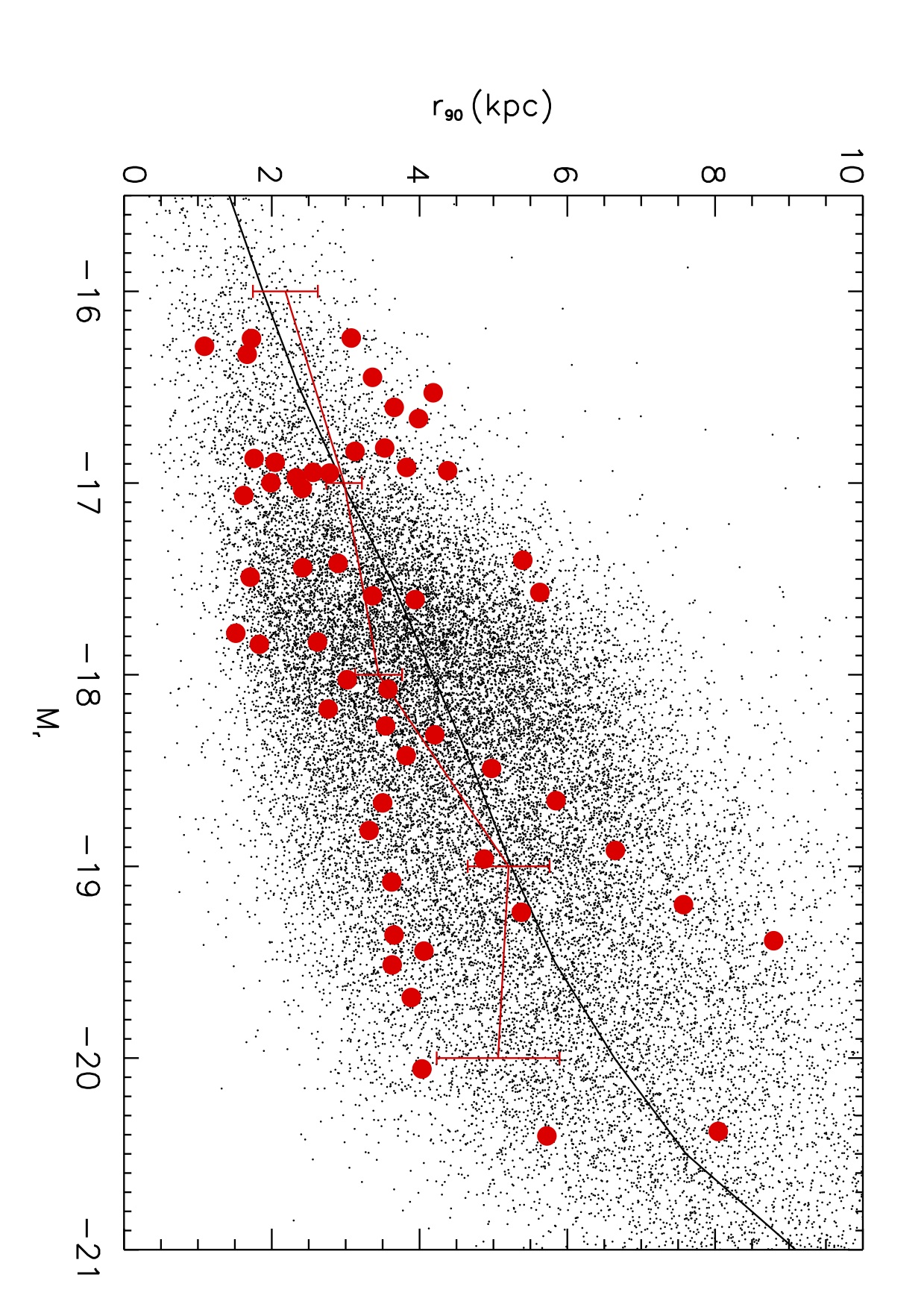}
\caption{Optical size of the late type ($r_{90}/r_{50} < 2.86$) disks of the VGS (circles) and a SDSS sample at $0.01 < z < 0.03$, with the mean values of each overplotted.  Unlike in the pilot sample, there is no systematic offset in the sizes of the stellar disks for the VGS galaxies.  There is a preference for smaller disks, however the distribution of sizes at fixed luminosity for the full VGS is in agreement with the larger SDSS sample, within the error in the mean and given the relatively small sample size.
\label{fig:r90}}
\end{figure}

In our pilot study we reported systematically smaller stellar disks in the galaxies compared with a volume limited sample of SDSS galaxies, however this systematic effect appears to be due largely to small number statistics within the original sample of 15 galaxies.  Figure \ref{fig:r90} shows a preference for smaller stellar disks in late type VGS galaxies as compared to a magnitude limited sample of late type SDSS galaxies ($z < 0.025$, m$_r < 17.77$).  Here we define late type galaxies as those having a light concentration where $r90/r50 < 2.86$ \citep{Shen2003}. One complication for assessing whether the VGS galaxies are smaller is the relatively large error in the mean.  We will consider this question in more detail in an upcoming paper employing deeper optical images.

\begin{figure}[t!]
\centering
\includegraphics[width=1.75in,angle=0]{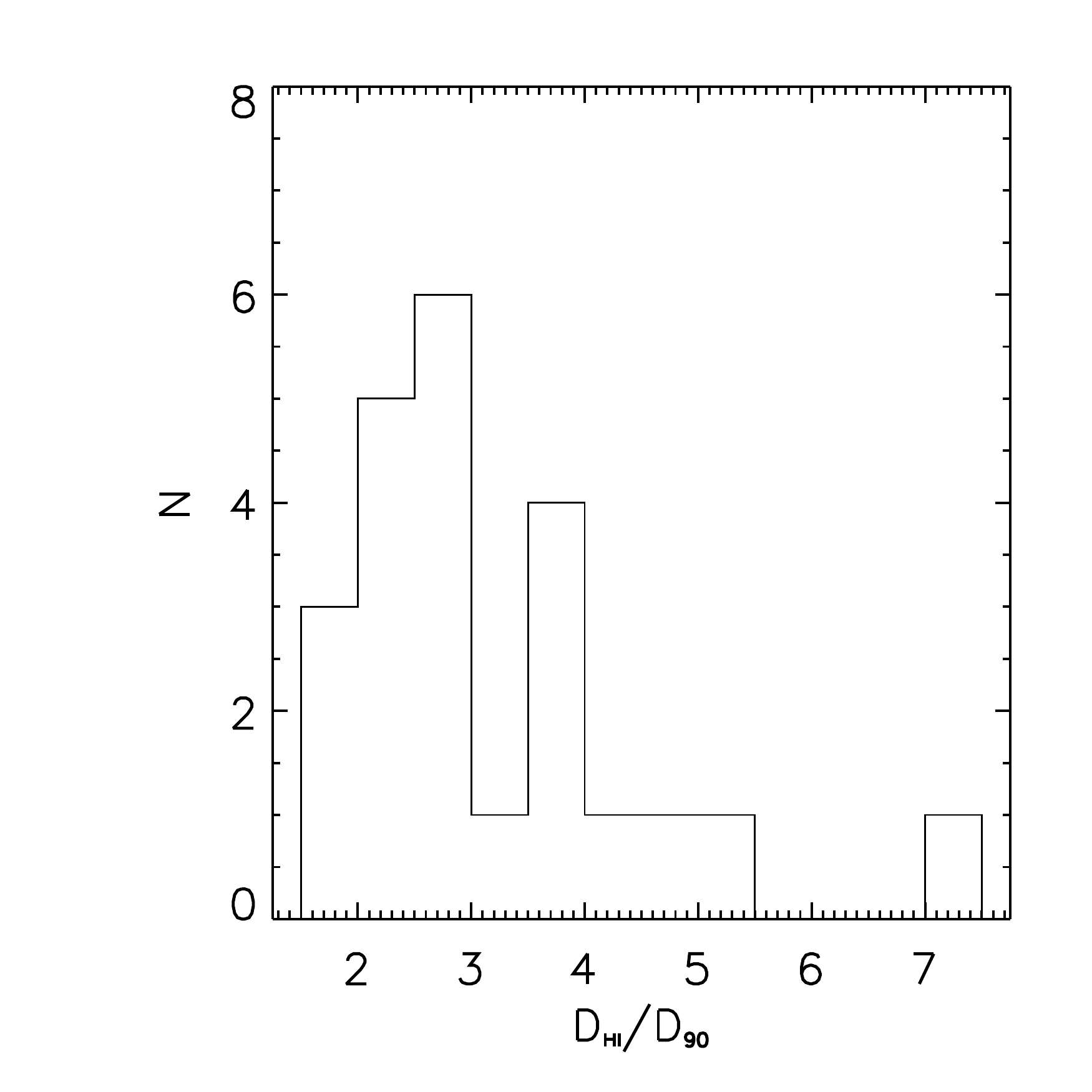}~\hspace{-.5cm}~\includegraphics[width=1.75in,angle=0]{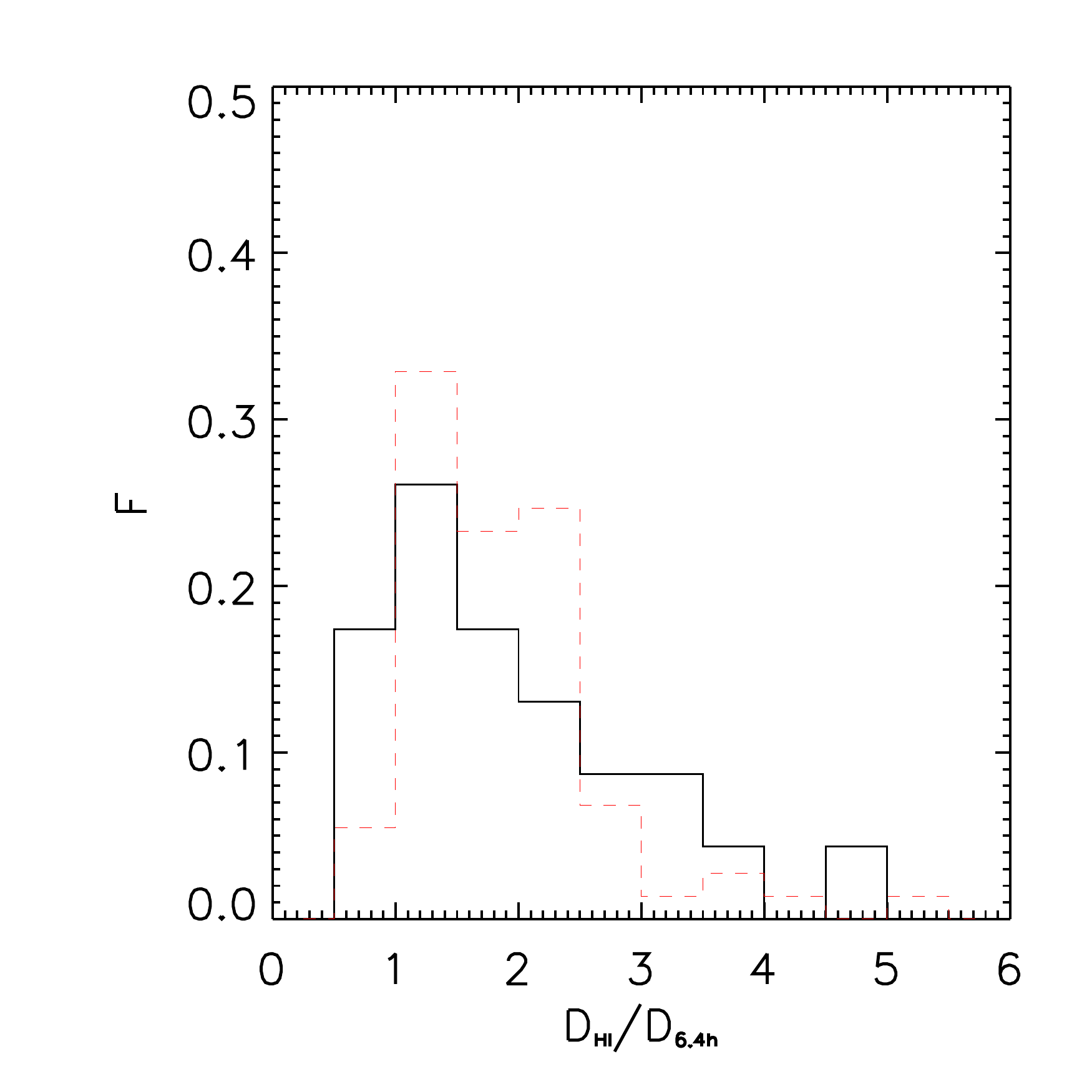}
\caption{Distribution of H \textsc{i} to optical diameter ratios.  The $r_{90}$ radius (left) and 6.4 times the exponential disk scale length (right) measure slightly different extents of the stellar component, however the galaxies appear consistent with the late-type galaxies studied by \cite{Swaters2002} (red dashed line).
\label{fig:hiopt}}
\end{figure}

The most robust measure of optical galaxy size available from the SDSS pipeline, r$_{90}$, differs significantly from the Holmberg, D25 or exponential disk scale length measures that are typically available in the literature for H \textsc{i} galaxy samples.  Using  D$_{\textrm{H \textsc{i}}}$/D$_{90}$, the VGS spans a reasonable range (Figure \ref{fig:hiopt}, left), with the H \textsc{i} disks two to three times more extended than the stellar disks.  The SDSS pipeline provides an estimate of the exponential disk scale length, which we find in good agreement with the WHISP sample (Figure \ref{fig:hiopt}, right).  
By this measure, some of the H \textsc{i} disks are comparable or smaller than the stellar disk, however in these cases the galaxies are edge-on disks and we suspect the SDSS pipeline is unable to accurately recover these scale lengths.

\subsection{HI non-detections}
\label{sec:nondetections}
\begin{figure}[t!]
\centering
\includegraphics[width=1.66in,angle=0]{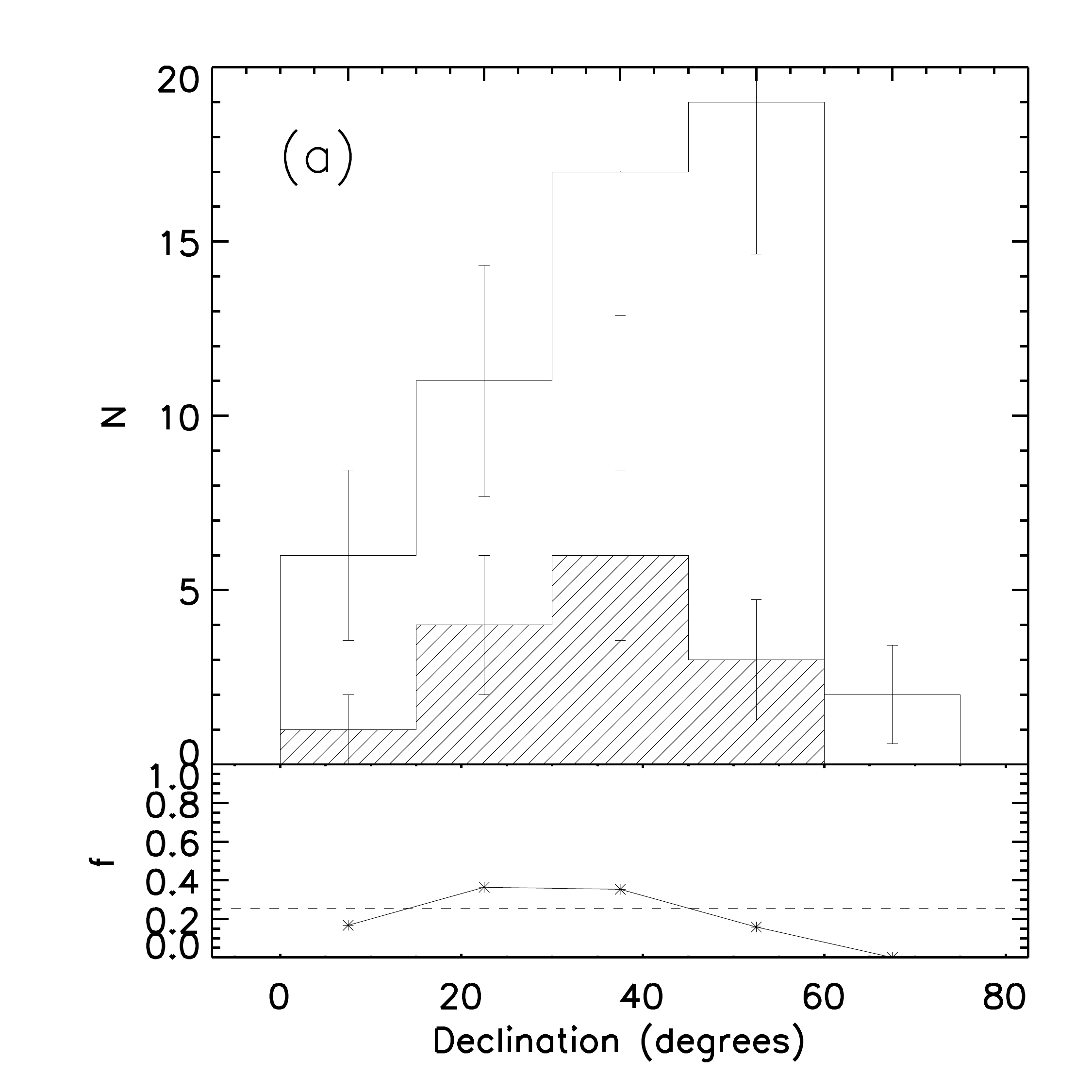}
\includegraphics[width=1.66in,angle=0]{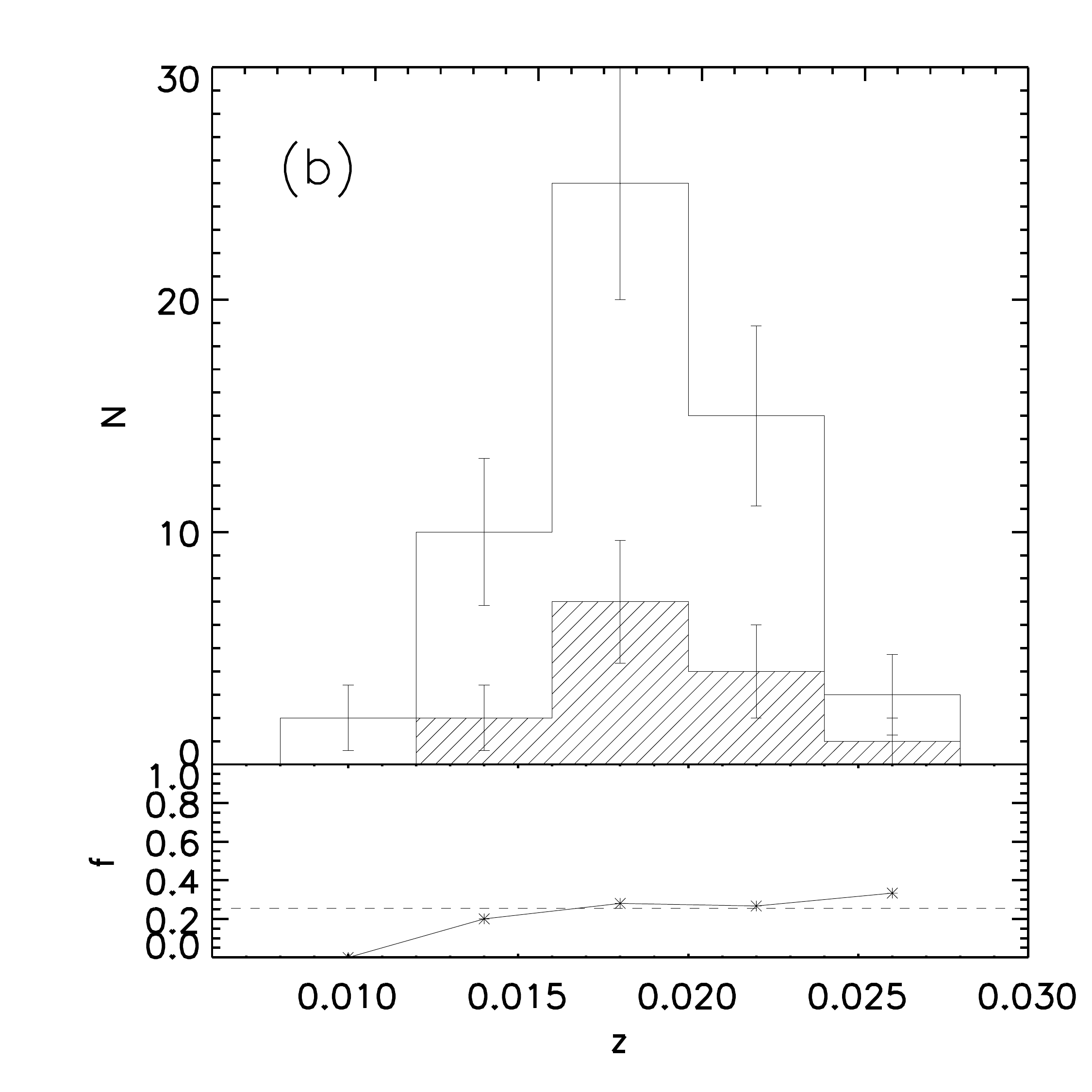}
\includegraphics[width=1.66in,angle=0]{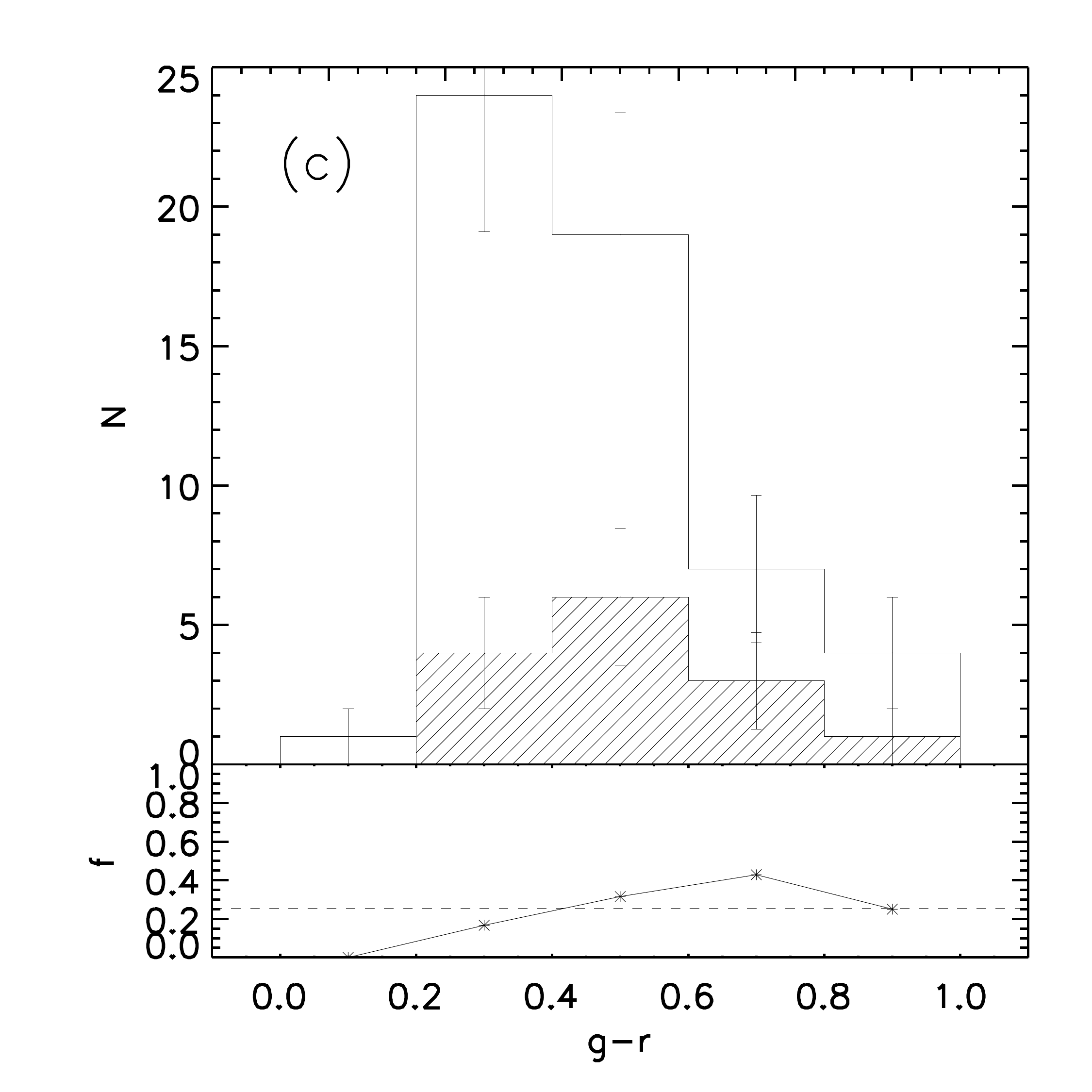}
\includegraphics[width=1.66in,angle=0]{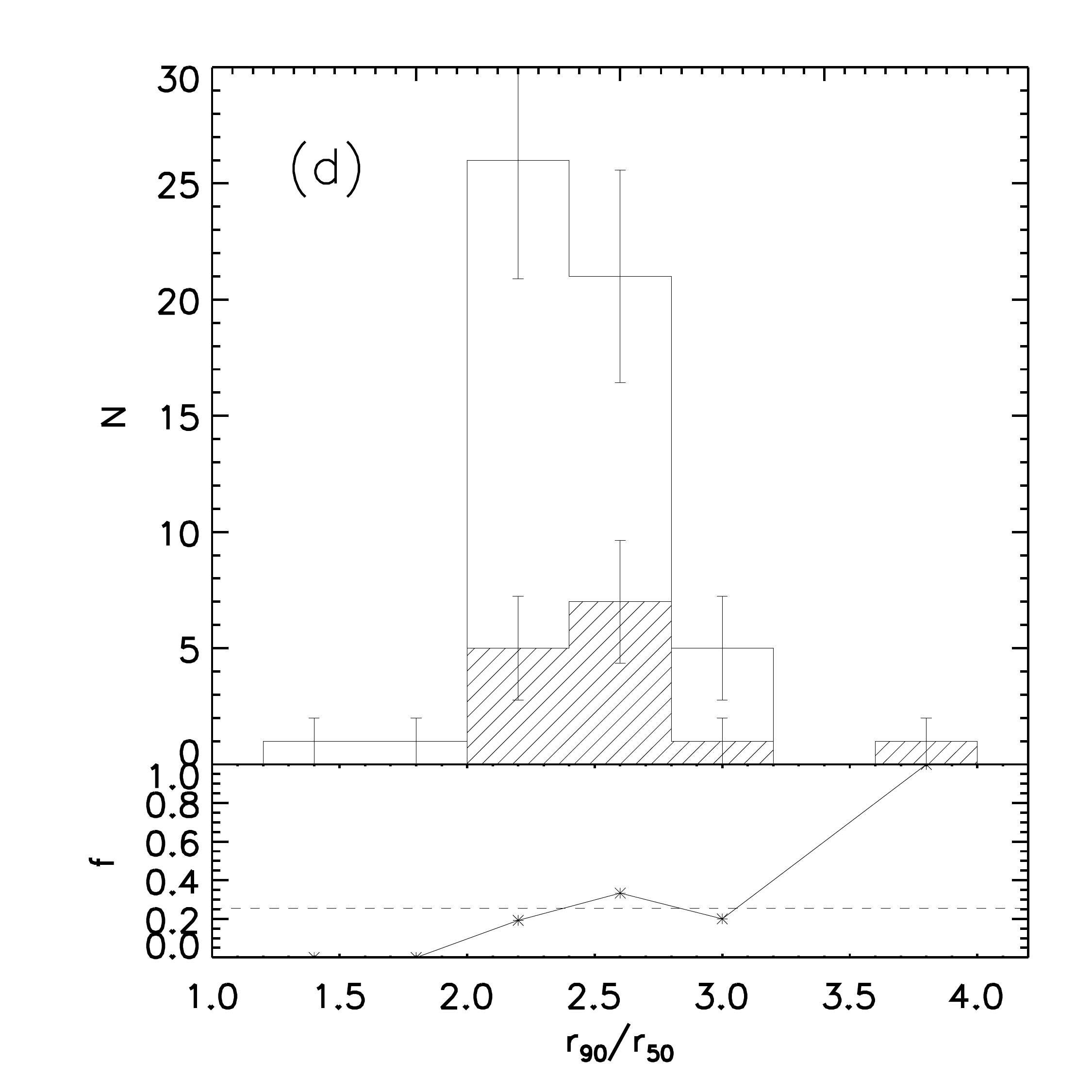}
\caption{Number of galaxies in the full VGS (unshaded region) compared to the H~\textsc{i} non-detections (shaded region) as a function of (a) declination, (b) redshift, (c) $g-r$ color, and (d) concentration index.  Error bars reflect Poisson counting errors. The fraction of non-detections is plotted at the bottom of each panel, with the dashed line indicating the overall non-detection fraction. There are very few red galaxies, with $g-r > 0.6$, or elliptical (highly concentrated) galaxies, with $r_{90}/r_{50} > 2.86$.   In general, there are no strong correlations, and the red and elliptical galaxies we observe are not found to be preferentially not-detected at a significant level.  We also note that we are not biased against detecting targets at low declination or high redshift.  
\label{fig:nod}}
\end{figure}

We detect 41 of the 55 galaxies observed at the WSRT in H \textsc{i}, a 75\% detection rate. 
There is no clear correlation with distance (Figure \ref{fig:nod}a) as might be expected if H \textsc{i} sensitivity was limiting, or with declination (Figure \ref{fig:nod}b), which can cause extremely elliptical beam shapes for east-west arrays like the WSRT. 
In general, we find no significant correlations in the non-detections (Figure \ref{fig:nod}). As already mentioned, color and concentration index ($r_{90}/r_{50}$) are not infallible indicators of optical morphology, which might be expected to correlate with the non-detections. There is a slight correlation with color though the errors are large due to the small sample size. Dividing the sample by color at $g-r=0.6$, 77 $\pm$ 13\% of blue galaxies were detected  while only 64 $\pm$ 24\% of red galaxies were detected.  We do note that the only 3 galaxies that look morphologically
like elliptical or S0 galaxies are not detected in H \textsc{i}.

Six of the non-detections (VGS\_04, VGS\_22, VGS\_16, VGS\_03, VGS\_41, VGS\_43) have very similar optical morphologies, with particularly small stellar components ($r_{90} < 2$ kpc) and blue, disk-like morphologies.  All are fainter, with $-16 < M_r < -18$, and presumably have an H \textsc{i} content that is slightly below the detection limits.  To test this we scaled the intensity of emission in each cube to a common distance and, assuming the spatial and velocity resolution remains roughly equivalent between targets, stacked the emission for these non-detected targets.  This increased the sensitivity to $\sim$ 0.24 mJy per beam and resulted in a (1.2 $\pm$ 0.3) $\times 10^8$ M$_\sun$ detection, just below the detection limits for the galaxies individually.  Stacking limits from combining the other eight non-detections are  considerably lower and not statistically significant, (4.9 $\pm$ 1.7) $\times 10^7$ M$_\sun$.  However, we note that the upper limits for the H \textsc{i} content in the non-detected galaxies in general are not inconsistent with the expected H \textsc{i} content based on the galaxy luminosities (see Section \ref{sec:hicontent}).

\subsection{HI morphology and kinematics}
\label{sec:morphkin}
H \textsc{i} imaging of the VGS reveals it to contain a diverse collection of H \textsc{i} rich galaxies, with about half showing signs of  strongly disturbed gas morphology or kinematics.  Five are clearly interacting with companions, four have some gas that does not follow the regular disk rotation, five are asymmetric at the 3$\sigma$ H \textsc{i} contours, six have a lopsided H \textsc{i} distribution, eight are kinematically lopsided, and three have warps (see Figure \ref{fig:vgs}, for the complete atlas).  Similar visual inspection of the WHISP galaxies in \cite{Swaters2002} reveal that nearly all their targets have irregularities in the gas morphology or kinematics with similar frequency.  This is also found in studies of integrated H \textsc{i} galaxy profiles \citep{Richter1994}. The Ursa Major cluster galaxies reside in a very different environment and appear significantly more regular, with only about a quarter having strong irregularities in the gas, however as they are larger they may also be affected by deeper gravitational potentials.  Both the WHISP and Ursa Major cluster galaxies have a significant number of H \textsc{i} rings, which we would be unable to resolve for the majority of the VGS galaxies. 

\begin{figure*}[t!]
\centering
\includegraphics[height=2.2in]{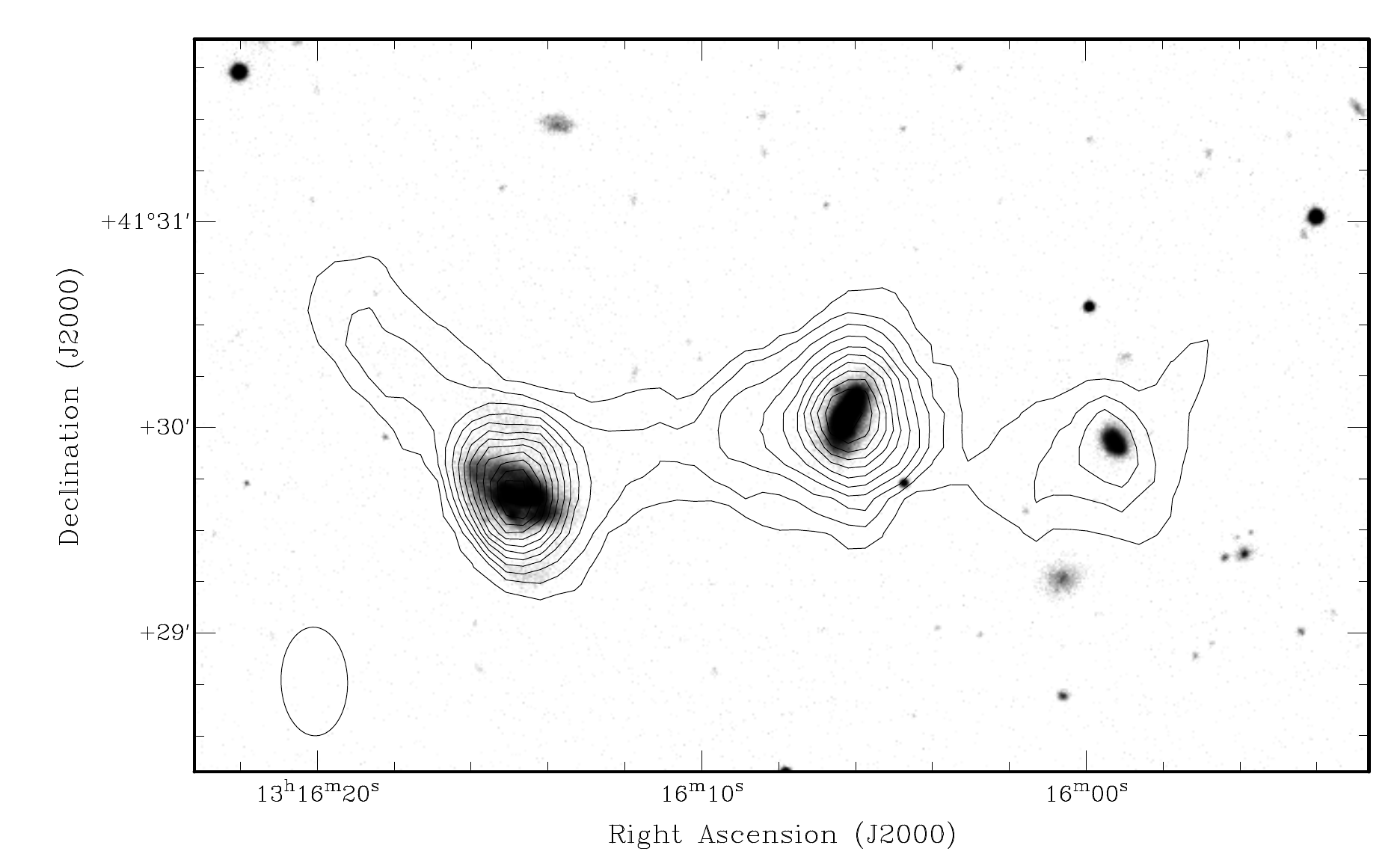}
\includegraphics[height=2.2in]{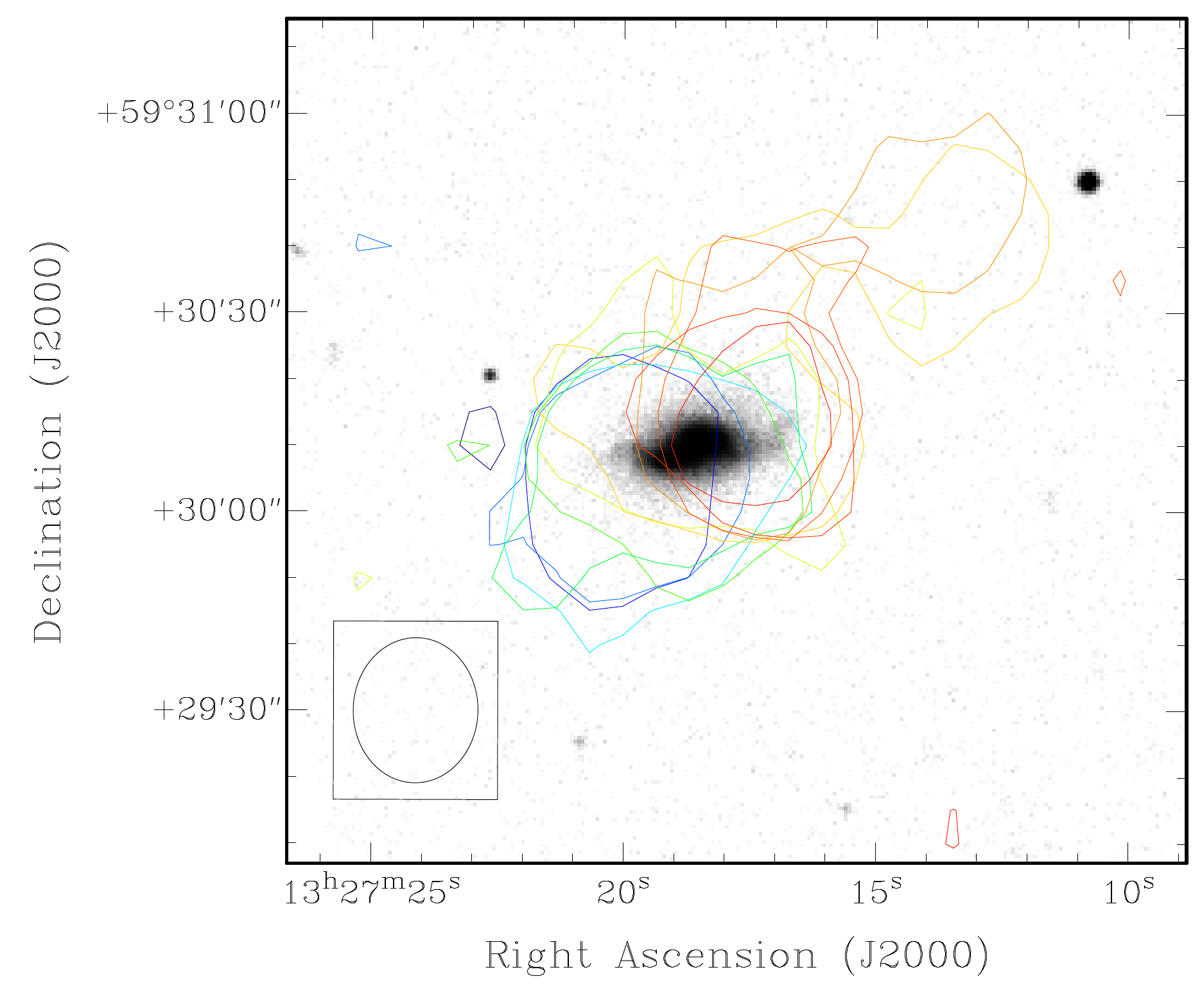} \\
\caption{Examples of void galaxies with gas outside the disk.  VGS\_31 (left) shows signs of interaction between three close galaxies connected by a low column density H \textsc{i} bridge.  VGS\_34 (right) shows disk rotation along the optical disk with a significant amount of H \textsc{i} at velocities inconsistent with regular rotation to the north-west side. VGS\_31 is imaged with natural weighting and contours at 5 $\times$ 10$^{19}$ cm$^{-2}$ plus increments of 10$^{20}$ cm$^{-2}$.  VGS\_34 overlays velocity coded contours at 1.2 mJy beam$^{-1}$ (2.5$\sigma$) from channels 24 km s$^{-1}$ apart. 
\label{fig:examples1}}
\end{figure*}

\begin{figure*}[t!]
\centering
\includegraphics[height=1.8in]{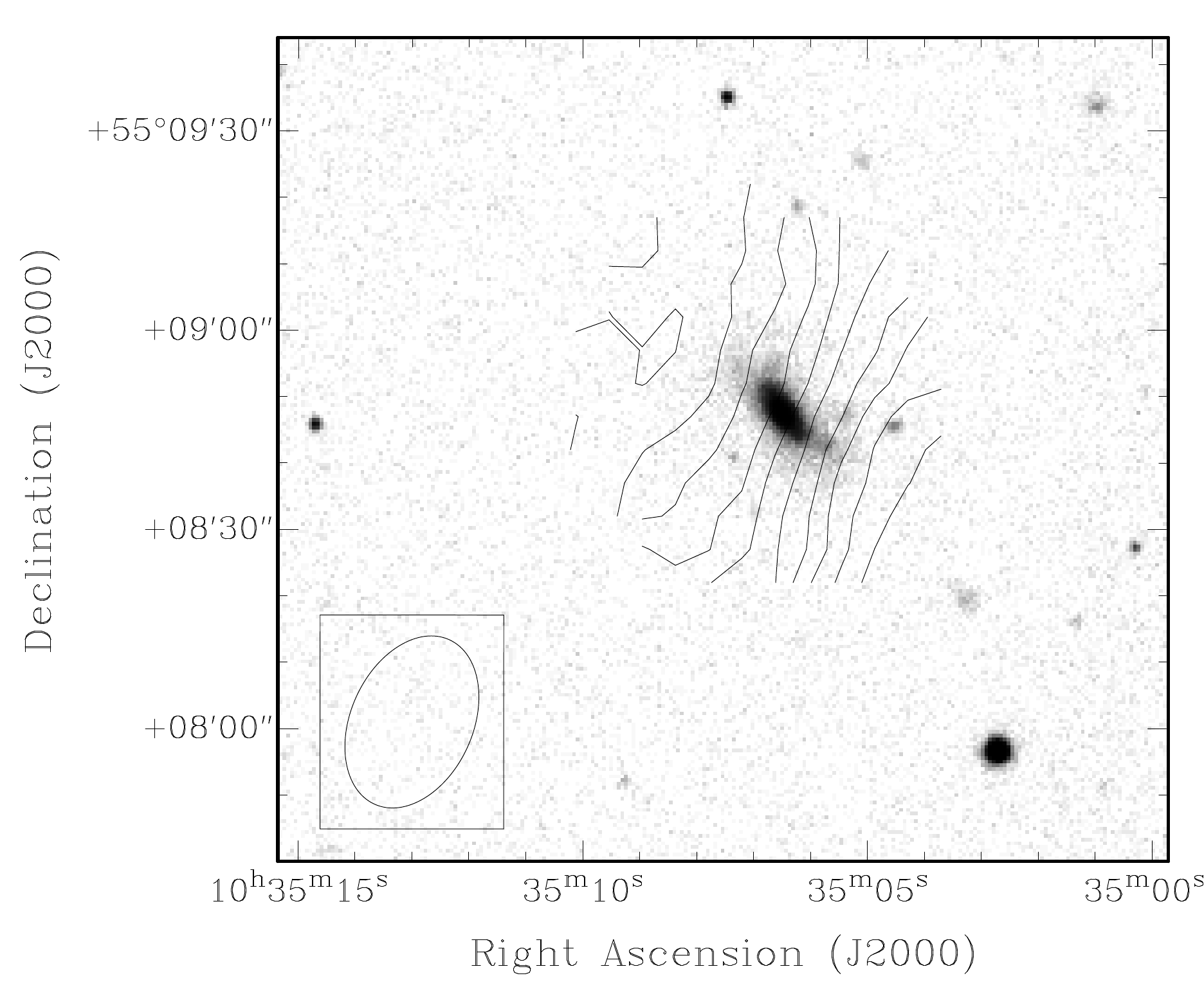}
\includegraphics[height=1.8in]{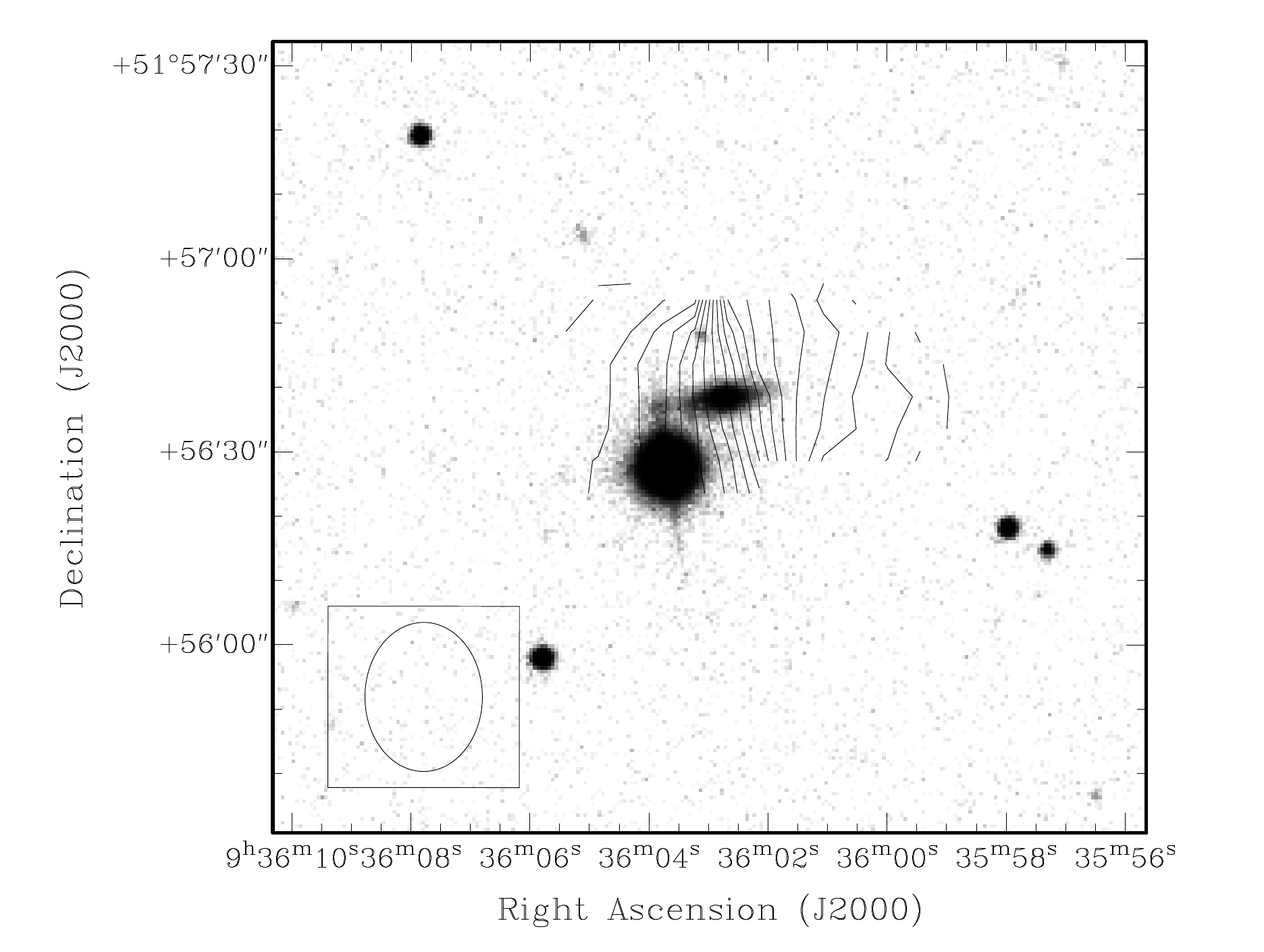}
\includegraphics[height=1.75in]{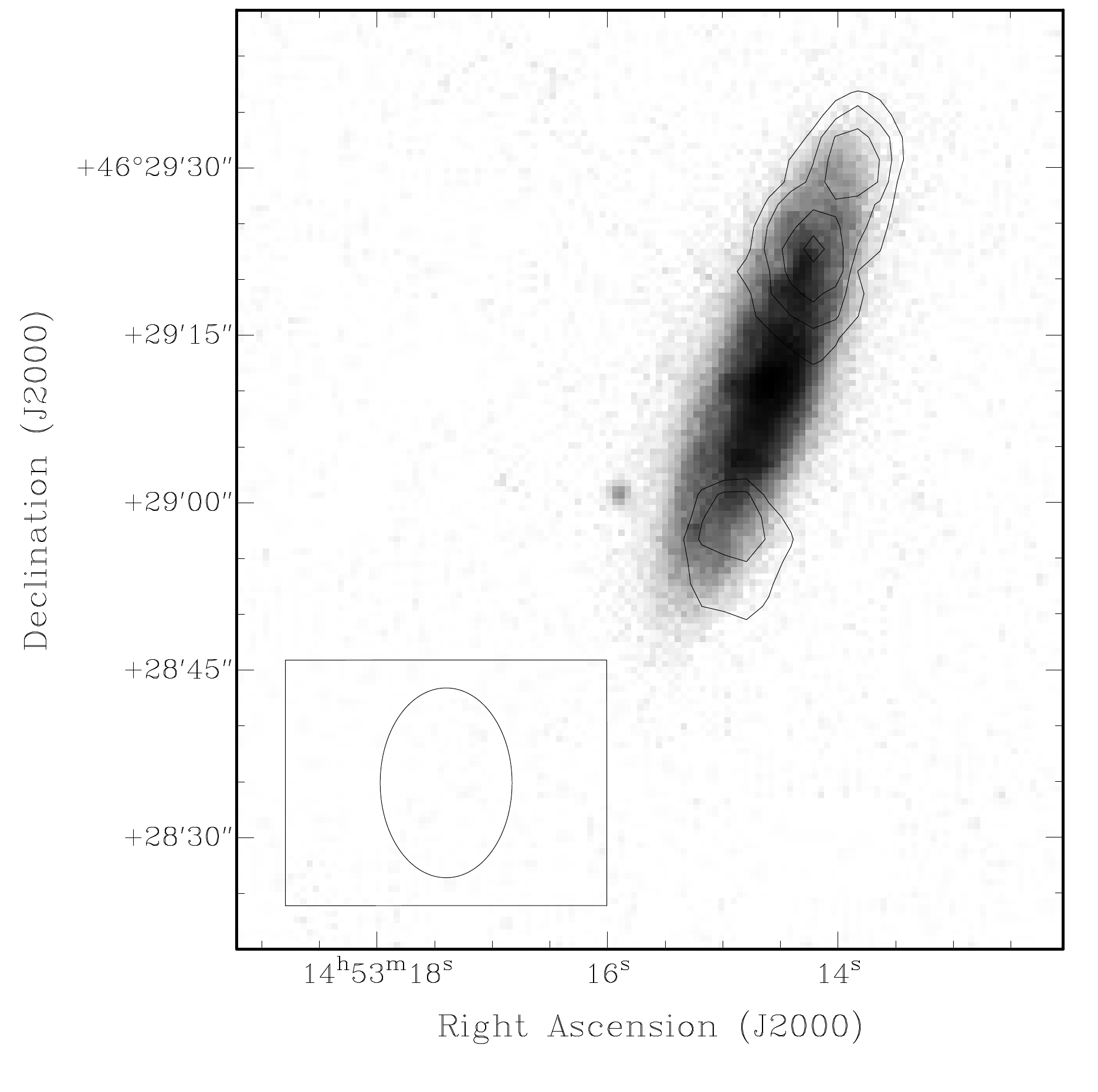} \\
\caption{Examples of void galaxies with irregular gas disks.  VGS\_14 (left) is kinematically lopsided, with a shallower velocity gradient on the north-east side of the optical disk.  VGS\_06 (center) is warped in its outmost extent, with velocity field lines that are perpendicular in the center to the disk major axis but warps strong to the west and more weakly to the east.  VGS\_47 (right) has a lopsided H \textsc{i} distribution, with significantly more H \textsc{i} to the north-east edge of the disk.   Velocity field lines in VGS\_14 and VGS\_06 are at intervals of 8 km s$^{-1}$.  VGS\_47 has been imaged using uniform weighting to increase the resolution, and velocity smoothed to 35 km s$^{-1}$ to improve the sensitivity.  Here the H \textsc{i} contours show lopsided distribution of the high column density gas, at 8 $\times$ 10$^{20}$ cm$^{-2}$ (2$\sigma$)  plus increments of 4 $\times$ 10$^{20}$ cm$^{-2}$.
\label{fig:examples2}}
\end{figure*}

\begin{figure}
\centering
\includegraphics[width=1.8in]{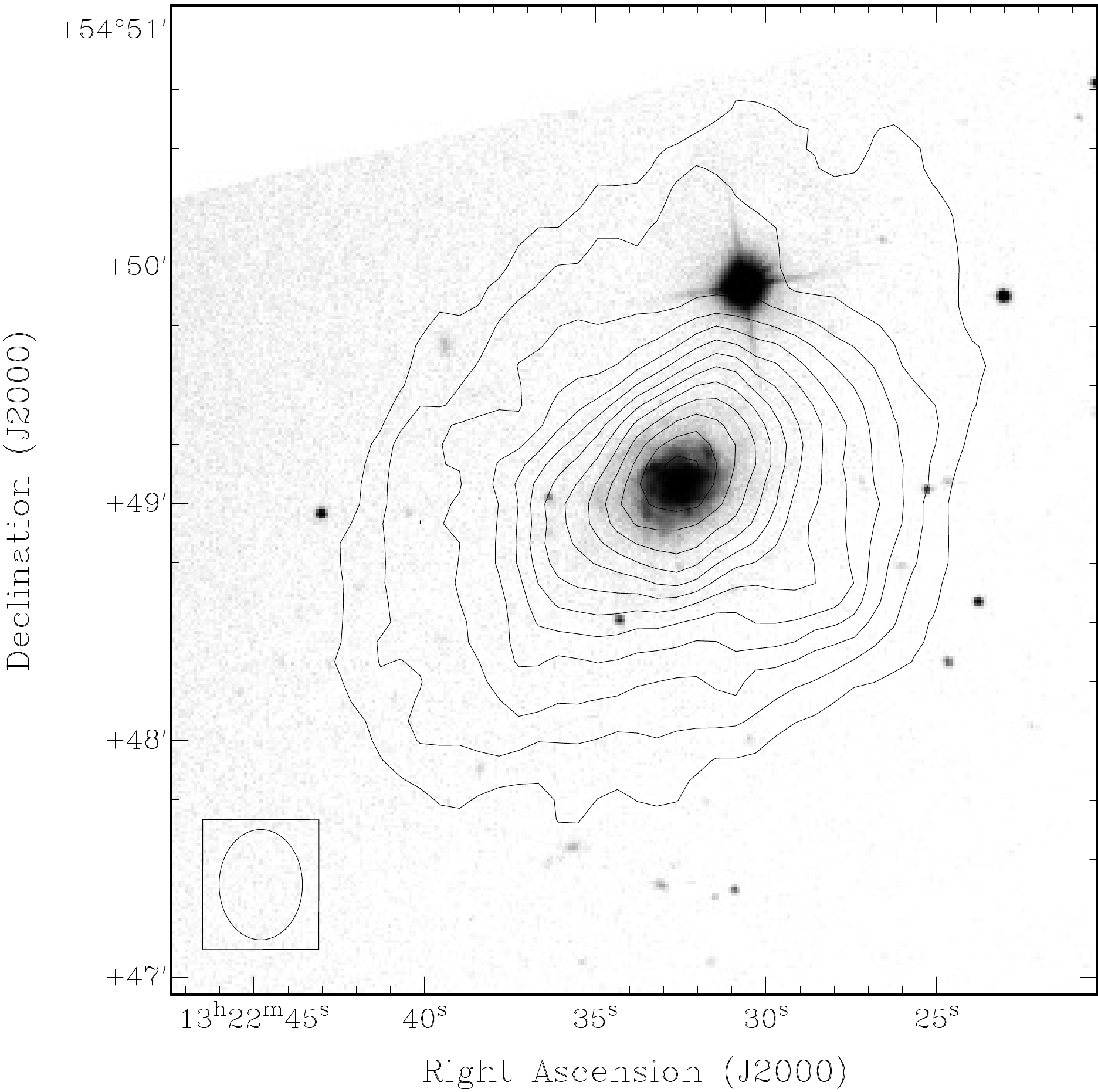} 
\includegraphics[width=1.45in]{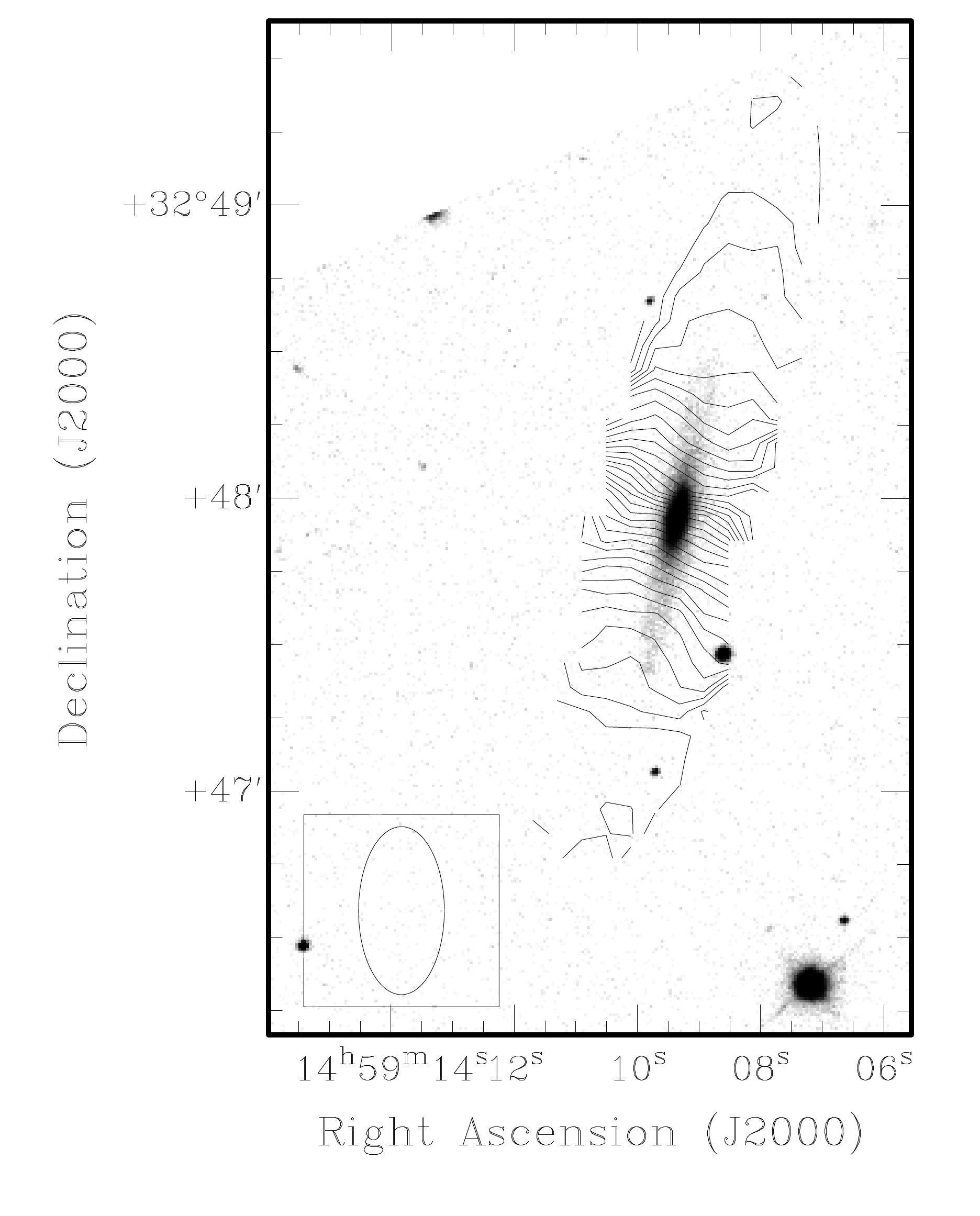}
\caption{Examples of void galaxies with regularly distributed gas disks. VGS\_32 (left) has a well resolved, fairly symmetric H \textsc{i} disk. VGS\_50 (right) is nearly edge on, and kinematically symmetric with no sign of a warp.  H \textsc{i} contours in VGS\_32 are at 9 $\times$ 10$^{19}$ cm$^{-2}$ (3$\sigma$) plus increments of 10$^{20}$ cm$^{-2}$ for a naturally weighted image.  The velocity field in VGS\_50 indicates intervals of 8 km s$^{-2}$.
\label{fig:examples3}}
\end{figure}

Some of the VGS galaxies are particularly striking.  Figure \ref{fig:examples1} shows two examples of galaxies with kinematically irregular gas located outside the disk with no clear optical counterpart.   VGS\_34 has an H \textsc{i} disk with regular rotation that is centered on the optical galaxy, however in addition it has low column density gas to the north-west that persists over multiple channels in velocity.  This gas is inconsistent with the disk rotation and has no clear optical counterpart, however the optical morphology is quite disturbed, suggesting a recent interaction. It also has a small companion galaxy about 100 kpc away at a nearly coincident velocity.  VGS\_31 is one of two VGS galaxies discovered to have two nearby companions, and in both cases the three galaxies are linearly aligned and joined within a low column density common envelope of H \textsc{i}.  Figure \ref{fig:examples2} shows three more examples of gas disks that are morphologically or kinematically lopsided.  VGS\_14 is kinematically lopsided, exhibiting a shallower velocity gradient to the north-east side.  VGS\_06 is warped in the outer extent, and VGS\_47 is very lopsided in the distribution of H \textsc{i} across the disk.  We also show for completeness in Figure \ref{fig:examples3} examples of galaxies that are fairly regular in the H \textsc{I} morphology (VGS\_32) and kinematics (VGS\_50).

The high fraction of strongly irregular H \textsc{i} disks is surprising if we consider void galaxies as evolving in relative isolation, where galaxies are observed to have a smaller fraction of asymmetric H \textsc{i} profiles \citep{Espada2011}.  However void galaxies are not by definition isolated. 
\cite{Szomoru1996} found that on small scales ($<$ 1 Mpc) the number of neighboring galaxies detected around targeted galaxies in the Bo\"{o}tes void is similar to that in average environments, a result supported by the VGS sample (see Section \ref{sec:clustering}).  The strong H \textsc{i} irregularities present clear evidence of ongoing interactions and gas accretion, and is fairly typical for gas rich galaxies.

\subsection{HI content}
\label{sec:hicontent}
\begin{figure}
\centering
\includegraphics[width=3.2in,angle=0]{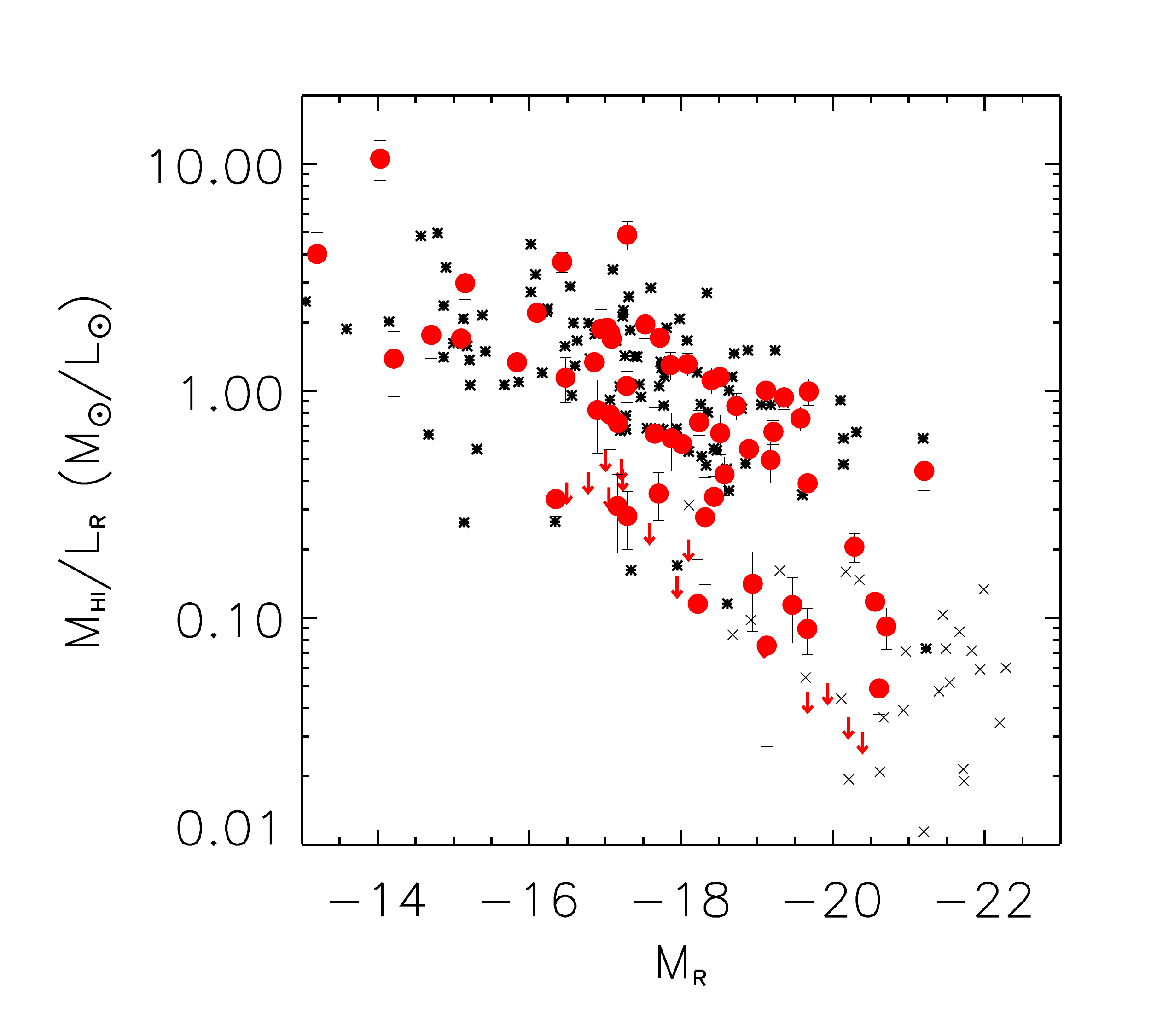}
\caption{H \textsc{i} mass to light ratio as a function of $R$-band absolute magnitude for the VGS and companion galaxies (circles) compared with WHISP galaxies (stars, \citealt{Swaters2002}) and Ursa Major cluster galaxies (crosses, \citealt{Verheijen2001}). Upper limits on the VGS non-detections are also indicated (arrows).  Our VGS follows the trend for fainter galaxies to have relatively higher H \textsc{i} mass to light ratios.
\label{fig:mhilight}}
\end{figure}

The VGS reproduces the trend for fainter galaxies to be relatively more gas rich than brighter galaxies, as is generally observed in disk galaxies.  In Figure \ref{fig:mhilight} we compare the H \textsc{i} mass to light ratio of the void and companion galaxies with H \textsc{i} imaged galaxies in the WHISP sample \citep{Swaters2002} and the Ursa Major cluster \citep{Verheijen2001}.  Here we have converted the SDSS $ugriz$ bands to Johnson-Cousins $R$-band magnitudes following the prescription of \cite{Jester2005}. The VGS generally agrees with the two samples within the errors, with the upper limits for the non-detections also falling within the range observed.  All three samples achieve similar H \textsc{i} column density sensitivities, though our sample is at significantly larger distances.  

Only one galaxy, VGS\_12, appears to have an unusually large H \textsc{i} mass to light ratio of 6.2 given its absolute magnitude M$_r$ = -17, and from the misalignment of its H \textsc{i} disk it is clear that the evolution of this system is progressing in an unusual fashion (see also Section \ref{sec:accretion}).  We observe no systematic increase in the H \textsc{i} mass to light ratio for the VGS, somewhat in contrast to previous observations and predictions from cosmological simulations.
\cite{Kreckel2011a}  report that simulated galaxies in the most underdense regions have slightly higher H \textsc{i} mass to light ratios at fixed luminosity.  An increased H \textsc{i} mass to light ratio has also been reported in observational H \textsc{i} studies of dwarf galaxies in voids \citep{Huchtmeier1997}, 
and a small sample of void galaxies selected from the Second Byurakan and Case surveys \citep{Pustilnik2002}.  The significantly larger sample size in this study allows more robust comparison with `average' galaxies, and provides no evidence for an increased H \textsc{i} mass to light ratio.

\subsection{Star formation}
\label{sec:sfr}
\begin{figure}
\centering
\includegraphics[width=3.2in,angle=0]{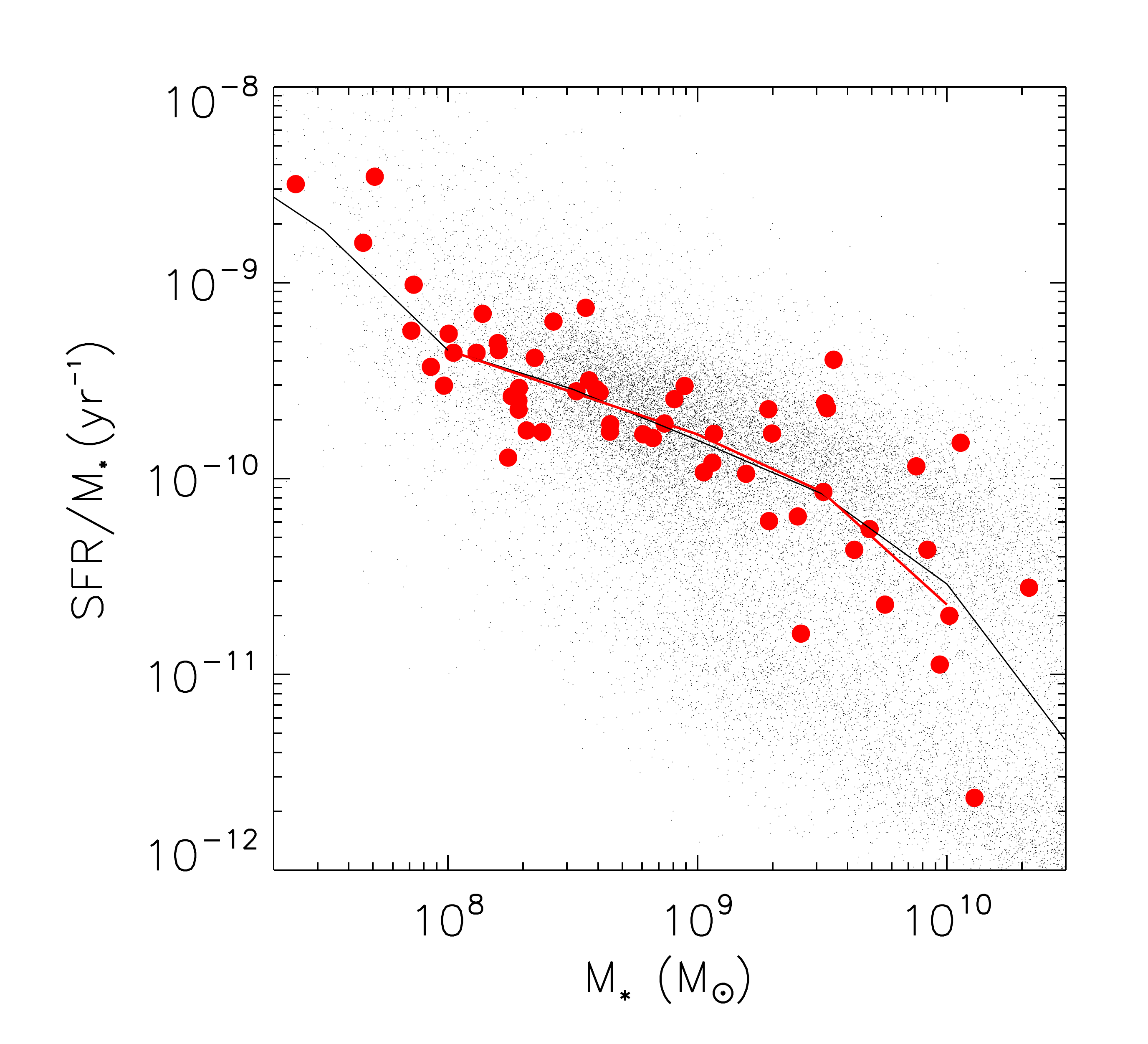}
\caption{SFR normalized by stellar mass as a function of stellar mass for the VGS (red) and a magnitude limited ($z < 0.03$, m$_r < 17.77$) sample of SDSS galaxies (black).  The medians of both samples (lines) agree quite well, suggesting that buildup of the stellar disks in the VGS is appropriate to their smaller sizes.
\label{fig:ssfrsdss}}
\end{figure}

None of the VGS galaxies have SFRs that would suggest they are undergoing a starburst, the highest having $\sim$2 M$_\sun$ yr$^{-1}$ but the majority falling below 1 M$_\sun$ yr$^{-1}$.
As the SFR is typically lower for smaller galaxies, it is useful to normalize the SFR by the stellar mass and consider the specific star formation rate (SSFR) for these systems. The SSFR has been found to be enhanced in void galaxies at fixed luminosity \citep{Rojas2005, BendaBeck2008}, which suggests that as a population they are still in the process of building their stellar disks. However, as this is generally the case for smaller galaxies \citep{Kauffmann2003c}, it is not clear if these results hold at fixed luminosity and morphological type \citep{Park2007}.  Comparing the VGS with a magnitude limited ($z < 0.03$, m$_r < 17.77$) sample of SDSS galaxies we find no trend for higher SSFR as a function of galaxy stellar mass (Figure \ref{fig:ssfrsdss}).  This discrepancy with past results may come from the different range of luminosities probed, as our sample consists of relatively faint void galaxies (Figure \ref{fig:hists}), whereas the most significant effect on SSFR is identified in void galaxies brighter than M$_r \sim -19$ \citep{Rojas2005, BendaBeck2008}.    

Recently the star formation efficiency (SFE), measured as the star formation rate normalized by the H \textsc{i} mass, has been considered as a way to judge the general effectiveness of turning gas into stars. 
The GALEX Arecibo SDSS Survey (GASS) compares the SFE for a large sample of galaxies with stellar masses above $10^{10}$ M$_\sun$, and finds it stays relatively constant at log$_{10}$(SFE) $\sim$ -9.5 when considered as a function of parameters such as stellar mass, color, and morphology \citep{Schiminovich2010}.  Only four galaxies in the VGS have stellar masses within this range, two of which are detected and have log$_{10}$(SFE) of  -8.6 and -9.1, higher than the average and suggestive that void galaxies are somewhat more efficient in their star formation. Two are not detected in H \textsc{i}, with lower limits on the log$_{10}$(SFE) of  -10  and -9.1.  We continue our investigation into the efficiency of star formation in void galaxies with lower stellar masses in Section \ref{sec:alfalfa}.

\section{Discussion}
\label{sec:discussion}
Given the observed H \textsc{i} and optical properties of the VGS, we consider what role the large scale environment might play in the evolution of these systems and the implications for cosmology.

\subsection{Evidence for ongoing gas accretion}
\label{sec:accretion}
The strong disturbances we observe in the H \textsc{i} morphologies and kinematics (see Section \ref{sec:morphkin}) present convincing evidence for ongoing gas accretion in these systems.   
None of the galaxies with lopsided H \textsc{i} disks highlighted in Figure \ref{fig:examples2} have companions detected in H \textsc{i}, however these galaxies are small enough, with H \textsc{i} radii of $\sim$10 kpc and rotational velocities of $\sim$150 km s$^{-1}$, that rotation timescales are only a few hundred million years and without recent gas accretion these disks are expected to settle in much less than a Hubble time \citep{Simonson1982}. 
However it is difficult to distinguish the effects of different gas accretion mechanisms, such as interactions, internal gas recycling, and cosmological inflow.
Simulations suggest that gas accreted from the intergalactic medium can either be shock heated to form a hot halo or can penetrate directly to the halo center as a cold flow, depending on the halo mass.  For low mass halos, found predominantly at higher redshifts and at $z = 0$ in low density environments, these cold flows may be the dominant mechanism for galaxy growth \citep{Birnboim2003, Keres2005, Dekel2006, Dekel2009, Keres2009}.   

This appears to be the case for VGS\_12, which was found to have an extremely extended and massive H \textsc{i} disk oriented perpendicular to the direction of the stellar disk, and is discussed at length in \cite{Stanonik2009}. The polar disk is devoid of stars, and the central stellar disk has retained its rotational support, suggesting a slow accretion of polar material.  This mechanism for the formation of polar disks has also been suggested for NGC 4650A \citep{Iodice2006}, and has been reproduced in simulations \citep{Maccio2006}. In the VGS we also find two systems, VGS\_38 and VGS\_31, with three galaxies linearly aligned and joined by low column density H \textsc{i}, that is very suggestive of filamentary formation.   Hierarchical void theory predicts that void galaxies will reside in relatively low density substructure within the void \citep{Dubinski1993, Sahni1994, Shethwey2004, Furlanetto2006, Einasto2011}, which has been observed in simulations \citep{Springel2005} as well as in nearby voids \citep{Popescu1997}, and the VGS systems present a unique location to search for surrounding low column density  intergalactic gas.

Further evidence of gas accretion in void galaxies can be found in the literature.  Individual studies of dwarf galaxies in voids suggest that they are uniquely unevolved, with lower metallicities and higher star formation rates than typical dwarf galaxies \citep{Corbin2005, Pustilnik2006}.  KK 246, a dwarf galaxy in the Local Void, has an extremely extended H \textsc{i} disk and evidence of gas at anomalous velocities compared to regular disk rotation \citep{Kreckel2011b}.  NGC 6946 has been imaged in H \textsc{i} because of its unusual gas kinematics and signs of gas infall \citep{Boomsma2008}, and was selected for these characteristics independent of its location within the Local Void \citep{Sharina1997}. Individually, these examples are interesting, but taken together they convincingly show that voids are a uniquely promising place to search for evidence of ongoing gas accretion.

\subsection{Void galaxy H I properties at fixed morphology, luminosity, and stellar mass}
\label{sec:alfalfa}
\begin{figure}
\hspace{-.5cm}~\includegraphics[width=1.9in]{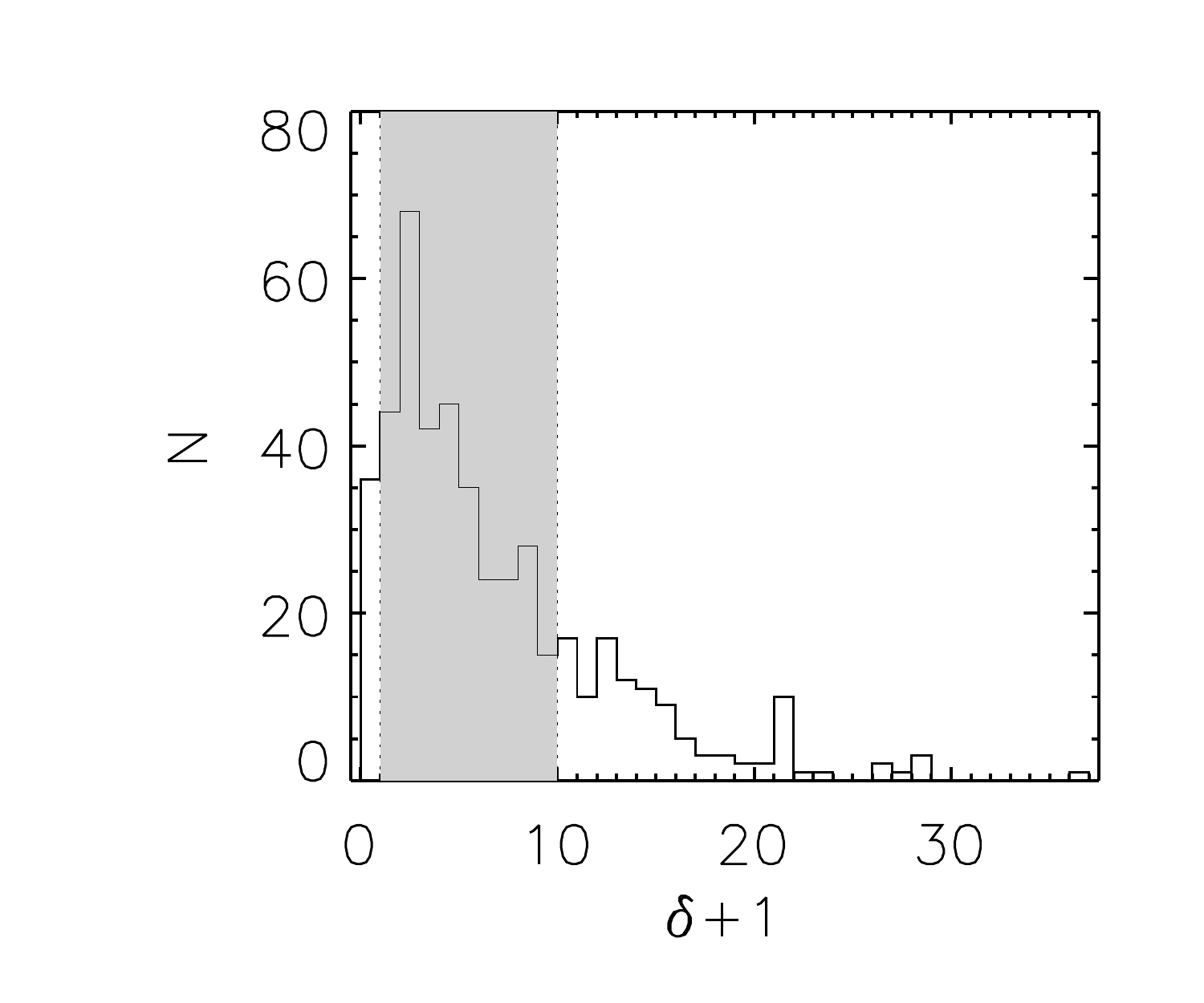}~\hspace{-.7cm}~\includegraphics[width=1.9in]{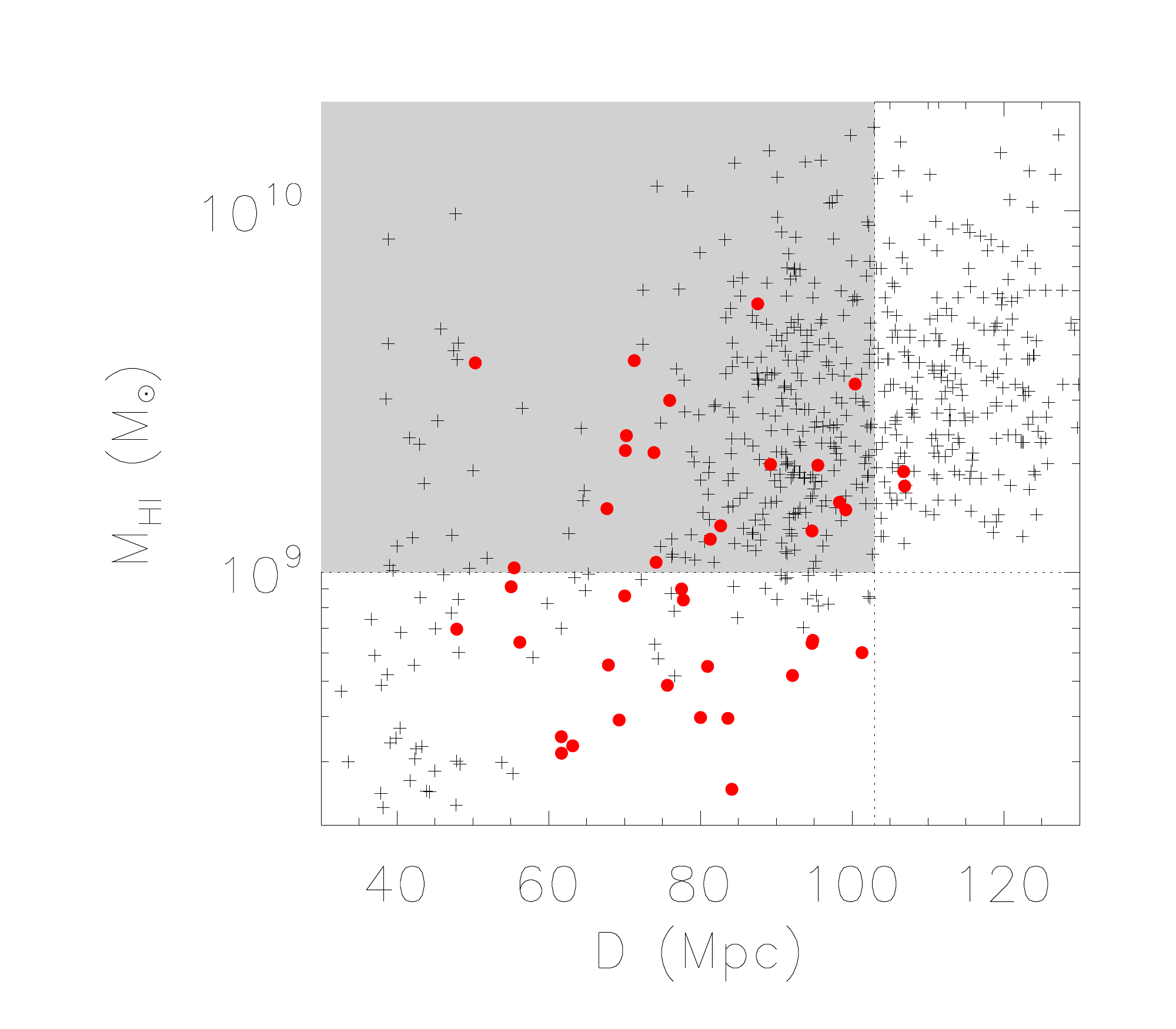}
\caption{Properties of the galaxies cross-matched between the ALFALFA and SDSS DR7 catalog.  The 447 nearby galaxies span a representative range of environments, including both void and cluster galaxies (left).   The Malmquist bias in the ALFALFA H \textsc{i} flux limit is apparent in the Spaenhauer diagram (right), and the VGS galaxies (circles) are detected at lower H \textsc{i} masses to greater distances compared to the ALFALFA galaxies (crosses).  As a control sample, we construct a volume limited (M$_{\textrm{H \textsc{i}}} > 10^9$ M$_\sun$, $z < 0.024$) sample in average ($1 < \delta +1 < 10$) environments (in grey) of 207 ALALFA galaxies.  
\label{fig:ctrlhist}}
\end{figure}

To carefully examine the integrated H \textsc{i} properties for our sample considering their biased morphology, luminosity and stellar mass distributions, we have constructed an environmentally constrained control galaxy sample by cross-matching galaxies in five publicly available catalogs of the Arecibo Legacy Fast ALFA  Survey (ALFALFA; \citealt{Giovanelli2007, Saintonge2008, Kent2008, Martin2009, Stierwalt2009}) with the Sloan Digital Sky Survey Data Release 7 (SDSS DR7; \citealt{Abazajian2009}). The ALFALFA catalog identifies H \textsc{i} detected galaxies with a signal-to-noise ratio greater than 6.5 sigma, as well as marginal detections  between 4 and 6.5 sigma that have clear optical counterparts.  The ALFA receiver's 3.5 arcminute beam allows sufficient resolution to uniquely determine a corresponding optical counterpart for almost all cataloged H \textsc{i} detections, though confusion of sources within the beam can be an issue for more distant targets.  We have restricted this control sample to a redshift range of $0.007 < z < 0.024$ and a volume limited subset with M$_{\textrm{H \textsc{i}}} \geq 10^9$ M$_\sun$ to approximately match the H \textsc{i} sensitivity limits to the WSRT observations (Figure \ref{fig:ctrlhist}, see also Section \ref{sec:observations}).  Following the cross-matching technique described in \cite{Toribio2011}, we require that the   optical counterpart centers agree in each catalog by less than 10$^{\prime\prime}$, a typical minimum size for the optical disks of galaxies in that redshift range.  We also require that the optical and H \textsc{i} redshift velocities agree by less than 300 km s$^{-1}$, though most source optical redshifts are within $\sim$30 km s$^{-1}$ of the H \textsc{i} detection.   We omit any targets flagged as suffering from confusion within the beam, and we further exclude 24 galaxies from this cross-matched catalog that are identified within the SDSS redshift survey to have close companions, within 2$^\prime$ and a velocity difference of (W$_{50}$/2 + 50 km s$^{-1}$).  We recalculate the distances for the ALFALFA detections and adjust the H \textsc{i} masses based on the Hubble flow velocity, instead of the cataloged CMB velocity, to remove any bias when comparing with the VGS.  Given the imposed lower limit on the redshift range this does not significantly affect our distance estimates.  We identify 447 galaxies for this control sample, with 207 in the volume limited subsample.  As expected, the bulk of the control sample galaxies are located at average densities, with a limited number in voids and clusters (Figure \ref{fig:ctrlhist}).

\begin{figure*}[t!]
\centering
\includegraphics[width=3in,angle=0]{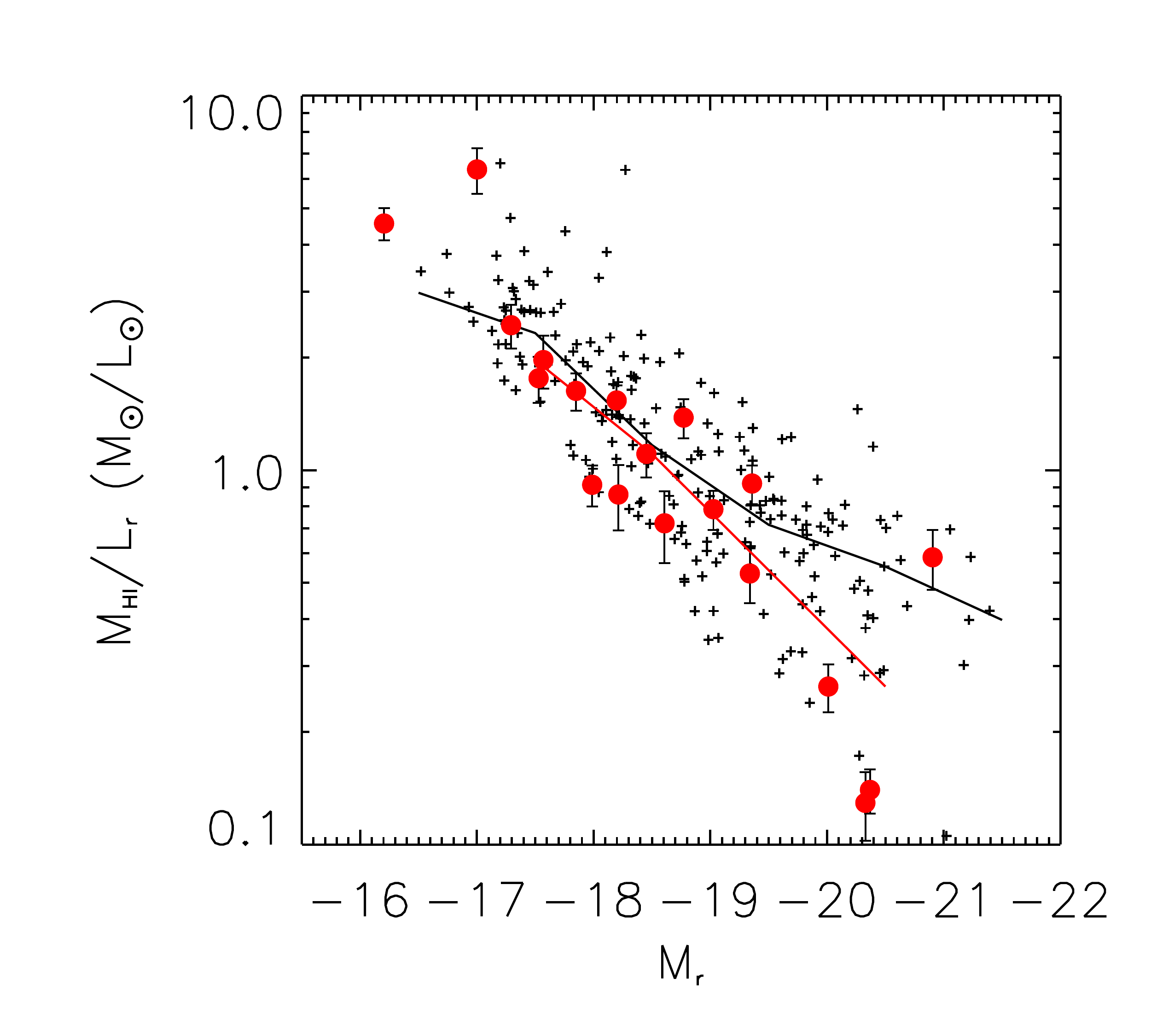}
\includegraphics[width=3in,angle=0]{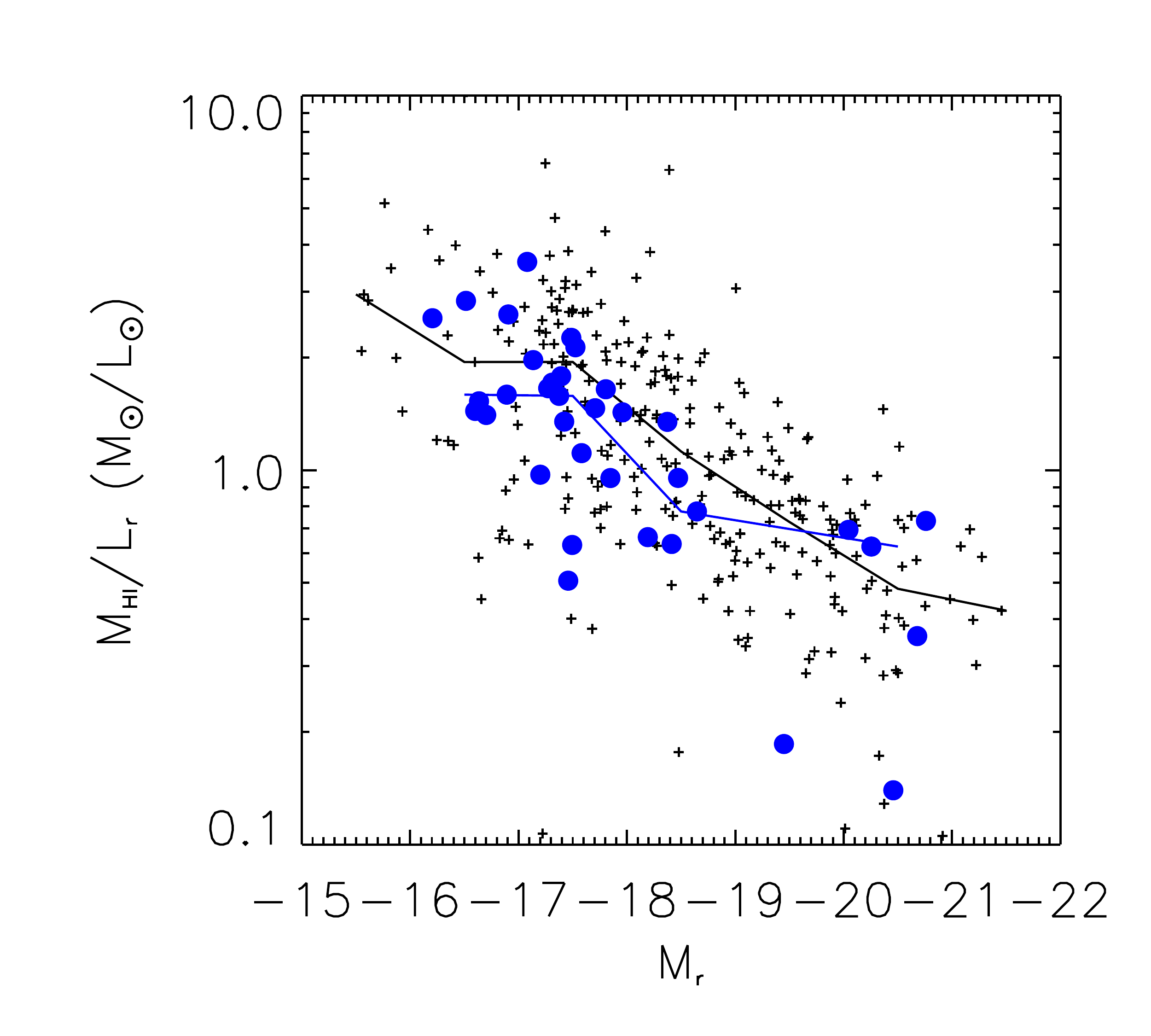}
\caption{Left: H \textsc{i} mass to light ratio for volume limited subsets (M$_{\textrm{H \textsc{i}}}$ $>$ 10$^9$ M$_\sun$) of late type galaxies in the VGS (circles) and ALFALFA control sample (crosses) in average ($1 < \delta + 1 < 10$) environments.  
Our void galaxies are slightly offset from the ALFALFA control sample.  
Right: H \textsc{i} mass to light ratio of all ALFALFA galaxies in average environments  (crosses) compared with low density  ($1 < \delta + 1 < 10$) environments (circles) as a function of $r$-band absolute magnitude.  A systematic offset is much more apparent, suggesting the offset found for the VGS galaxies may be real.
In all figures, errors on ALFALFA are omitted as they are typically smaller than the symbol. Lines show the median of each sample.  
\label{fig:mhilight2}}
\end{figure*}

We find that late type (r$_{90}$/r$_{50} < 2.86$) void galaxies have approximately the same median H~\textsc{i} mass to light ratios at fixed luminosity as compared to late type galaxies in the volume limited ALFALFA control sample (Figure \ref{fig:mhilight2}, left).  Here the void sample has been subjected to the same volume limited (M$_{\textrm{H \textsc{i}}}  > 10^9$ M$_\sun$) restrictions to avoid a Malmquist bias.
The slight systematic offset to lower values is approximately equal to the typical errors due to uncertainty in the H \textsc{i} mass, however the lack of high H \textsc{i} mass to light ratio galaxies in the VGS brighter than M$_r = -17$ is striking. As this effect is not apparent when comparing with the WHISP and Ursa Major cluster galaxies (Figure \ref{fig:mhilight}), we consider whether the total H \textsc{i} masses may be affected by instrumental differences between the WSRT and Arecibo observations.

Three galaxies in the VGS sample overlap with the ALFALFA control, and for these we find H \textsc{i} masses that agree within the quoted errors.
It is also possible that the H \textsc{i} mass mesaurements are strongly contaminated by confusion within the ALFALFA beam, which would result on average in an overestimate of the total H \textsc{i} mass.  At a redshift of $z = 0.024$ the 3.5$^\prime$ beam includes any emission within a physical distance of $\sim$50 kpc, approximately the distance from the Milky Way to the Large Magellanic Cloud. This is a relatively small separation, for example five of the target galaxies in the VGS sample when observed by Arecibo would potentially suffer from confusion with their nearby companions, resulting in mass increases of up to a factor of two.

There are further concerns with the absolute mass measured in the ALFALFA sample.  \cite{Toribio2011} examined a sample of cross-matched SDSS and ALFALFA galaxies, and for a subsample  of well resolved galaxies with known morphological type and optical diameter they calculated the expected H \textsc{i} mass following \cite{Haynes1984} and \cite{Solanes1996}.  They reported a negative H \textsc{i} deficiency of 20-30\% which increases with redshift, suggesting that their H \textsc{i} masses may be overestimated, however they attribute this to the omission of H \textsc{i} deficient galaxies below the survey detection limits, the inclusion of more gas rich late type galaxies, as well as a difference in the methods used compared to previous work.
  
While these concerns affect the comparison of absolute H \textsc{i} flux measurements, they should not affect relative comparisons within the ALFALFA sample.  In fact, we observe a similarly decreased H \textsc{i} mass to light ratio when comparing the low density ($\delta +1 < 1$) subset of ALFALFA control galaxies with average density ALFALFA galaxies (Figure \ref{fig:mhilight2}, right).  For both the low density ALFALFA galaxies and the VGS, the galaxies in low density regions appear to have an H \textsc{i} mass to light ratio that is approximately 10-20\% lower than that of galaxies in average density environments.  We consider such a large contribution to the total H \textsc{i} mass by confusion unlikely but possible, as this would require a strongly environmental effect, with gas-rich galaxies in low density regions less clustered.  This is in contradiction with our own findings (see Section \ref{sec:clustering}).

\begin{figure}[b!]
\centering
\includegraphics[width=3.2in]{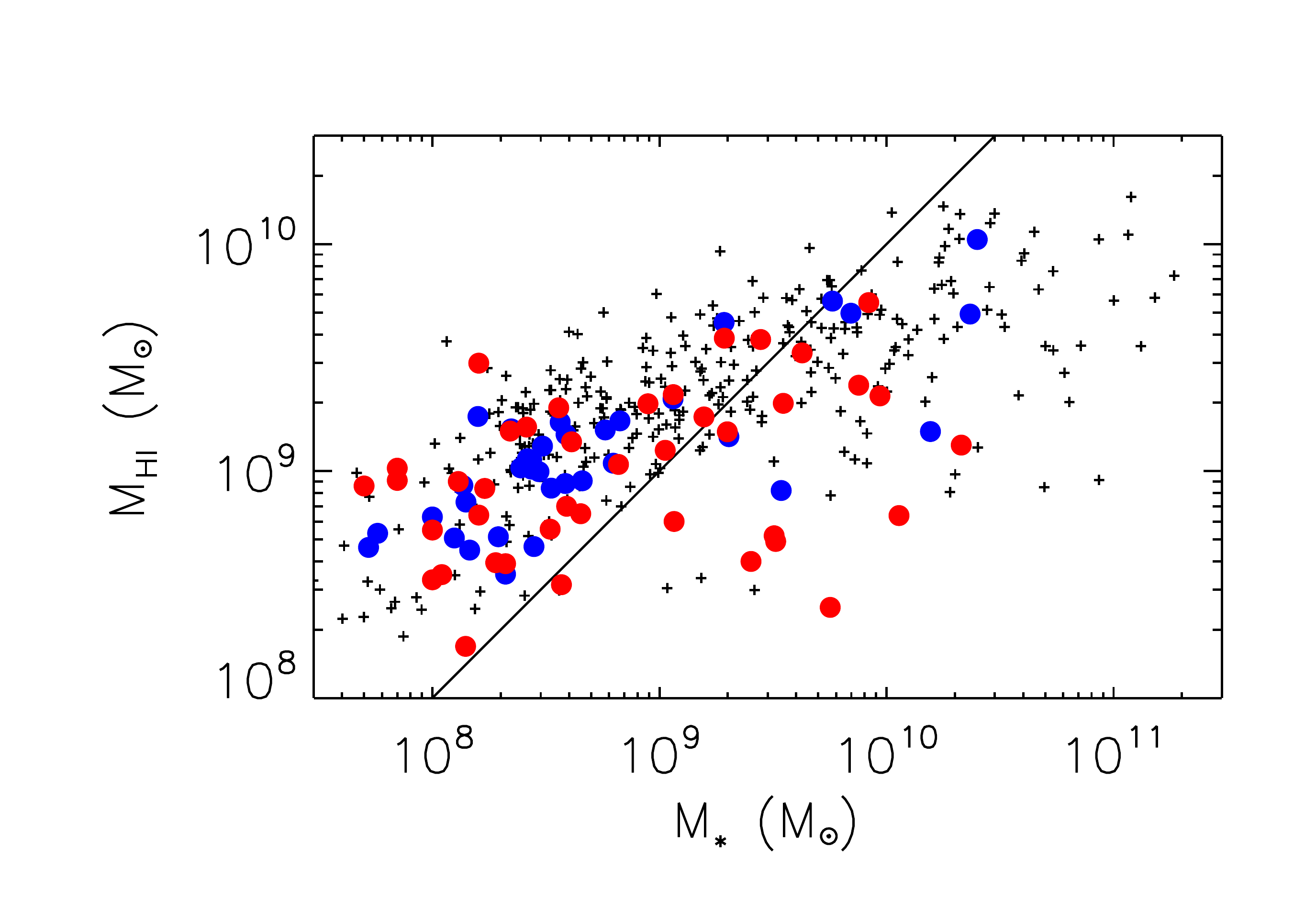}
\caption{The total H \textsc{i} mass as a function of stellar mass for the VGS (red circles) and ALFALFA low density (blue circles) and average density (crosses) galaxies. The line is drawn to indicate equal amounts of stellar and H \textsc{i} gas mass.  For all galaxies, the range of H \textsc{i} mass spans only two orders of magnitude while the stellar mass spans more than three orders of magnitude.  Note that the ALFALFA samples are strongly biased towards detecting targets with H \textsc{i} masses above 10$^9$ M$_\sun$.  We find very few void galaxies with high H \textsc{i} gas masses, and at the high stellar mass end we find no VGS galaxies with H \textsc{i} mass above $5 \times 10^9$ M$_\sun$. We also note the rather sharp transition in all environments between gas dominated galaxies below M$_* = 10^9$ M$_\sun$ and stellar dominated galaxies above M$_* = 10^{10}$ M$_\sun$.  
\label{fig:jplot}}
\end{figure}

The H \textsc{i} mass to light ratio is in some ways a rough proxy for comparing the H \textsc{i} and stellar masses of a system.  As the SDSS spectra allow us to more directly estimate the total mass in stars, we can in this case make an explicit comparison.  Figure \ref{fig:jplot} shows the H \textsc{i} mass as a function of stellar mass, as taken from the MPA-JHU catalog (see Section \ref{sec:sdssprop}), for the VGS and ALFALFA average and underdense samples, with a line indicating equal amounts of mass in stars and gas.  
We note that in all environments there is a smaller range in H \textsc{i} masses than stellar masses, and a rather sharp transition between gas dominated and stellar mass dominated galaxies.
Almost no galaxies with stellar masses below 10$^9$ M$_\sun$ have less H \textsc{I} than stars, and almost no galaxies with stellar masses above 10$^{10}$ M$_\sun$ have more gas than stars.
Given the range of stellar masses probed by these void galaxies, there appear to be very few with  high H \textsc{i} masses above $\sim$2 $\times$ 10$^9$ M$_\sun$, and none above $5 \times 10^9$ M$_\sun$, though this is routinely found for galaxies in average density environments.  The agreement between all samples appears somewhat better at low $\sim$10$^8$ M$_\sun$ stellar masses. Unfortunately, as ALFALFA is biased towards detecting targets with higher H \textsc{i} masses above $\sim10^9$ M$_\sun$, it is misleading to compare the mass distribution of our VGS targets directly with ALFALFA.  

\begin{figure}
\centering
\includegraphics[width=3.2in]{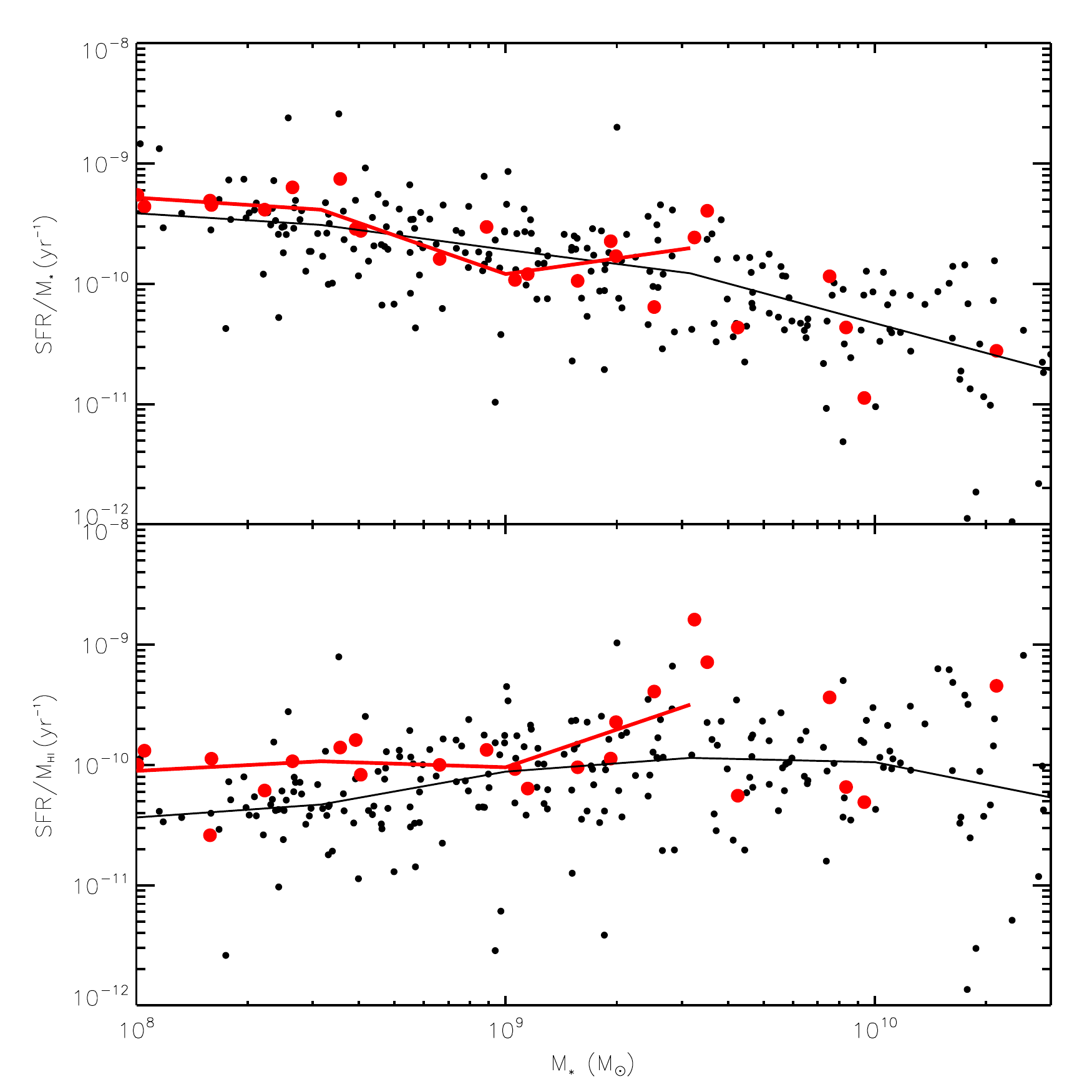}
\caption{SFR normalized by stellar mass (top) and H \textsc{i} mass (bottom) as a function of stellar mass for the VGS (red) and the ALFALFA average density sample (black).  The medians of both samples are overplotted (lines), revealing no strong trends in the specific star formation rate (top) but a systematically slightly higher star formation efficiency (bottom) for the void galaxies.
\label{fig:sfe}}
\end{figure}

We also compare the normalized star formation of the VGS galaxies with the average density ALFALFA galaxies (Figure \ref{fig:sfe}, top).  We find good agreement between the SSFRs when comparing the volume limited VGS and average density ALFALFA samples, which we also find when comparing with a magnitude limited sample of SDSS galaxies (see Section \ref{sec:sfr}). This  suggests that star formation per stellar mass is the same in voids as in higher density environments.  However the SFE appears systematically higher (Figure \ref{fig:sfe}, bottom), as was already suggested by our pilot study.  Again, the ALFALFA H \textsc{i} masses are susceptible to contamination within the beam, which would have the effect of decreasing the ALFALFA SFEs, but this observed trend is intriguing if true.

\subsection{Baryon content}
In Figure \ref{fig:tf} we show the $I$-band and baryonic Tully-Fisher relation for the void galaxy sample overplotted on the fit derived by \cite{Mcgaugh2000} and \cite{Geha2006}.  W$_{20}$ values have been corrected for instrumental velocity broadening and inclination.  Error bars reflect a 10\% uncertainty in the inclination angle.  In considering the $I$-band relation, we convert the SDSS $ugriz$ bands to the Johnson Cousins $I$-band using the transformation equations of \cite{Jester2005}, and find very good agreement with the general relation.  We calculate the total baryons by adding the stellar mass and 1.4 times the H \textsc{i} mass, to correct for the contribution from Helium and other metals.  While the baryonic relation has been found to typically have less scatter and more reliably fit lower mass galaxies where there is a more significant gas fraction, we find close agreement with relations from the literature but a slight systematic scatter towards higher velocities or lower baryon masses.

\begin{figure}[b!]
\hspace{-.5cm}~\includegraphics[width=1.8in,angle=0]{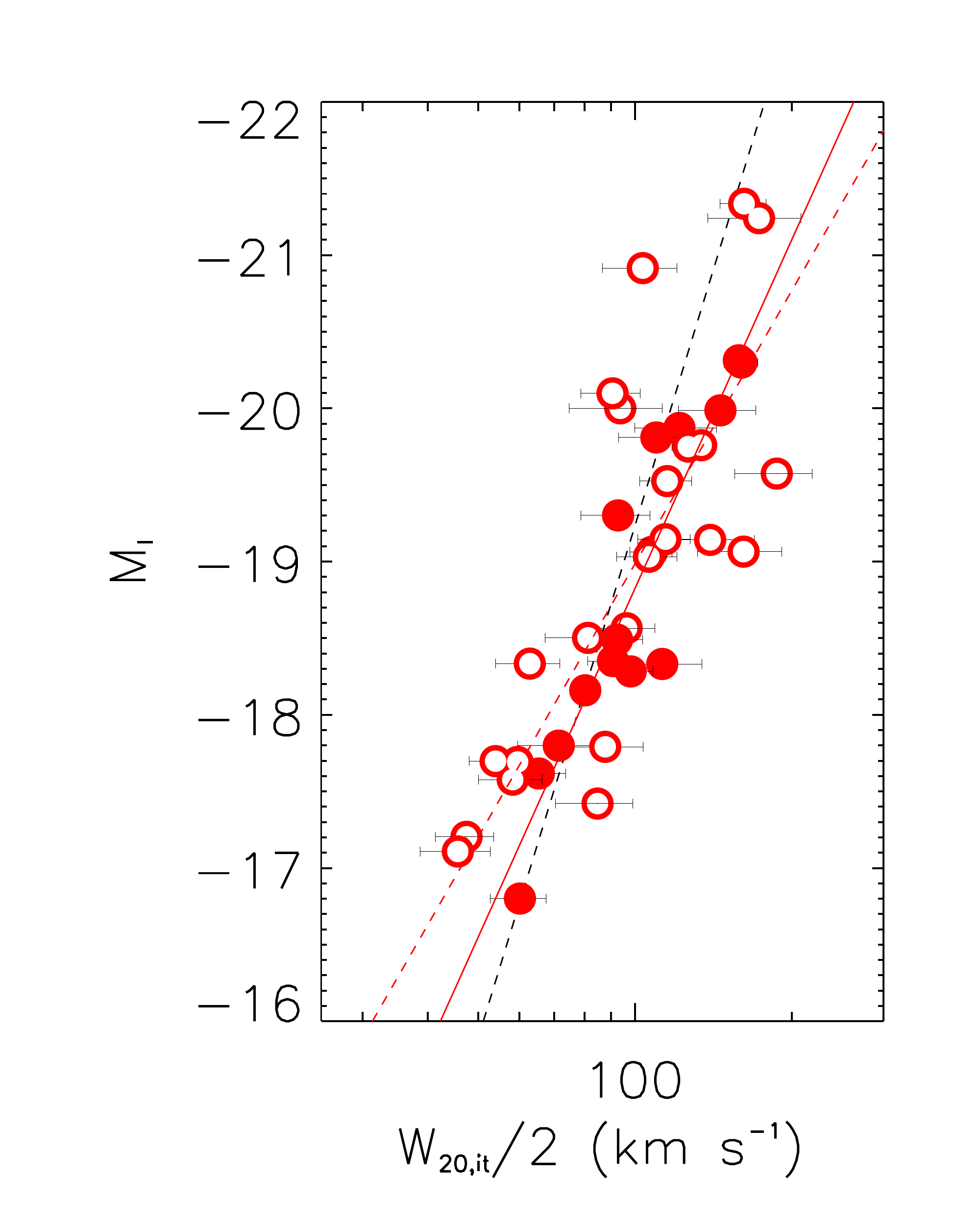}~\hspace{-.5cm}~\includegraphics[width=1.8in,angle=0]{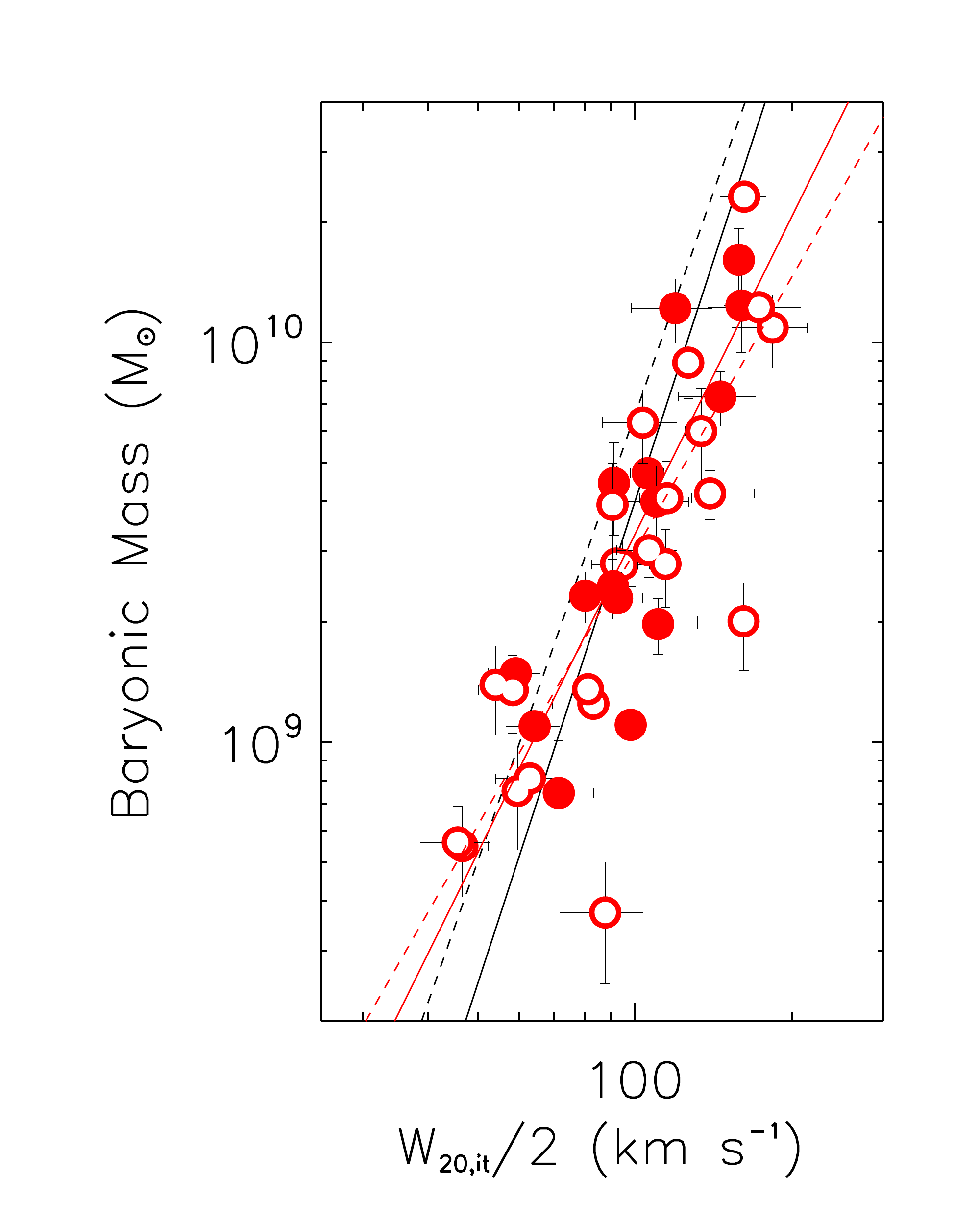}
\caption{$I$-band (left) and baryonic (right) Tully-Fisher relations.  Open points indicate those galaxies where we see no flattening of the curve, and may be underestimating the H \textsc{i} line width, W$_{20,it}$. The dashed black line is the observationally determined fit from \cite{Geha2006}, and the solid black line is from \cite{Mcgaugh2000}. The dashed red line is a fit to all the VGS, and the solid red line is fit to just those filled points where we see flattening of the rotation curve.  While the sample is in good agreement with the $I$-band relation, it is diverges somewhat from the baryonic relation for high rotation velocities and high baryon masses.
\label{fig:tf}}
\end{figure}

This is particularly interesting as it appears most pronounced for the higher mass galaxies, where the Tully-Fisher relation is typically the most robust, while the low mass systems are generally consistent with average galaxies.  All of the targeted void galaxies have inclination corrected velocity widths, W$_{20,it}$, of more than 40 km s$^{-1}$, the threshold below which baryon deficiency due to UV photoheating is predicted \citep{Hoeft2010}. Excluded from calculations of the baryon mass of these systems is the molecular gas mass, which is not generally considered to contribute significantly \citep{Schombert1990, Leroy2005}, however considering the possible increase in the efficiency of star formation in these galaxies perhaps this assumption does not hold.  Molecular gas has been found to contribute significantly more to the baryonic gas fraction in galaxies at higher redshift, in an earlier phase of their evolution \citep{Tacconi2010}.

It is possible to over-estimate the rotational velocity in systems that are not spatially resolved, as rotation curves can be somewhat centrally peaked, and in fact the agreement is much better when we consider only galaxies with flattened rotation curves resolved (Figure \ref{fig:tf}, filled symbols). For those galaxies without a turnover observed in the rotation curve, turbulent motions in the disk or beam smearing of central bulge effects on the kinematics may result in overestimation of the rotational velocity width.

 \subsection{Emptiness of the voids and void galaxy clustering}
 \label{sec:clustering}
Though void galaxies have very few nearby galaxies on the large $\sim$20 Mpc scale size of the void, they are not necessarily isolated on small $\sim$1 Mpc scales.  \cite{Szomoru1996} found that the clustering of H \textsc{i} detected galaxies within 1 Mpc of galaxies in the void compared to galaxies in regions of higher global density is the same within a factor of two.  \cite{Abbas2007} similarly observed that galaxies in the void centers are more clustered than galaxies in the void outskirts, which in turn are less clustered than in moderately overdense regions, and find that this arises naturally from the development of perturbations in a primordial Gaussian density field.  In our survey, we are  sensitive to H \textsc{i} rich neighboring galaxies, many of which are optically quite faint (M$_r > -14$).  We detect galaxies up to 25$^\prime$ (roughly 600 kpc at the average distance of $\sim$85 Mpc) and 600 km s$^{-1}$ away, however we find that the majority of galaxies are within 100 kpc and 100 km s$^{-1}$. 

Because of the pencil beam shape and limited volume of space probed by our observations, calculating nearest neighbor statistics that include the H \textsc{i} detected companions is difficult.  Instead, to compare the number of galaxy companions with typical small scale galaxy clustering we examine a volume of the SDSS from $0.001 < z < 0.0135$ containing redshift information for a volume limited sample of all galaxies brighter than M$_r < {-16}$. 
Within this sample we consider how many would have companions detectable given our observational constraints: sky separations of 600 kpc and velocity differences of 600 km s$^{-1}$.   Additionally, given that the observed VGS velocity separations are strongly biased towards smaller values, and noting a strong threshold in velocity separation at 200 km s$^{-1}$ for the SDSS sample, we limit our comparison to close companions within 600 kpc and 200 km s$^{-1}$.   Though a 200 km s$^{-1}$ difference in Hubble flow velocity corresponds to a distance of nearly 3 Mpc, it is within the expected galaxy pairwise velocity dispersion in low density regions \citep{Strauss1998}.
As we are comparing with an optically selected redshift survey, we note that the clustering of H \textsc{i} selected galaxies is observed to be weaker than for optically selected samples \citep{Meyer2007}.
From a sample of 3,323 SDSS galaxies, we find that 11\% have close neighbors observable by our survey, with 8\% having one neighbor and 2\% having two.
For similar clustering in our sample of 55 galaxies, we expect to find about six galaxies with neighbors, four having one neighbor and one having two. In fact, we find eight that have neighbors with M$_r > -16$, and one with two neighbors.  This remarkable agreement shows that even to relatively faint limits the small scale clustering of galaxies in voids is surprisingly similar to average environments.

Aside from the detection of nearby neighbors, the wide bandwidth of the WSRT observations in our survey also serves as a blind search in H \textsc{i} for galaxies throughout the voids. There has been much debate on the expected distribution of galaxies in the void given a CDM cosmology, which arguably predicts a large population of faint galaxies that fill the void but has not been detected (see \citealt{Peebles2001, Tinker2009}).  We probe with the WSRT a total volume in voids of 485 Mpc$^3$ around the targeted void galaxies, and are sensitive to targets down to $\sim$10$^8$ M$_\sun$.  This is significantly smaller than the 1100 Mpc$^3$ probed by \cite{Szomoru1996} in the Bo\"{o}tes void, but an order of magnitude deeper in H \textsc{i} mass sensitivity.  It is equivalent to the volume and sensitivity reached with H \textsc{i} Blind searches within the Pisces-Persius void, where no void galaxies are detected \citep{Weinberg1991, Saintonge2008}. Our sample is inherently biased towards finding at least one void galaxy with each observation, our chosen target, and the many companions we detect are strongly clustered around the target, even though the volume probed away from it is significantly larger.  Only one detection, VGS\_26a, is at a significant distance from the target and in the void outskirts, however with M$_r$ = -21 it is not a representative of the missing faint void galaxy population.  

\subsection{Red and dead galaxies within the voids}
\label{sec:hubbletypes}
With the VGS we sample a truncated stellar mass range, where all galaxies have less than $3 \times 10^{10}$ M$_\sun$.  This transition mass is observed to be the point where bulges and spheroids begin to dominate and SFRs decrease \citep{Kauffmann2003c}.  Yet within the voids we do observe galaxies representative of the entire Hubble sequence, including three gas-poor early type galaxies (Figure \ref{fig:optmorph}).  It is interesting to consider how these formed, and how they lost their gas.  
This small ($\sim$5\%) fraction is suggestive of an extension of the morphology-density relation to the deepest underdensities of the voids, as was also found by \cite{Park2007}.  
  We note that in our sample of 60 galaxies none have radio AGNs, which has been suggested as a secular mechanism for halting star formation in void galaxies \citep{Croton2008}. Seven of the VGS galaxies could potentially be classified optically as AGNs, and of these one appears to be an early type galaxy, forming an interesting population for further study.

\section{Conclusion}

We have identified 60 nearby (d $\sim$ 85 Mpc) void galaxies within the SDSS to form a new Void Galaxy Survey, and imaged 55 of them in H \textsc{i} with the WSRT.  Reconstruction of the galaxy density field allows us to carefully select our targets to be situated well inside of geometrically identified voids.  We find that optically the VGS galaxies span the entire color range, but they are uniformly small, with no stellar masses in excess of 3 $\times$ 10$^8$ M$_\sun$.  They are generally low luminosity, though not dwarf, blue disk galaxies, yet three appear to be early type galaxies not detected in H \textsc{i}.  

41 VGS galaxies are detected in H \textsc{i}, and we find that they are gas rich with regular rotation, about half of which with strongly disturbed gas morphologies and kinematics. This is consistent with galaxies of similar luminosity and optical morphology in average environments. Six of the galaxies not detected in H \textsc{i} appear to have very similar small blue disks, and are presumably gas rich with an H \textsc{i} mass below the detection limits. The H \textsc{i} masses are consistent with galaxies of the same luminosity located in higher density environments.   We find no evidence for an increased H \textsc{i} mass to light ratio in low density regions, and in general find very few void galaxies with elevated H \textsc{i} mass to light ratios.  
The  star formation rates normalized by the stellar mass is average given their stellar mass, however there is a hint that the star formation rates normalized by the H \textsc{i} mass is higher in the voids, making them slightly more efficient at forming stars.

The small scale clustering of galaxies in voids is very similar to that in higher density regions, and all but one of the neighboring galaxies detected  in H \textsc{i} in the voids appear to be companions of the targeted void galaxy, with most detected within 100 kpc and 100 km s$^{-1}$.  Though we probe 485 Mpc$^3$ within the voids with H \textsc{i} sensitivity of $\sim$10$^8$ M$_\sun$ we find no evidence in H \textsc{i} of the missing low luminosity void galaxy population.

In summary, we find that the large scale underdensities do appear to affect the growth of galaxies, as we find only low mass systems, but the properties of these small galaxies are not different from small galaxies in higher density environments.  Through the irregularities in the H \textsc{i} morphology and kinematics of these systems, void galaxies show many signs of ongoing interactions and gas accretion.  VGS\_12 in particular shows evidence of ongoing cold accretion of gas \citep{Stanonik2009}.  In general, void galaxies form a very interesting population of gas rich disk galaxies in which to study galaxy evolution.

\acknowledgments
We thank Jim Peebles for his ongoing interest in this project, his probing questions about the nature of void galaxies, and his enlivening discussions. KK also thanks Greg Bryan, Mary Putman and Tony Wong for their valuable comments.   We further thank the anonymous referee for their careful reading and comments.  This work was supported in part by the National Science Foundation under grant \#1009476 to Columbia University.  We are grateful for support from a Da Vinci Professorship at the the Kapteyn Institute.

    Funding for the SDSS and SDSS-II has been provided by the Alfred P. Sloan Foundation, the Participating Institutions, the National Science Foundation, the U.S. Department of Energy, the National Aeronautics and Space Administration, the Japanese Monbukagakusho, the Max Planck Society, and the Higher Education Funding Council for England. The SDSS Web Site is http://www.sdss.org/.

    The SDSS is managed by the Astrophysical Research Consortium for the Participating Institutions. The Participating Institutions are the American Museum of Natural History, Astrophysical Institute Potsdam, University of Basel, University of Cambridge, Case Western Reserve University, University of Chicago, Drexel University, Fermilab, the Institute for Advanced Study, the Japan Participation Group, Johns Hopkins University, the Joint Institute for Nuclear Astrophysics, the Kavli Institute for Particle Astrophysics and Cosmology, the Korean Scientist Group, the Chinese Academy of Sciences (LAMOST), Los Alamos National Laboratory, the Max-Planck-Institute for Astronomy (MPIA), the Max-Planck-Institute for Astrophysics (MPA), New Mexico State University, Ohio State University, University of Pittsburgh, University of Portsmouth, Princeton University, the United States Naval Observatory, and the University of Washington.

\appendix 


\section{Image catalog and discussion of individual systems}
\label{sec:appendix}
This Appendix presents an image atlas of the 60 galaxies comprising the Void Galaxy Survey, as well as an H \textsc{i} atlas presenting an overview of our observations of 55 of the VGS targets and the VGS companions detected in H \textsc{i}.  These are followed by a brief discussion of some individually interesting galaxies in our sample. 

\subsection{Atlas of VGS targets}

Figure \ref{fig:poststamps} presents the \textit{ugriz} combined color image from the SDSS catalog for each of the 60 VGS galaxies, with the VGS catalog name given above each image.  All are scaled to the same physical size, such that each image shows approximately a 16 kpc by 16 kpc region of the sky at the distance of the target.

Figure \ref{fig:vgs} provides an H \textsc{i} atlas of the 41 void galaxies detected with the WSRT.  On the left, H \textsc{i} contours from the total intensity map are overlaid on the SDSS $g$-band optical image. In the center, the intensity weighted velocity field is colored to emphasize any rotation in the gas, with contours to guide the eye.  On the right, we present a position velocity slice that is aligned with the H \textsc{i} kinematic major axis.  All images are created following the uniform reduction described in Section \ref{sec:observations}, and are shown with uniformly defined contours to aid comparison between the targets.

\begin{figure}
\centering
\includegraphics[width=.99\textwidth]{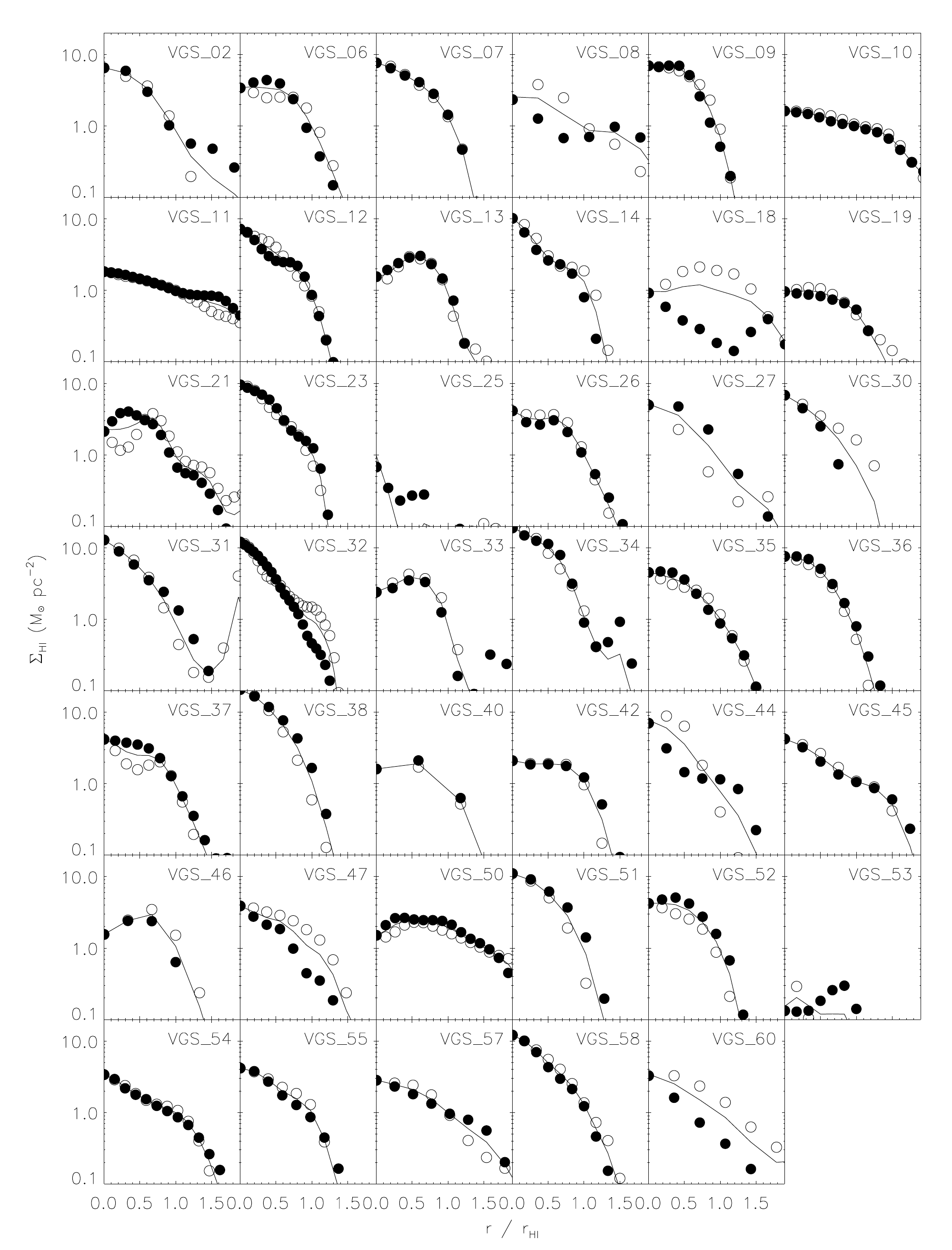}
\caption{Radial surface density profiles for the east (open circles) and west (filled circles) sides of each disk, with the average overdrawn.
\label{fig:surfdens}}
\end{figure}

Figure \ref{fig:surfdens} shows the H \textsc{i} radial surface density profiles for the 41 galaxies detected with the WSRT.  The technique used to determine these profiles is described in Section \ref{sec:results}.

\subsection{Atlas of VGS companions}

\begin{figure}
\centering
\includegraphics[width=0.85\textwidth]{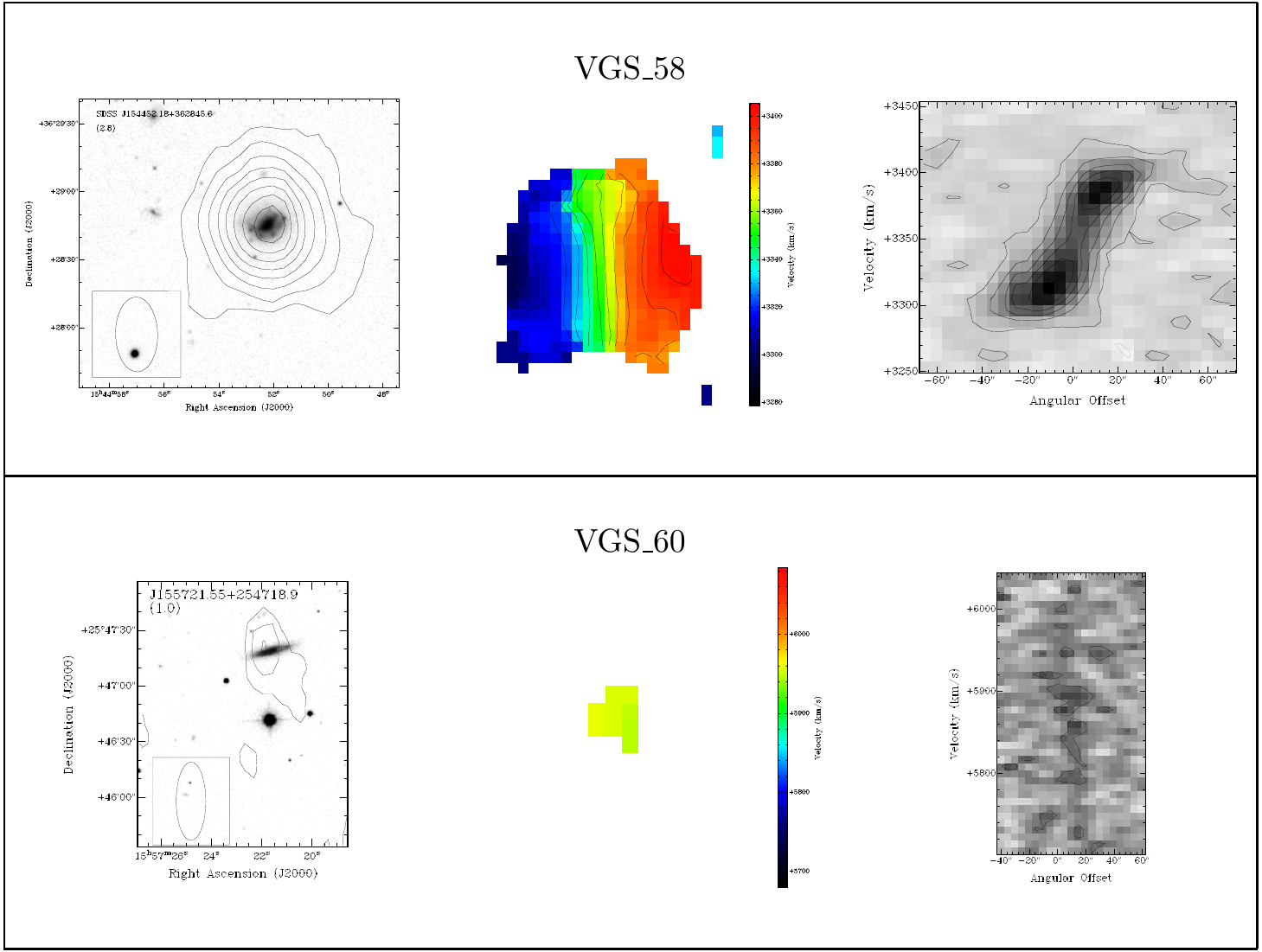}
\caption[]{Targeted void galaxies. Contours in the total intensity maps (left) are at $5 \times 10^{19} cm^{-2}$ plus increments of $10^{20} cm^{-2}$. Confidence level ($\sigma$) of the lowest contour is given in the top left corner of each image. Lines in the velocity field images (center) indicate increments of 10 km s$^{-1}$. Position-Velocity diagrams (right) are along the kinematic major axis, contours are at increments of -1.5 (grey), 1.5 (black) + increments of 3$\sigma$ .
\label{fig:vgs}}
\end{figure}

\begin{figure}[h!]
\centering
\begin{tabular}{l l l}
VGS\_07a & VGS\_09a & VGS\_10a \\
\includegraphics[width=2in]{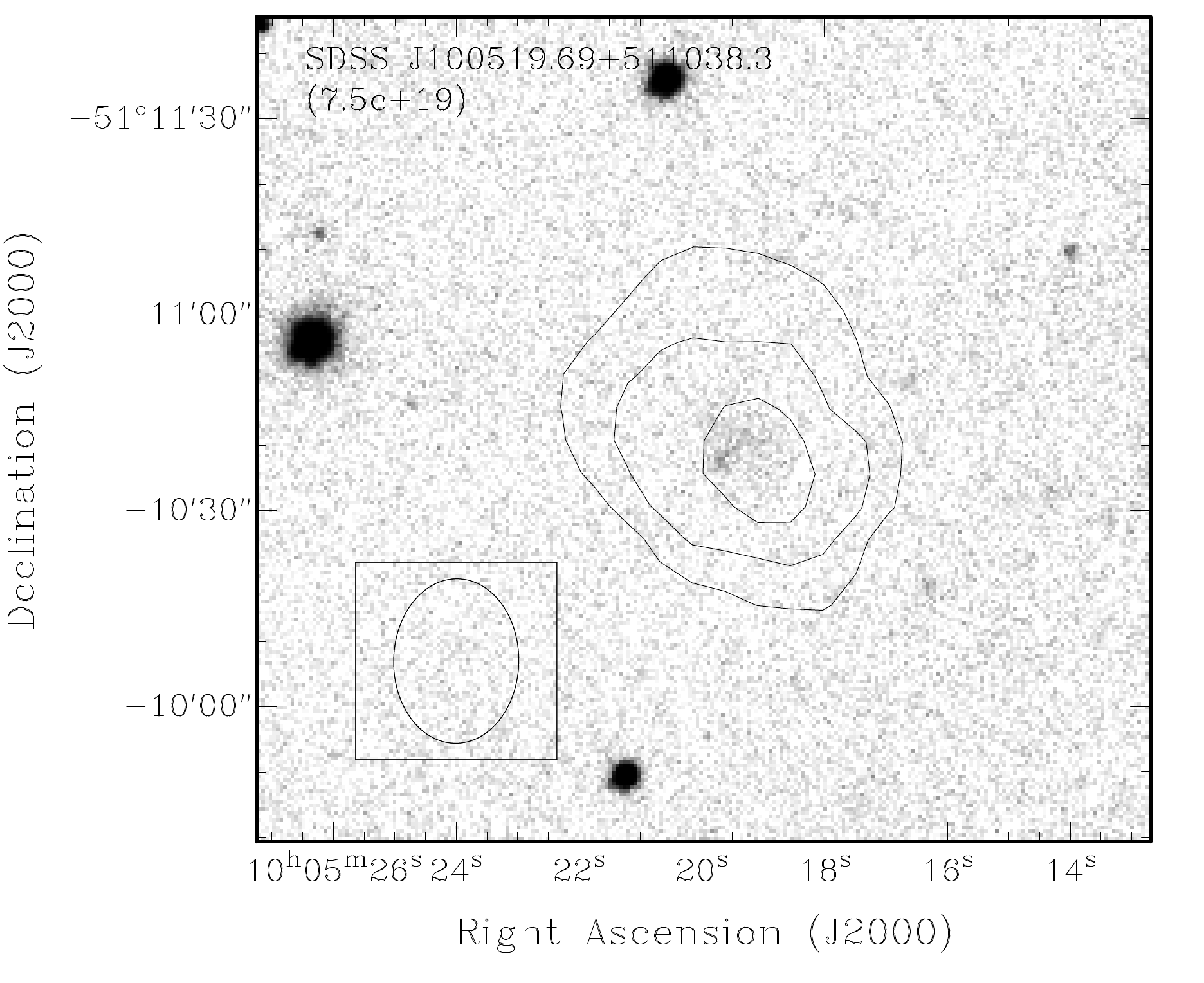} & 
\includegraphics[width=2in]{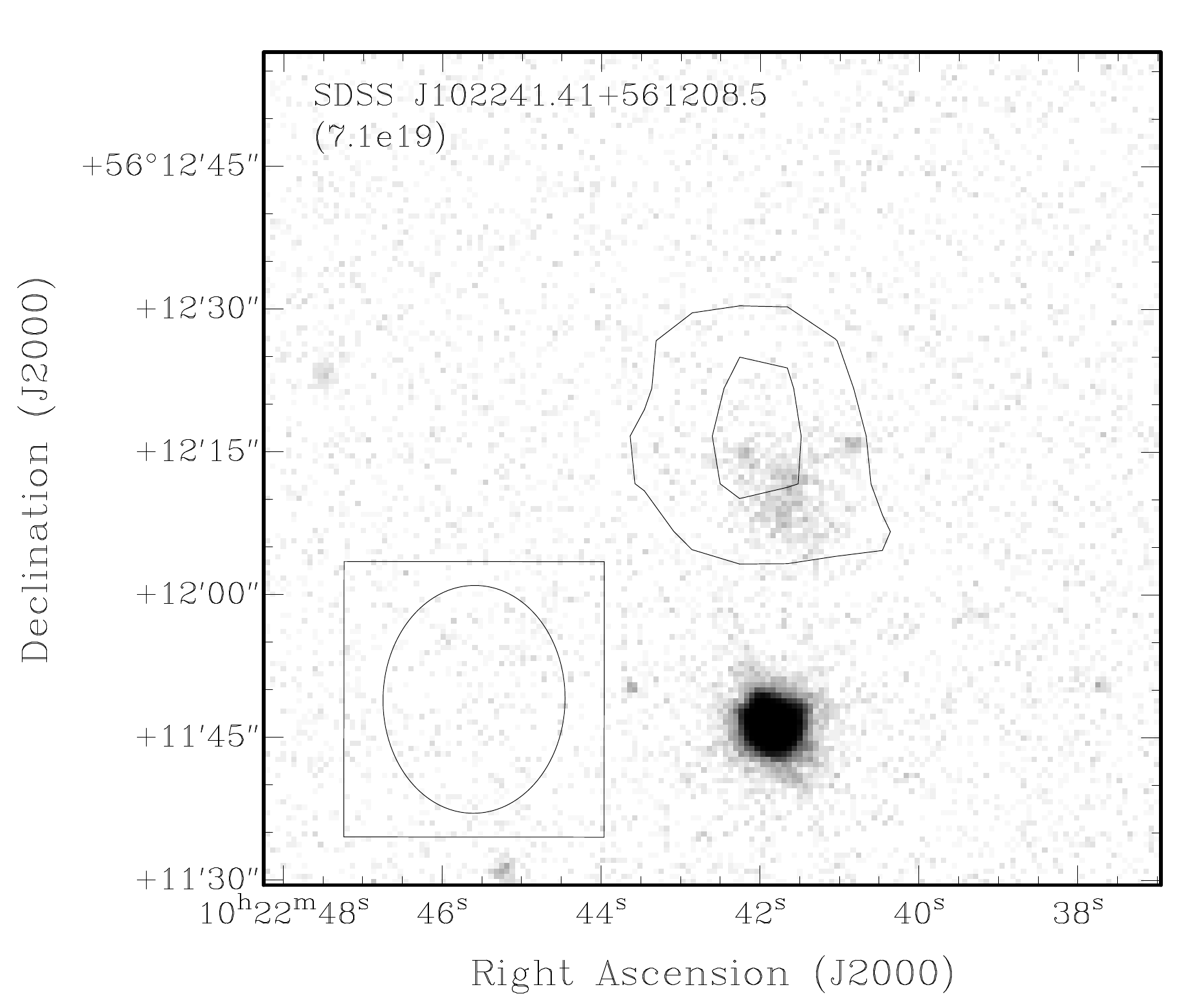} &
\includegraphics[width=1.5in]{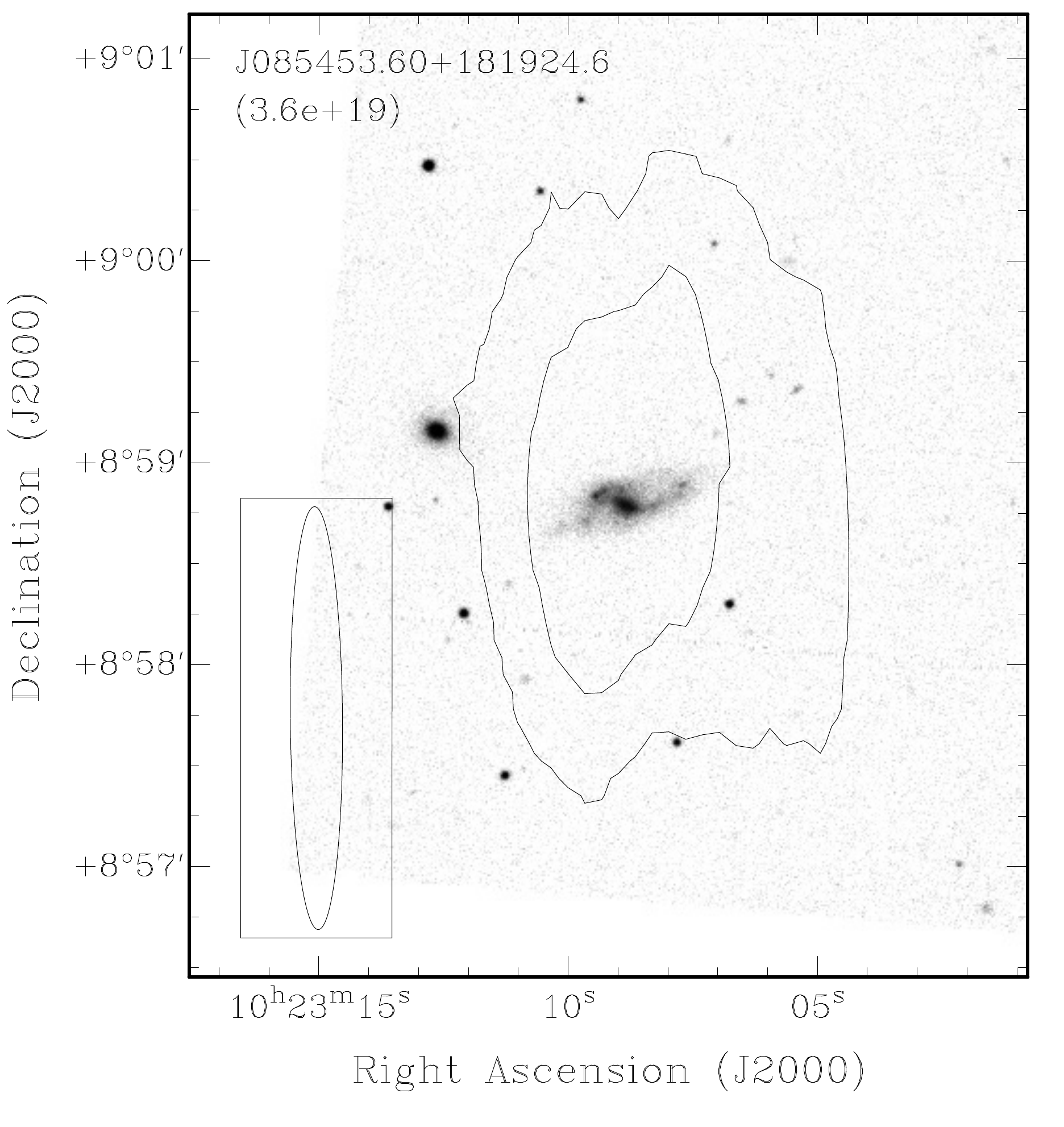} \\

VGS\_26a & VGS\_34a & VGS\_37a \\
\includegraphics[width=2in]{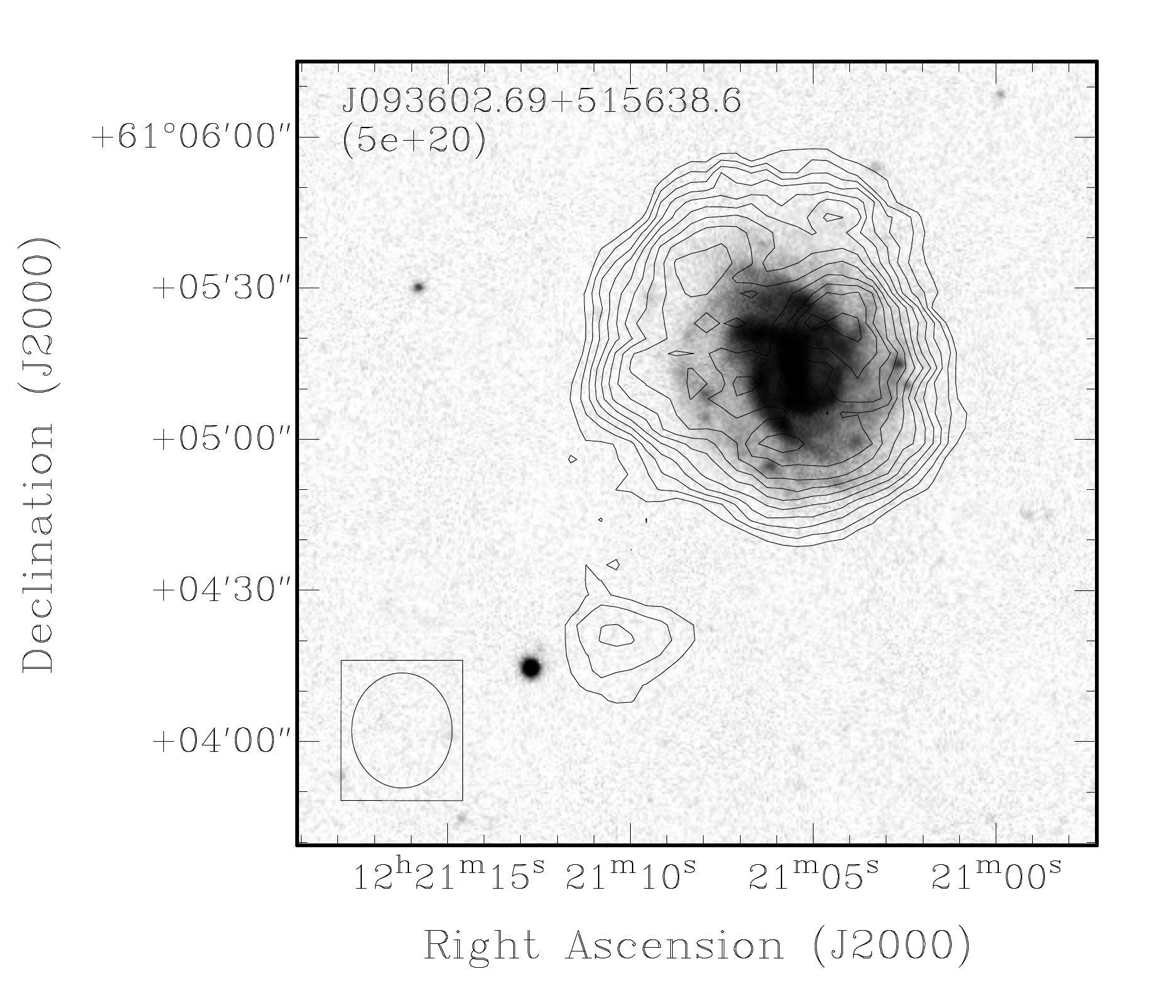} &
\includegraphics[width=2in]{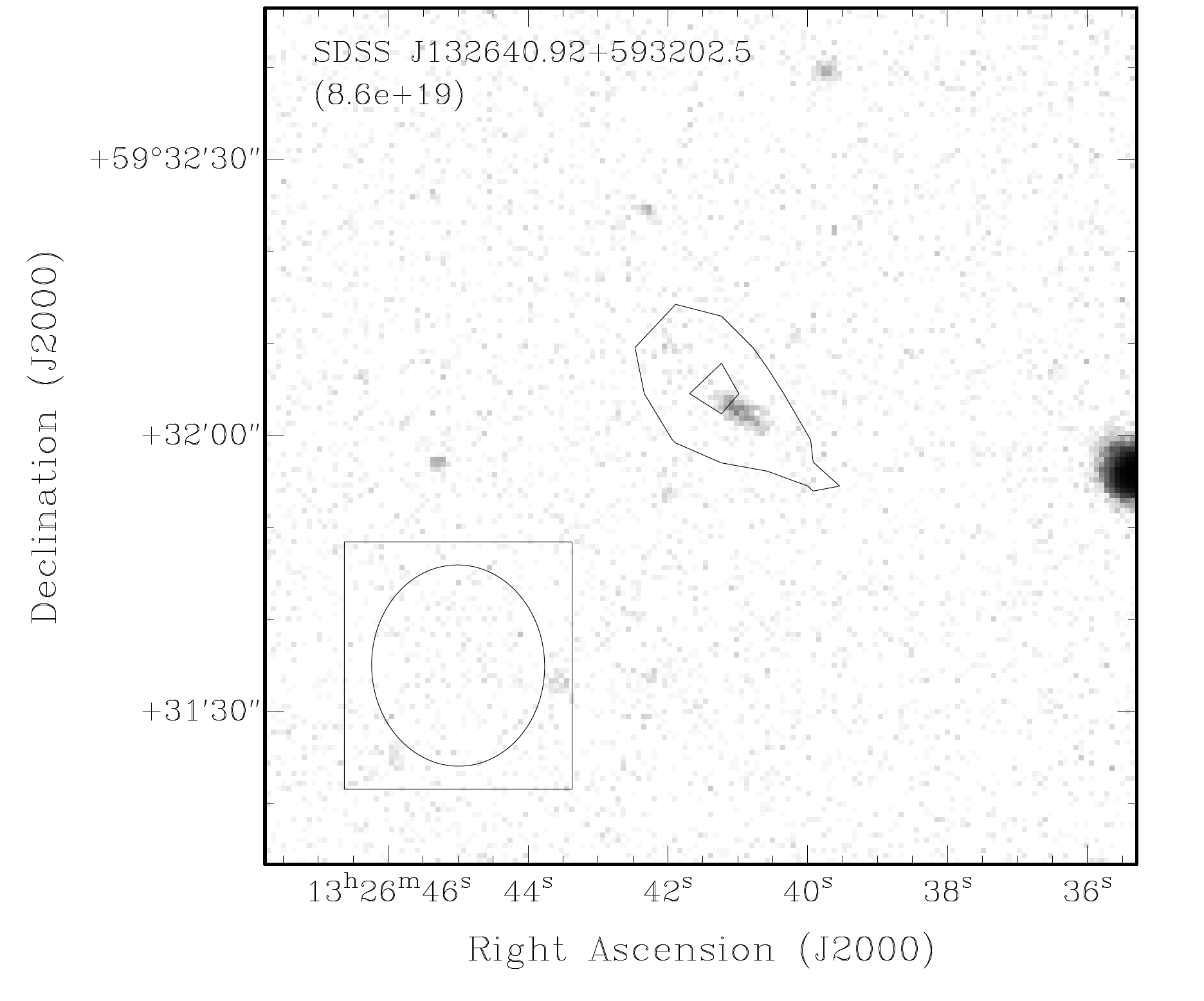} &
\includegraphics[width=2in]{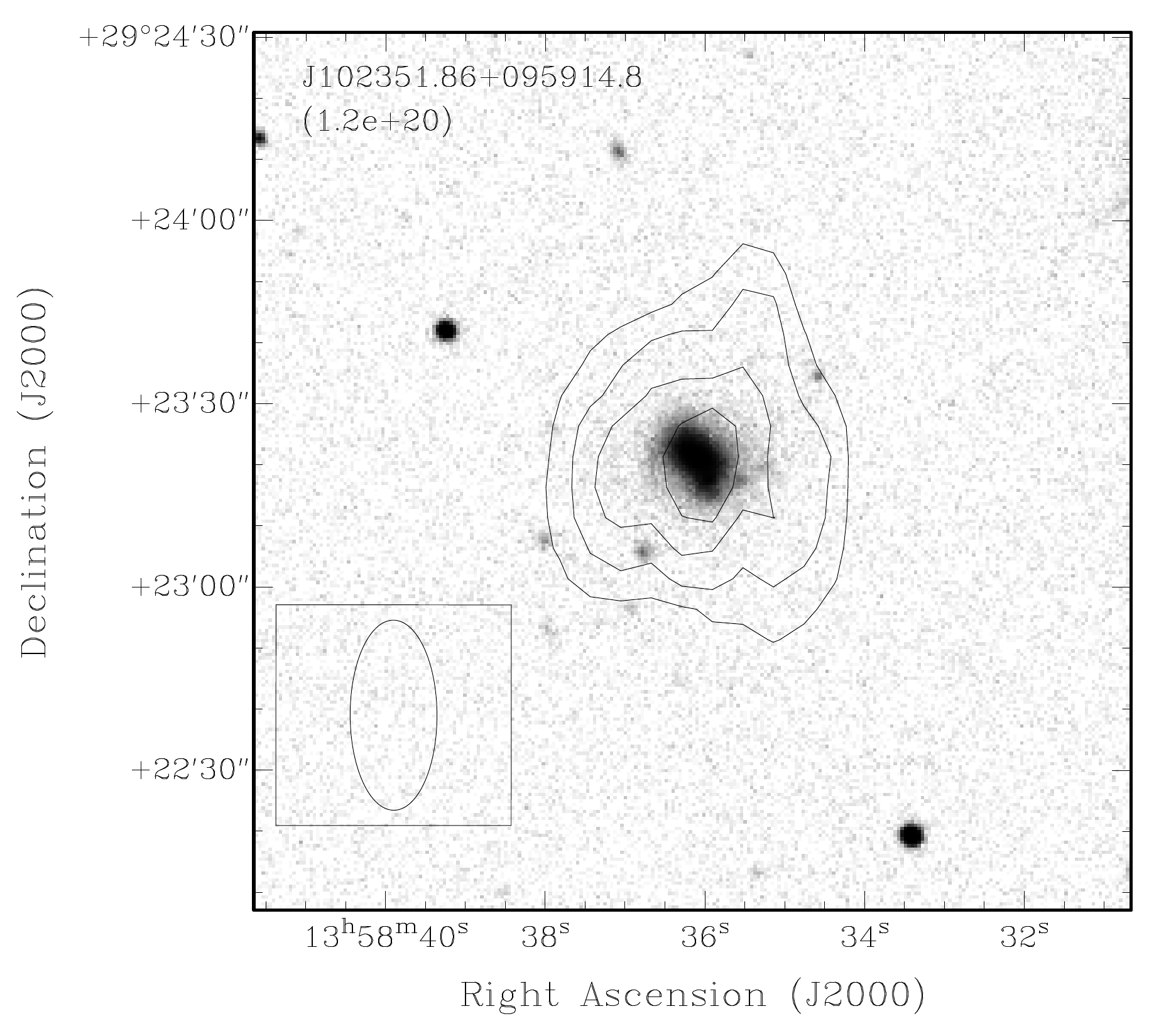} \\

VGS\_39a & VGS\_51a & VGS\_53a \\
\includegraphics[width=2in]{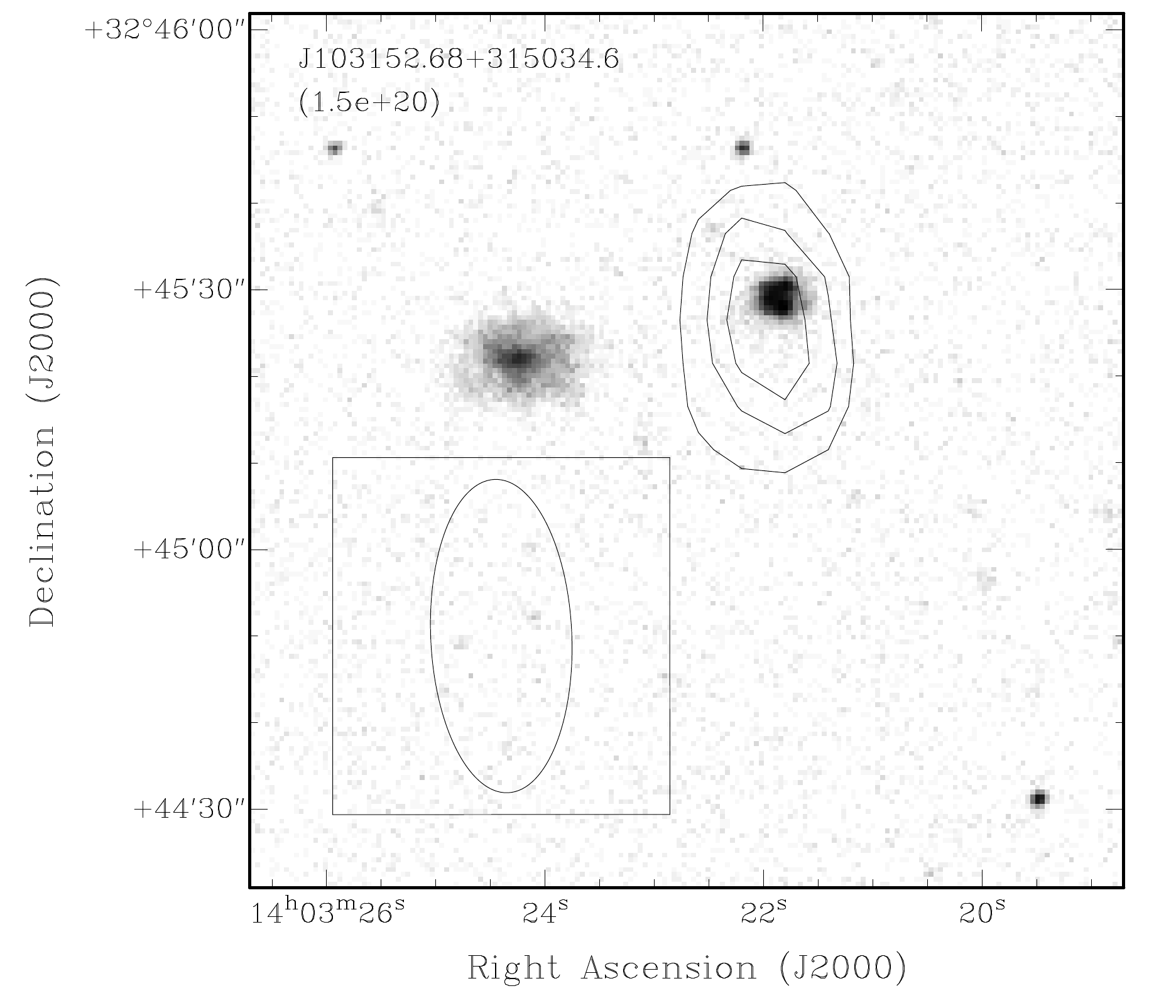} &
\includegraphics[width=1.8in]{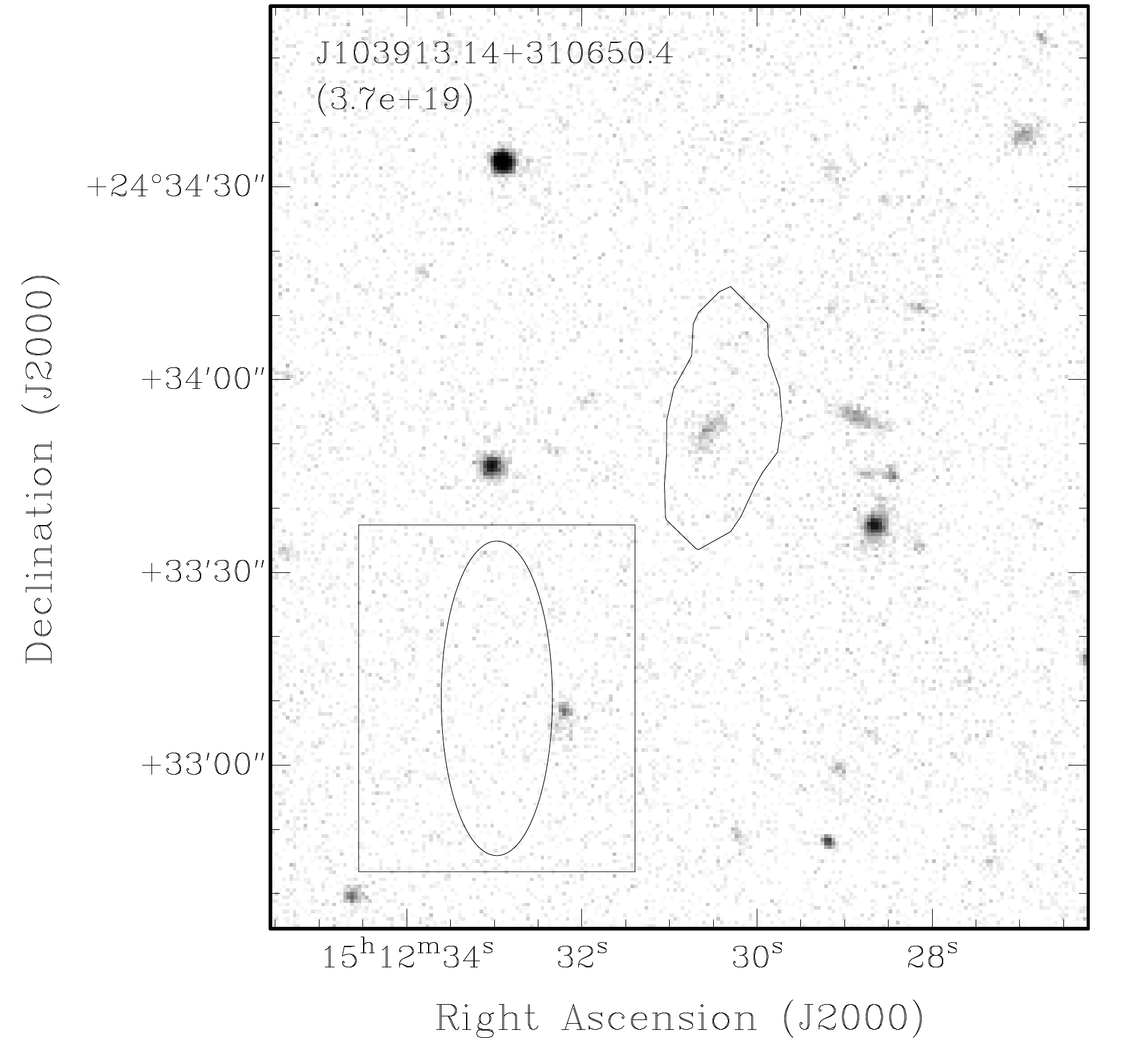} &
\includegraphics[width=2in]{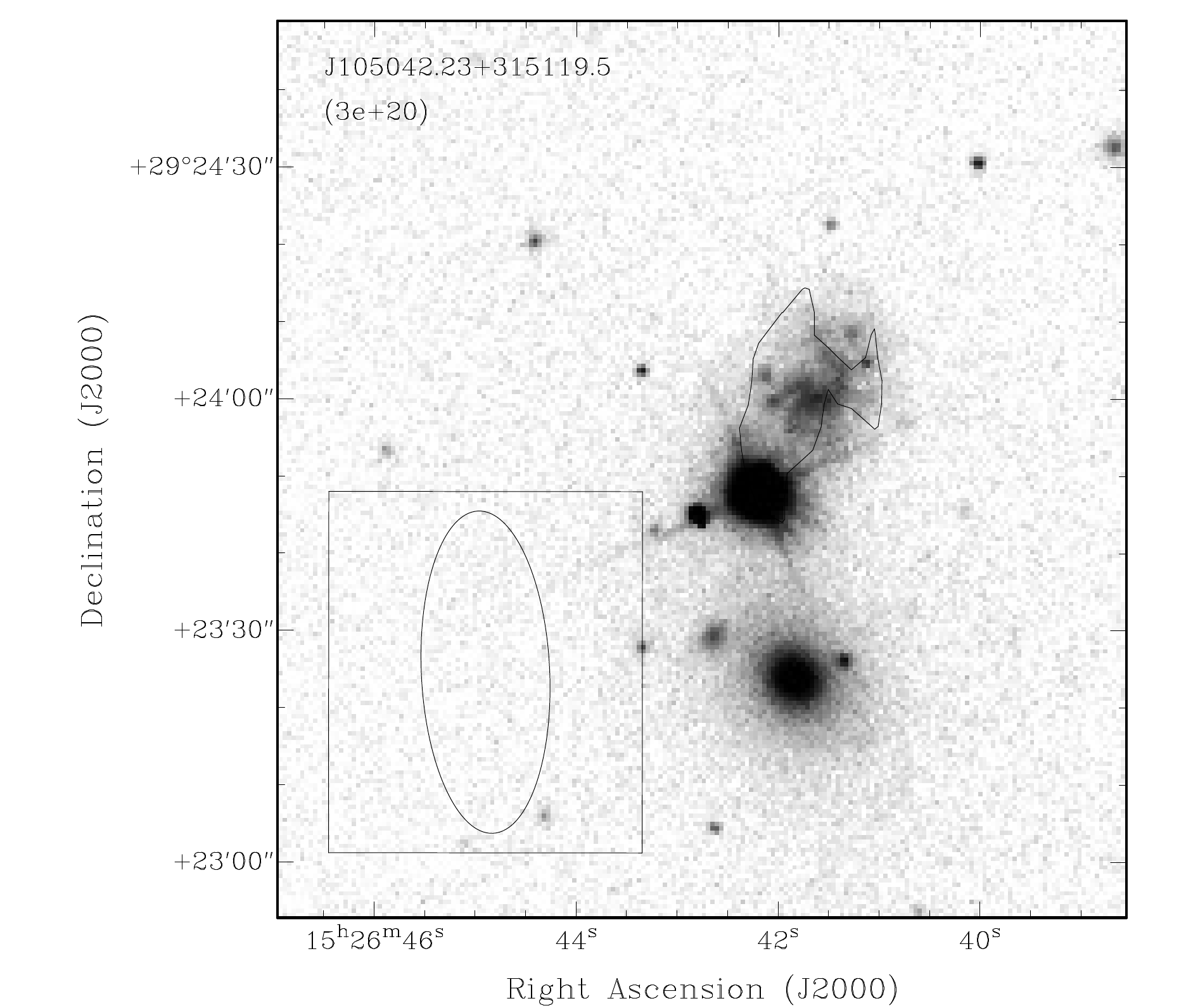} \\

VGS\_56a & VGS\_57a & \\
\includegraphics[width=2in]{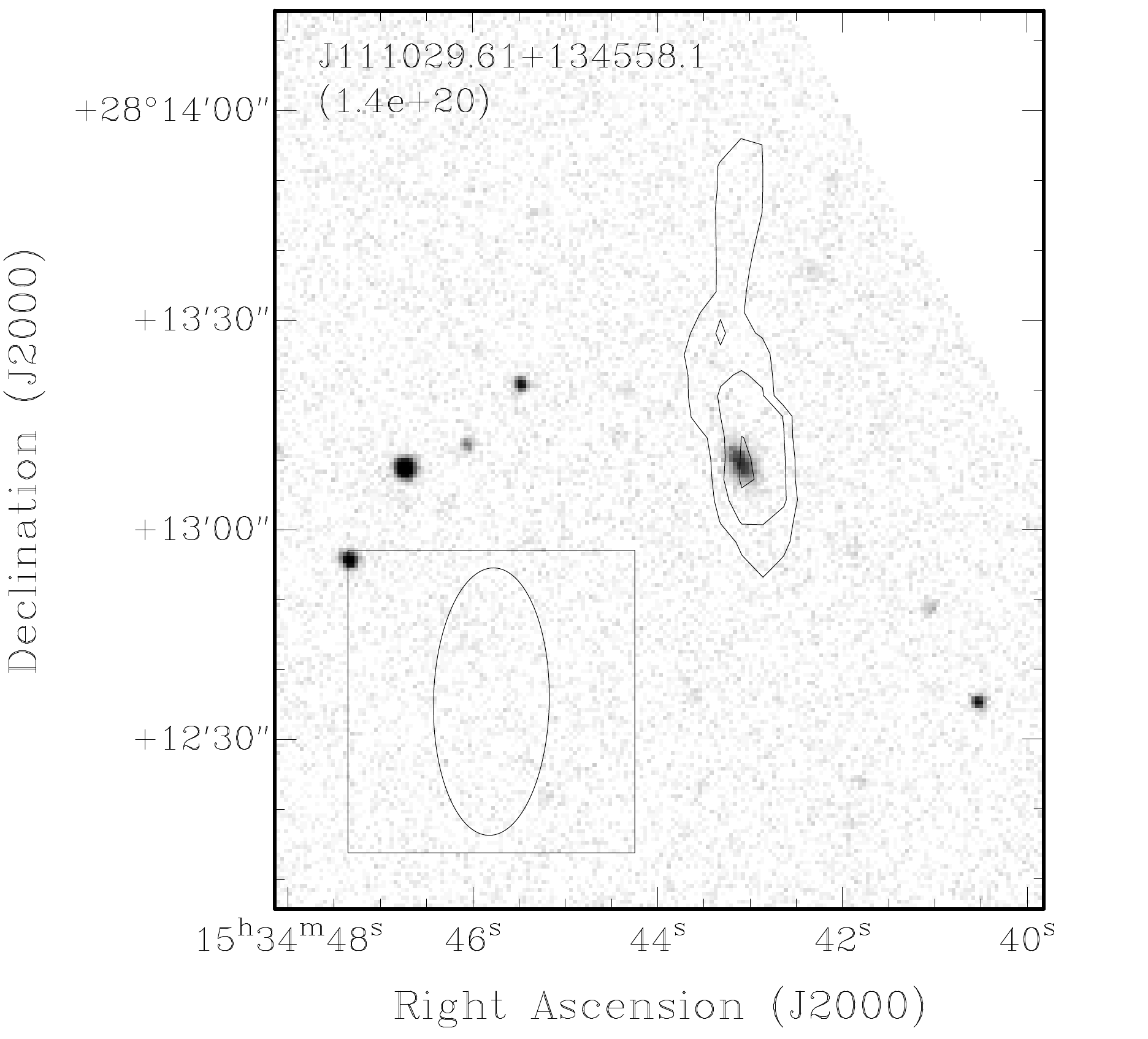} & 
\includegraphics[width=2in]{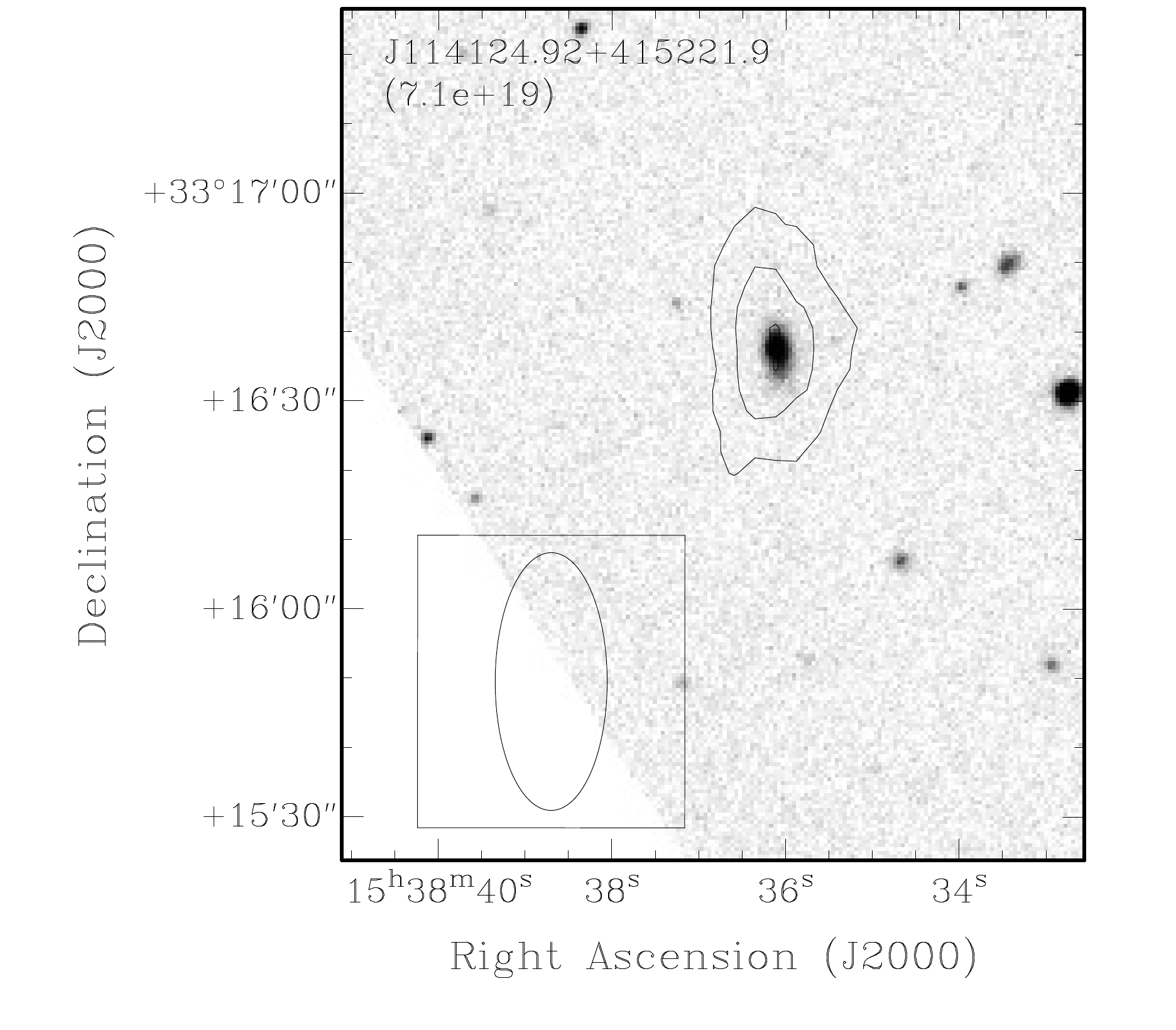} & \\
\end{tabular}
\caption{Companions.  Note that VGS\_30a, VGS\_31a, VGS\_31b, VGS\_36a, VGS\_38a, VGS\_38b and VGS\_54a are shown in Figure \ref{fig:vgs} along with the targeted void galaxy. Contours in the total intensity maps are at 3$\sigma$ plus increments of $10^{20} cm^{-2}$. Column density at the lowest contour, in atoms per cm$^{-2}$, is given in the top left corner of each image. 
\label{fig:comps}}
\end{figure}

In addition to the detection of 41 of the 55 VGS targets imaged with the WSRT, we detect  in H \textsc{i} 18 neighboring galaxies within the void.  Seven of these are shown in Figure \ref{fig:vgs} along with the target galaxies, and the remaining 11 are shown in Figure \ref{fig:comps}.   As many of these targets are observed within the primary beam but removed from the beam center, the H \textsc{i} sensitivity is typically somewhat worse for these detections than for the primary targets.  Thus we show here the H \textsc{i} contours for these detections with a lowest contour at 3$\sigma$, and indicate the corresponding column density in the upper right corner of each image.

\subsection{Notes on individual systems}
\label{sec:anotes}

In addition to the very interesting objects previously described in our pilot study \citep{Kreckel2011}, including interacting systems (VGS\_30, VGS\_34, VGS\_38), misaligned gas disks (VGS\_12, VGS\_35), optically unusual morphologies (VGS\_07) and undisturbed gas morphologies (VGS\_32), we highlight here some additional interesting galaxies identified within the VGS.

\textbf{VGS\_05} is not detected in H \textsc{i}, and has an early type optical morphology. Red in color ($g-r = 0.76$), it does not present an entirely uniform elliptical morphology, with the light distribution in the center suggesting the presence of a disk or bar.  It has the lowest specific star formation rate of the VGS.

\textbf{VGS\_24} is also not detected in H \textsc{i}, and has an optically very smooth, elliptical morphology but a slightly blue ($g-r = 0.55$) color. 

\textbf{VGS\_25} has a small stellar disk and low H \textsc{i} mass, consistent with the small blue disk galaxies not detected that have very uniform optical properties (see Section \ref{sec:nondetections}).  The H \textsc{i} mass detected in this galaxy is just above the stacked detection limit for the six other undetected targets described in Section \ref{sec:nondetections}, and we expect that this galaxy is analogous to those just below the limits.

 \textbf{VGS\_31} is the second void galaxy in our sample to have two H \textsc{i} rich companions, with all three sharing a common envelope in H \textsc{i}. The first, VGS\_38, was described in the pilot survey, and presents a very similar linear alignment of the three galaxies, suggesting filamentary structure within the void.  The H \textsc{i} bridge connecting all galaxies in VGS\_31 is more apparent in Figure \ref{fig:examples1}, where we have imaged with natural weighting to increase our sensitivity.  This system shows further signs of ongoing interactions in the left most galaxy, where a gas poor stellar ring is present just at the boundary of the H \textsc{i} disk.  This interesting system is discussed in greater detail in Beygu et al. (in prep.).
 
 \textbf{VGS\_47} has an undisturbed stellar disk but extremely lopsided H \textsc{i} morphology.  This is not strongly reflected in the gas kinematics, though some evidence of warping may be seen.
 
 \textbf{VGS\_50} is identified as having an extremely extended H \textsc{i} disk, with r$_{\textrm{H \textsc{i}}}$/r$_{90} \sim 5$, though it appears that the r$_{90}$ optical radius includes only the brighter, redder inner disk and excludes the faint blue, more extended disk observed in the combined color image.  With an edge-on gas disk and fairly regular kinematics with no indication of a warp, its 32 kpc physical extent is one of the largest in this sample.  The H \textsc{i} surface density profile shows some evidence for a central hole in the gas distribution.
 
 \textbf{VGS\_51} is located the deepest within the void, being at a density of less than a tenth the mean.  Optically, it presents a disk morphology with many distinctive blue knots presumably related to areas of high star formation, and the galaxy as a whole does have a relatively high specific star formation rate.  The HI distribution appears quite regular both morphologically and kinematically, though the gas disk may be slightly offset from the center of the stellar disk.  This is difficult to judge as the optical disk center may be confused by the irregular morphology.  It has a very faint (M$_r$ = -14) dwarf companion at nearly coincident velocity and a sky separation of 133 kpc.  
 
 \textbf{VGS\_57} has the highest star formation efficiency of the VGS.  It is one of the most luminous (M$_r$ = -20.4) galaxies in our sample with one of the highest SFRs, almost 2 M$_\sun$ yr$^{-1}$.  Optically, it appears to be a large, face-on strongly barred spiral, however the H \textsc{i} appears as a fairly concentrated disk with a significant warp at the position of the optical spiral arms.  It has a smaller companion with a systemic velocity only $\sim$70 km s$^{-1}$ different and a separation on the sky of 175 kpc.

\clearpage

\begin{figure}
\centering
\includegraphics[width=0.99\textwidth]{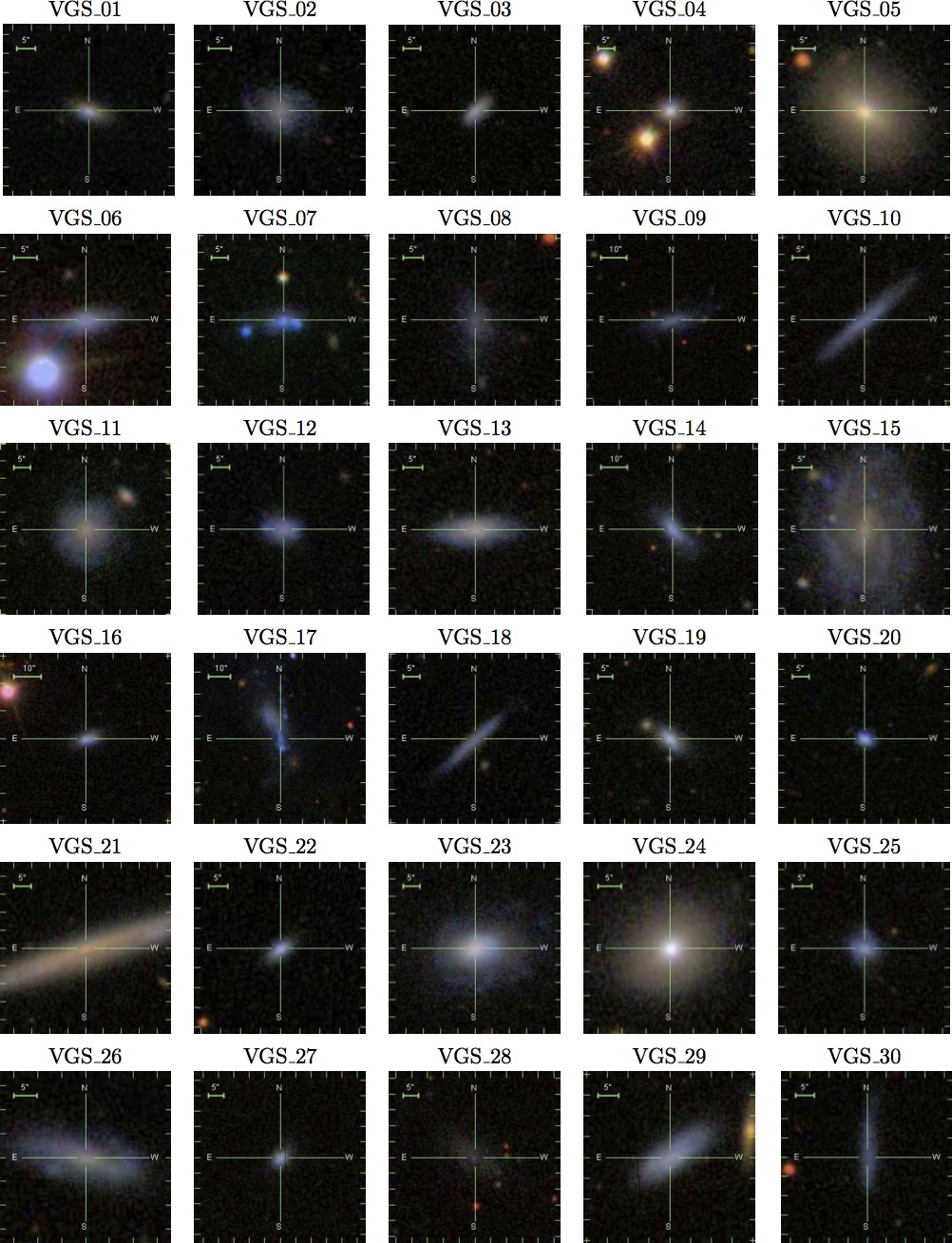}
\caption[]{(continued)}
\label{fig:poststampsa}
\end{figure}

\begin{figure}
\centering
\includegraphics[width=0.99\textwidth]{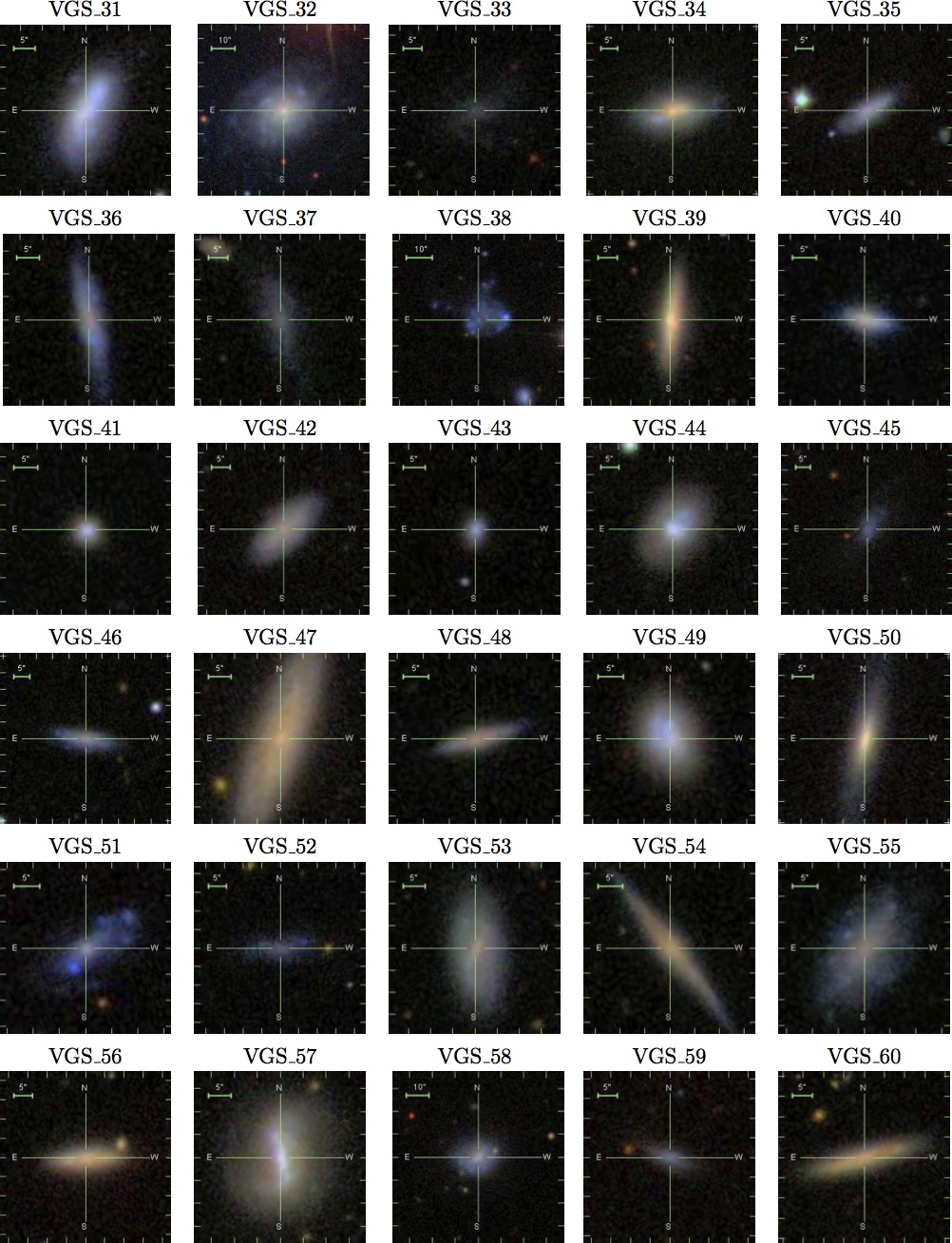}
\caption{Our sample of void galaxies, scaled to the same physical size. Composite color images are taken from the online 
SDSS Finding Chart tool. 
\label{fig:poststamps}}
\end{figure}

\begin{figure}
\centering
\includegraphics[width=0.85\textwidth]{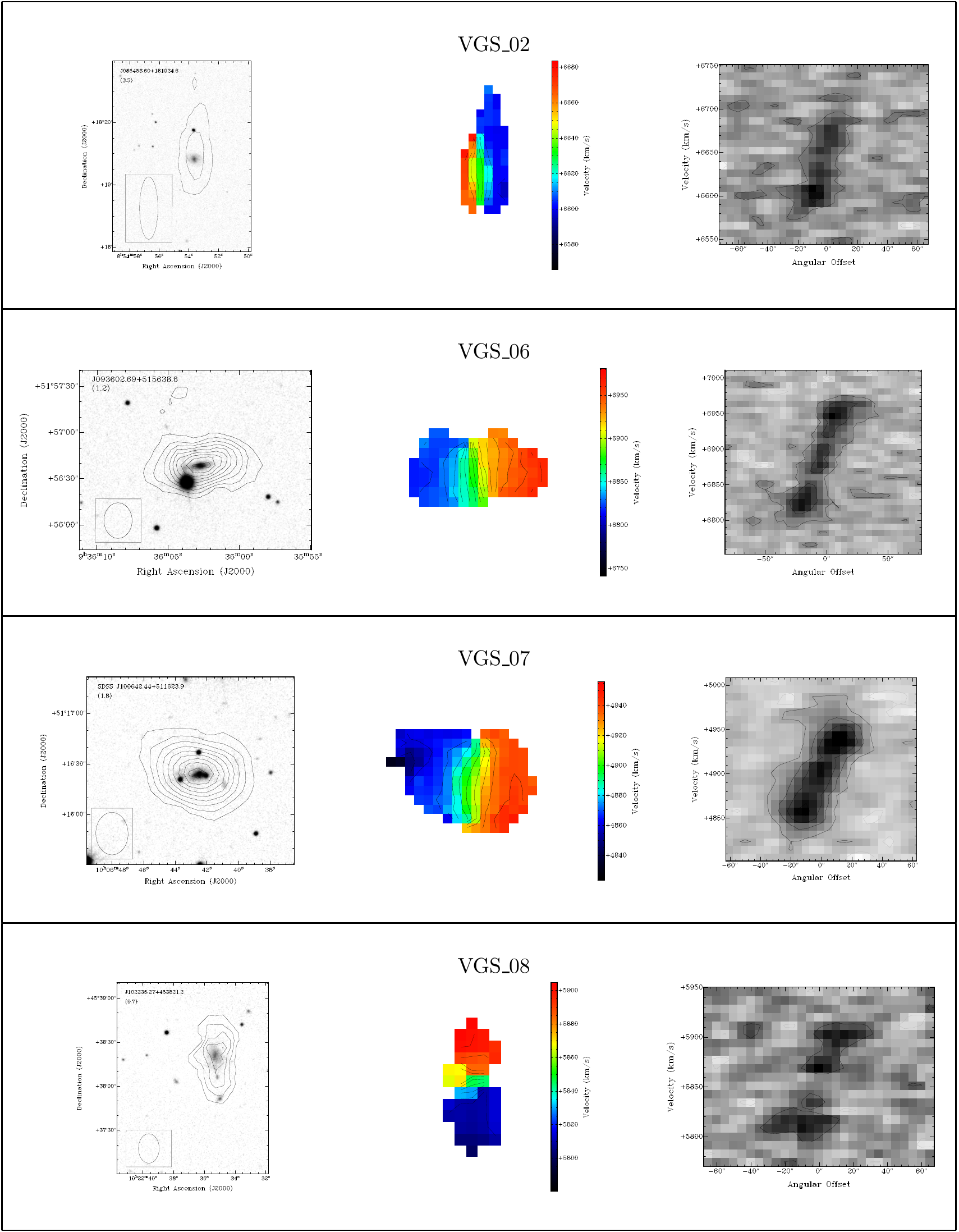}
\caption[]{(continued)}
\label{fig:voidgala}
\end{figure}

\begin{figure}
\centering
\includegraphics[width=0.85\textwidth]{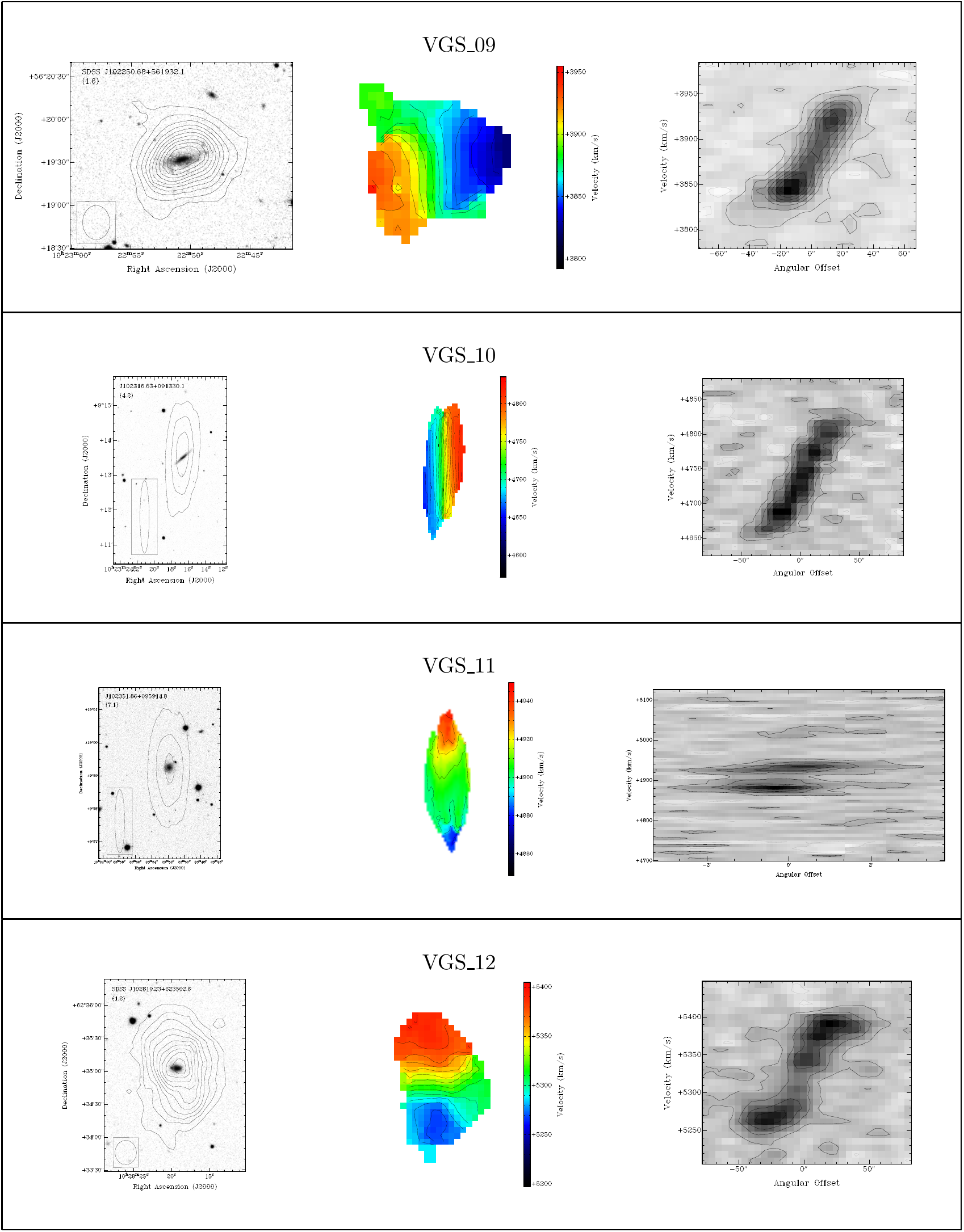}
\caption[]{(continued)}
\label{fig:voidgalb}
\end{figure}

\begin{figure}
\centering
\includegraphics[width=0.85\textwidth]{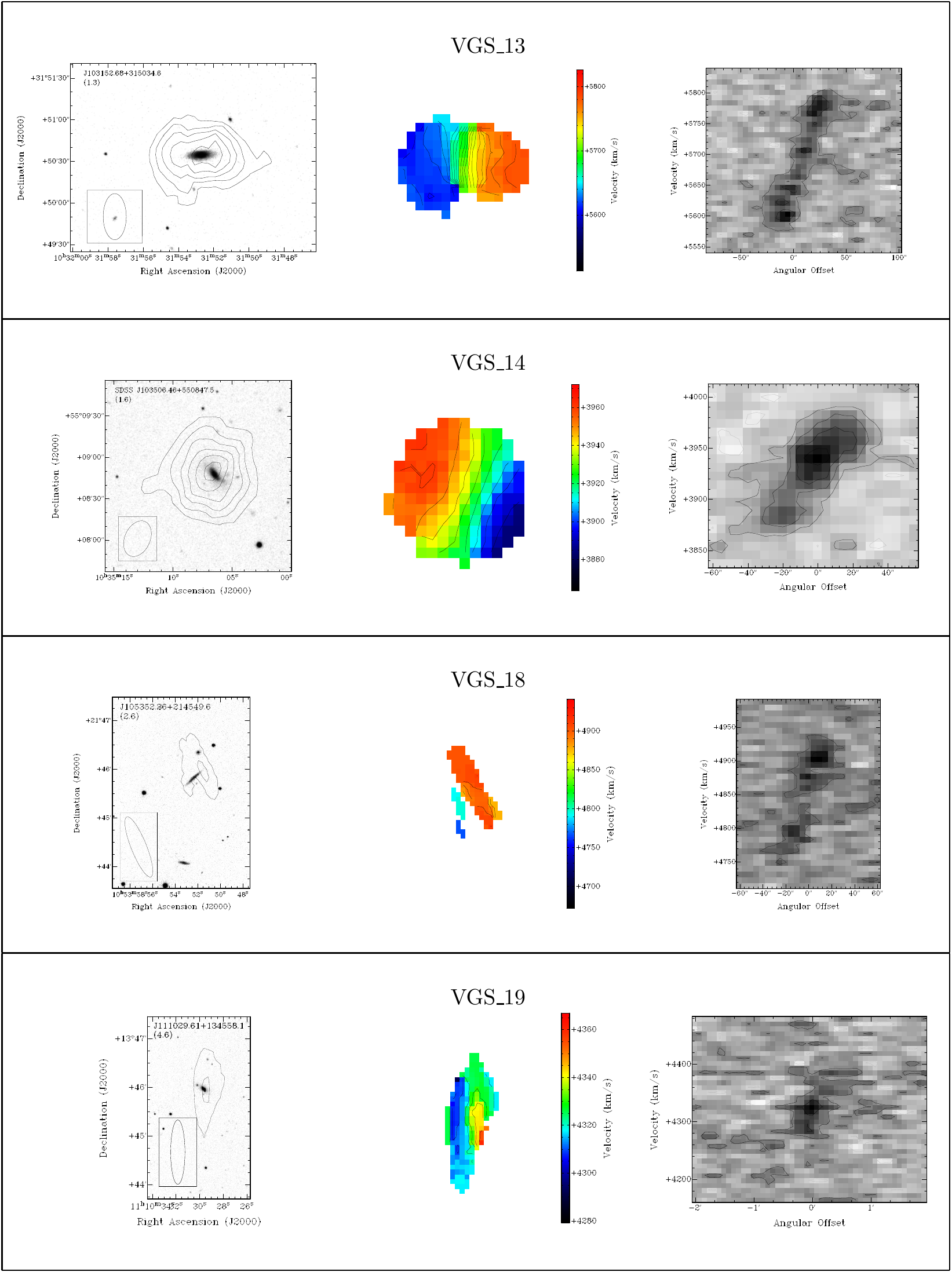}
\caption[]{(continued)}
\label{fig:voidgalc}
\end{figure}

\begin{figure}
\centering
\includegraphics[width=0.85\textwidth]{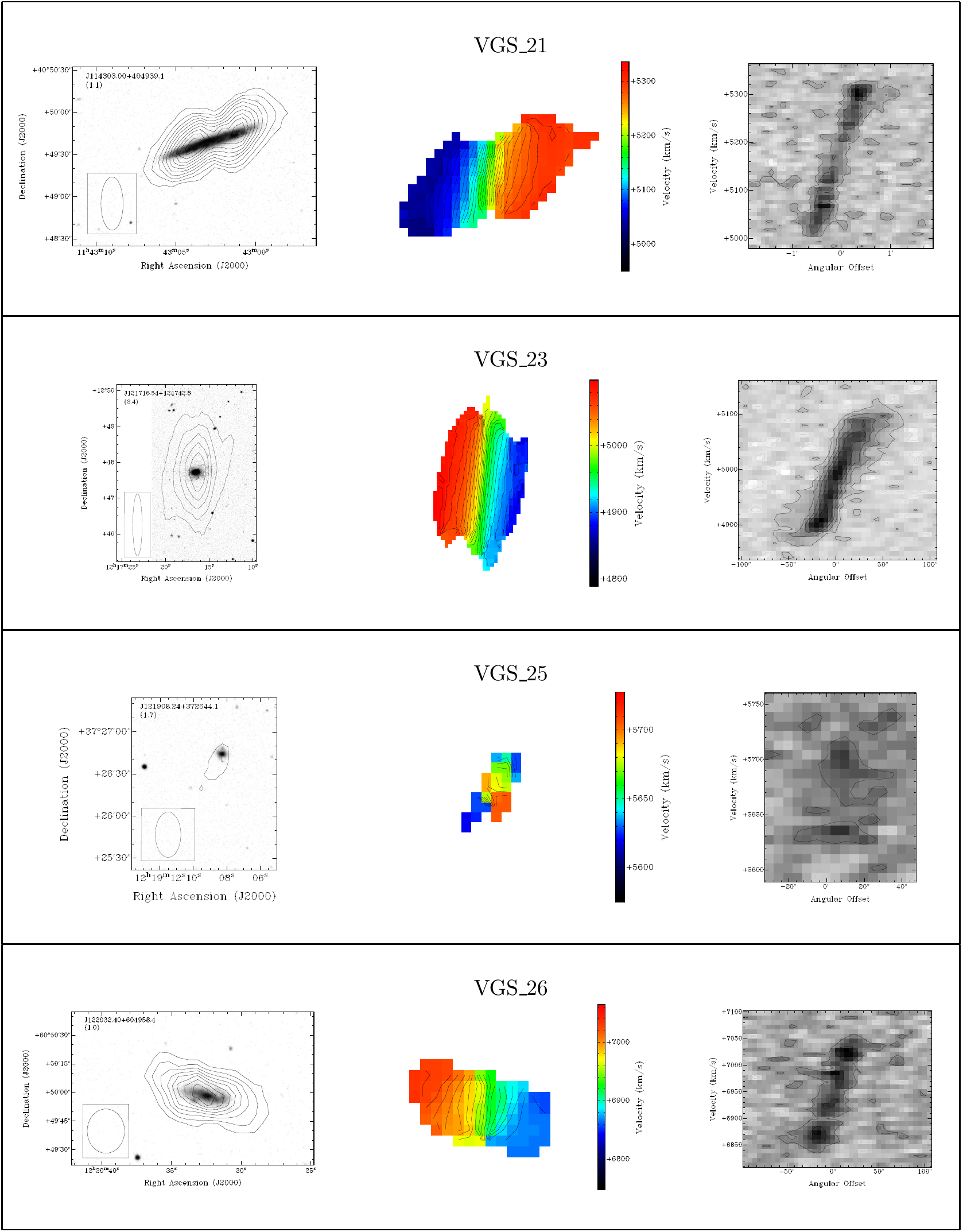}
\caption[]{(continued)}
\label{fig:voidgald}
\end{figure}

\begin{figure}
\centering
\includegraphics[width=0.85\textwidth]{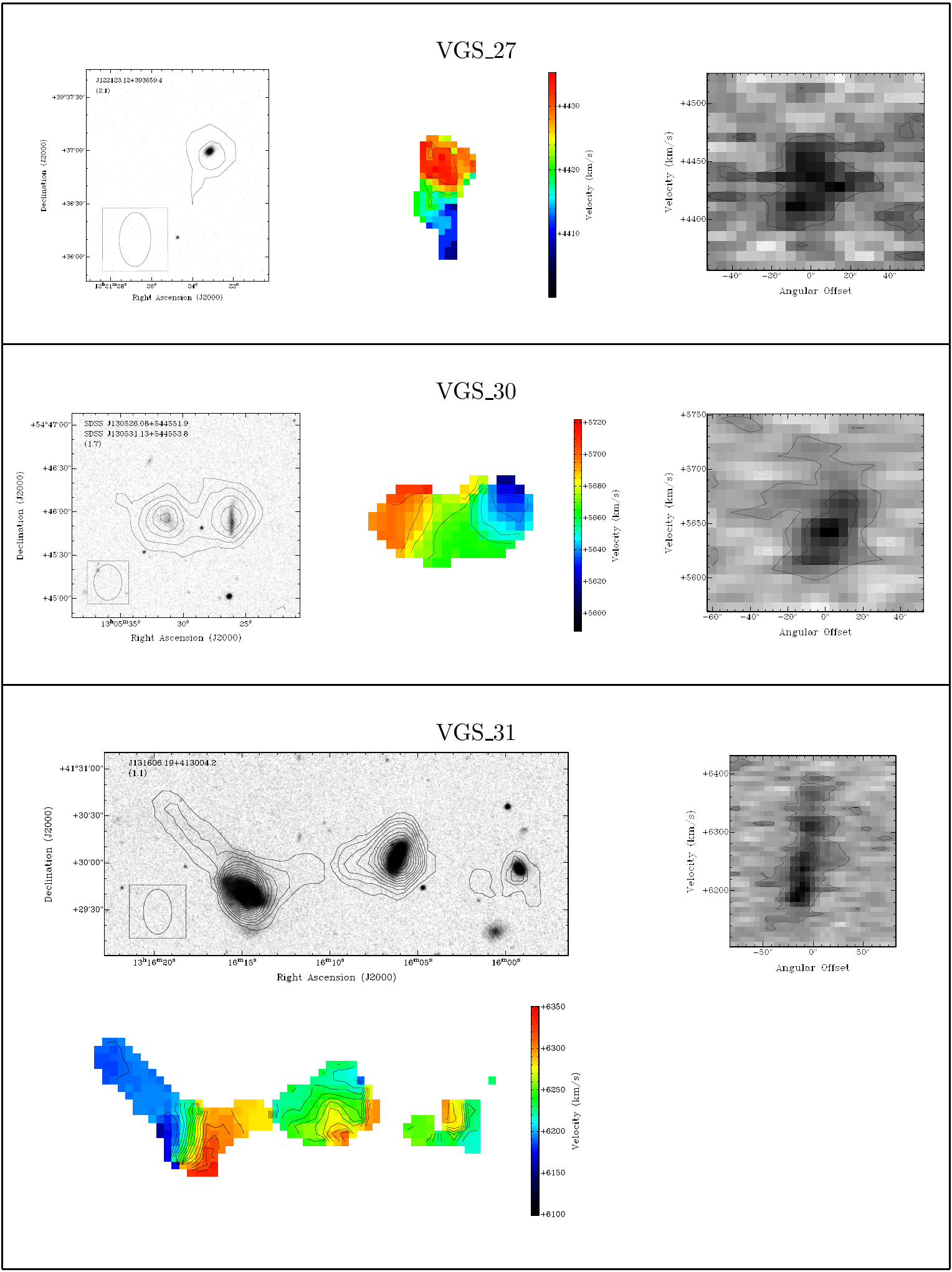}
\caption[]{(continued)}
\label{fig:voidgale}
\end{figure}

\begin{figure}
\centering
\includegraphics[width=0.85\textwidth]{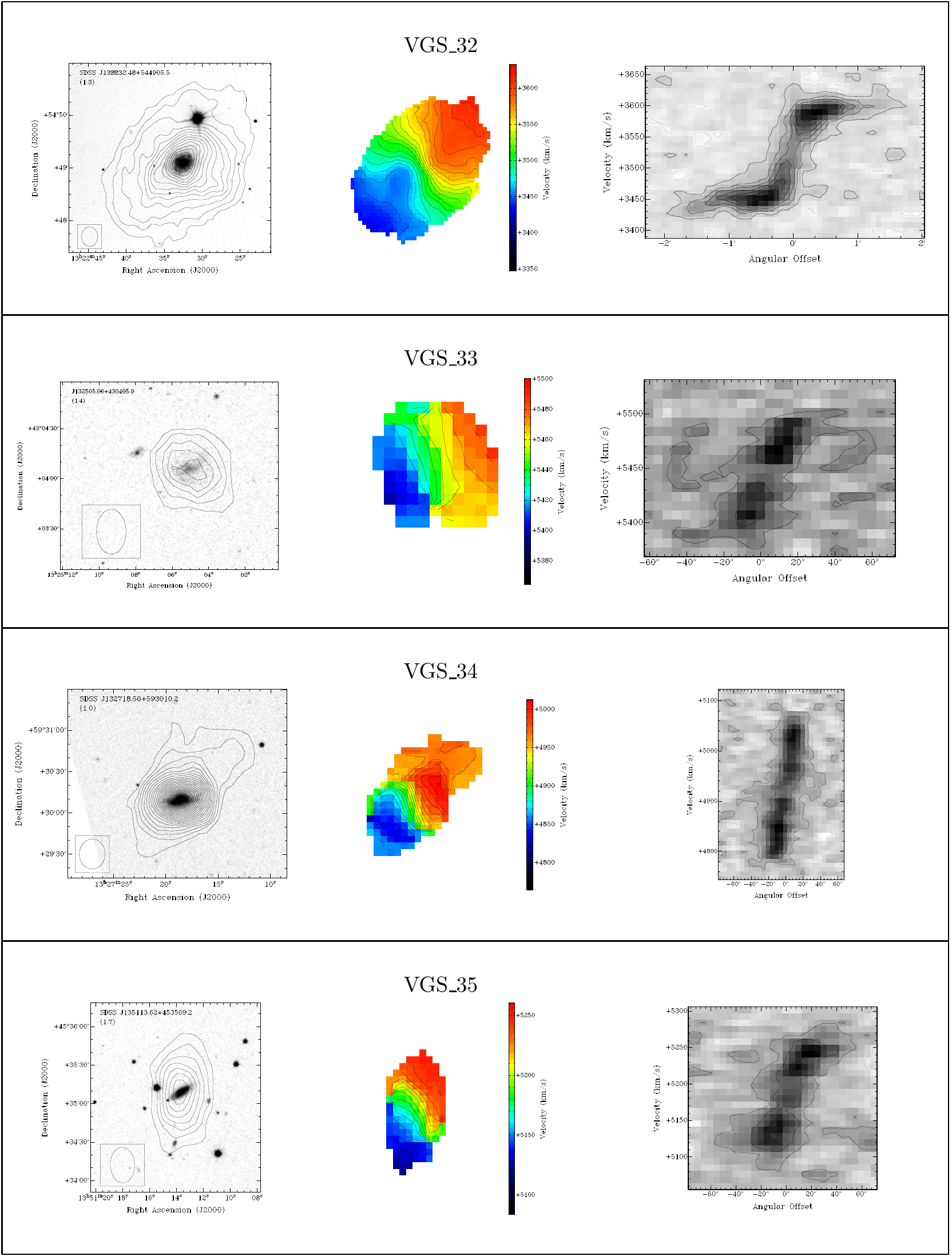}
\caption[]{(continued)}
\label{fig:voidgalf}
\end{figure}

\begin{figure}
\centering
\includegraphics[width=0.85\textwidth]{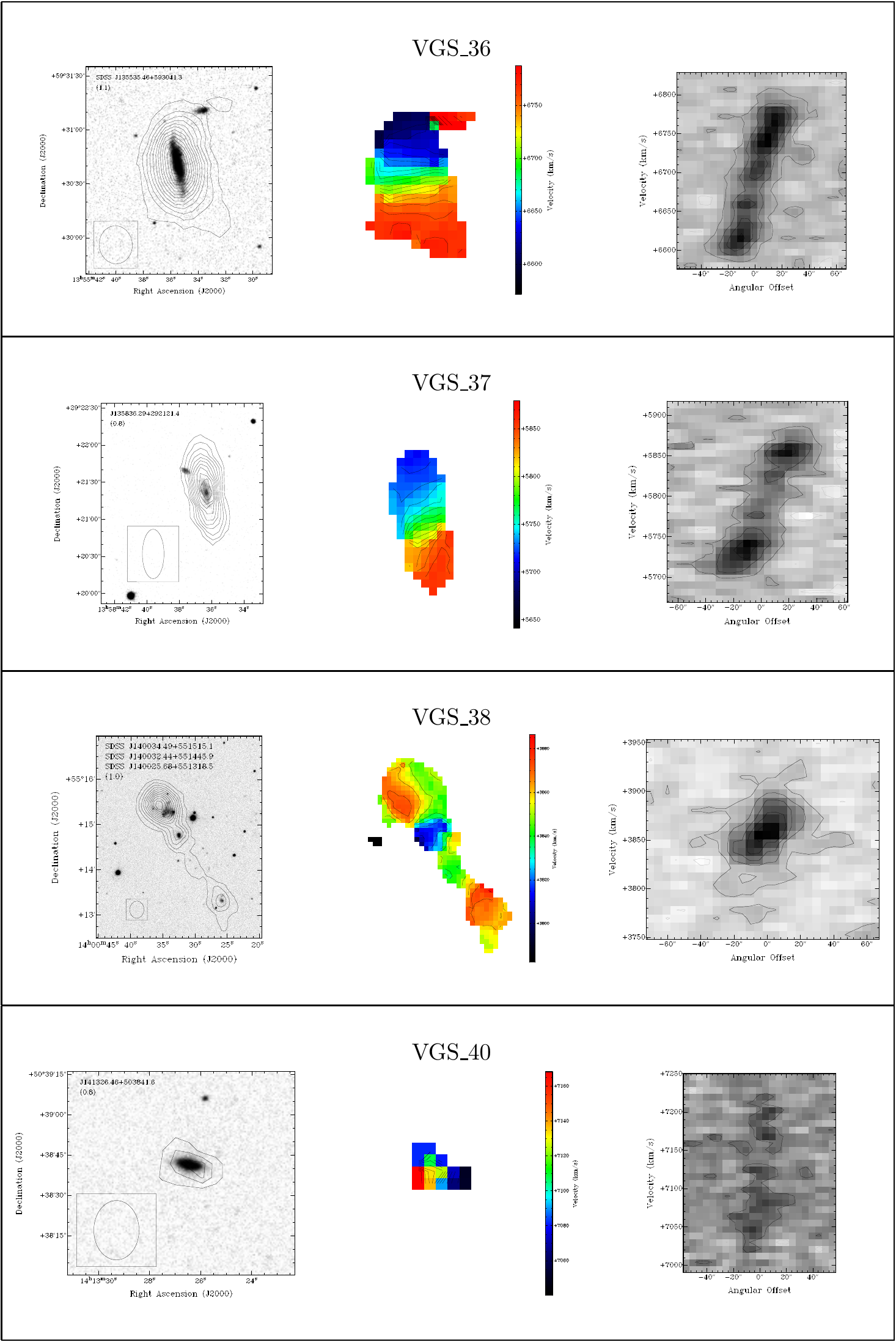}
\caption[]{(continued)}
\label{fig:voidgalg}
\end{figure}

\begin{figure}
\centering
\includegraphics[width=0.85\textwidth]{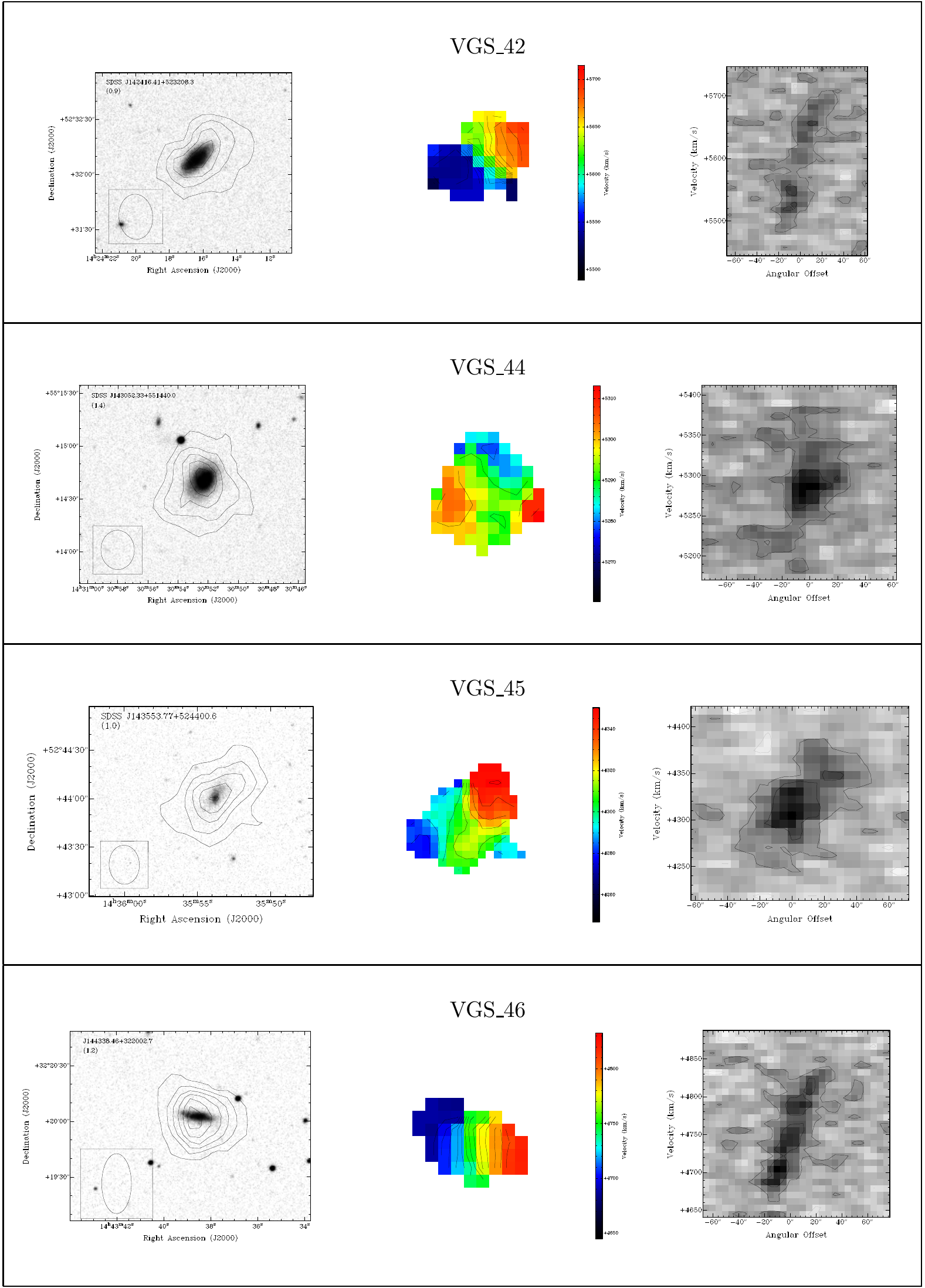}
\caption[]{(continued)}
\label{fig:voidgalh}
\end{figure}

\begin{figure}
\centering
\includegraphics[width=0.85\textwidth]{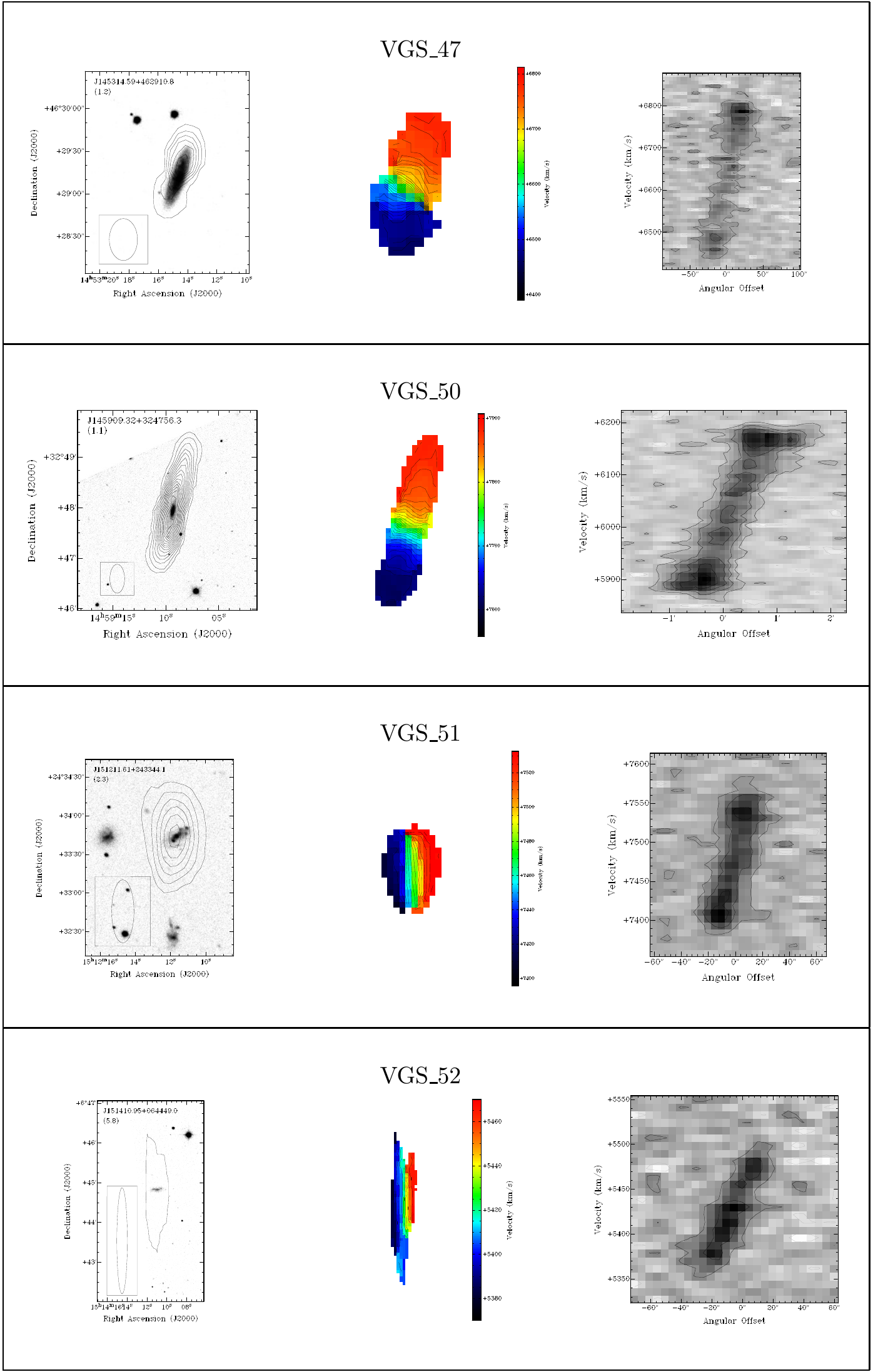}
\caption[]{(continued)}
\label{fig:voidgali}
\end{figure}

\begin{figure}
\centering
\includegraphics[width=0.85\textwidth]{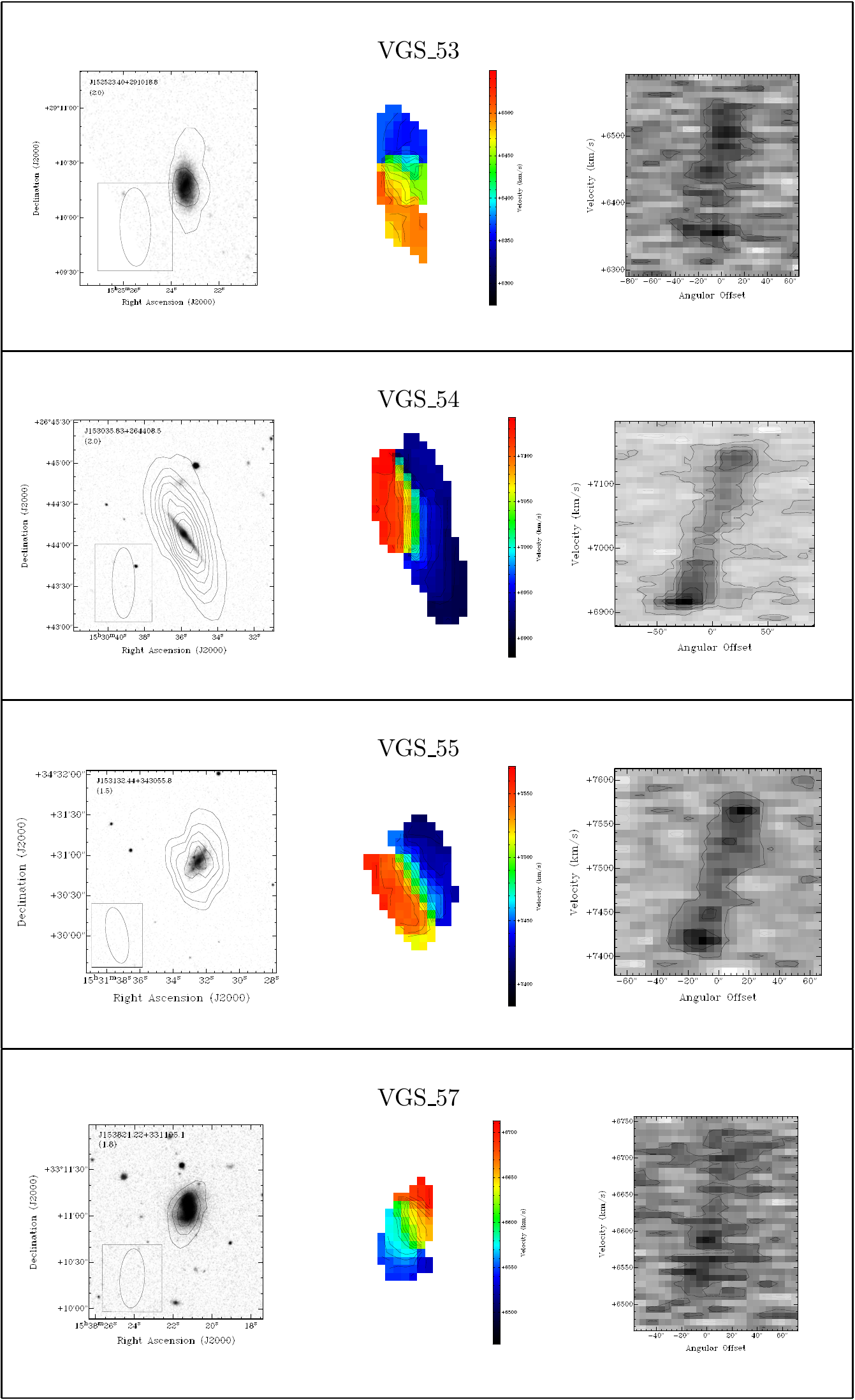}
\caption[]{(continued)}
\label{fig:voidgalj}
\end{figure}

\end{document}